\documentclass[12pt]{article} 
 \usepackage{amsmath}
\usepackage{amssymb}
\usepackage{graphicx}
\usepackage{tikz}
\usepackage{xcolor}
\usepackage{framed}
\usepackage{bm}
  \usepackage{subfigure}
  \usepackage{caption}
 \usepackage{mciteplus}
 \usepackage{bbold}
   \definecolor{darkblue}{rgb}{0.1,0.1,.7}
 \usepackage[colorlinks, linkcolor=darkblue, citecolor=darkblue, urlcolor=darkblue, linktocpage,hyperfootnotes=false]{hyperref} 
 
 \numberwithin{equation}{section}
 
  \newlength{\abstractwidth}
  \newcommand{\be}{\begin{equation}}
  \newcommand{\bea}{\begin{eqnarray}}
  \newcommand{\eea}{\end{eqnarray}}
  \newcommand{\ea}{\end{eqnarray}}
  \newcommand{\beq}{\begin{equation}}
  \newcommand{\ee}{\end{equation}}
  \newcommand{\eeq}{\end{equation}}

\DeclareMathOperator{\Tr}{Tr}

\newcommand*{\ZZ}{\mathbb{Z}}

\newcommand*{\QQ}{\mathbb{Q}}

\newcommand*{\rat}{\mathcal{Q}}
\newcommand*{\lb}{\langle}
\newcommand*{\rb}{\rangle}

\begin{document}

\begin{titlepage}
  \bigskip

  \bigskip\bigskip

  \bigskip

\begin{center}
{\Large \bf {The path integral of 3D gravity near extremality;}}\\
    \bigskip\bigskip
    \textit{\Large or,}\\
    \bigskip\bigskip
    {\Large \bf {JT gravity with defects as a matrix integral}}
\bigskip

\end{center}

\bigskip
  \begin{center}

Henry Maxfield\footnote{\texttt{hmaxfield@ucsb.edu}} and Gustavo J. Turiaci\footnote{\texttt{turiaci@ucsb.edu}}
  \bigskip \rm

{\small Department of Physics, University of California, Santa Barbara, CA 93106, USA}  \\
\rm

  \bigskip \rm
\bigskip

\rm

\bigskip
\bigskip

  \end{center}

\vspace{2.5cm}
  \begin{abstract}
  
  We propose that a class of new topologies, for which there is no classical solution, should be included in the path integral of three-dimensional pure gravity, and that their inclusion solves pathological negativities in the spectrum, replacing them with a nonperturbative shift of the BTZ extremality bound.  We argue that a two dimensional calculation using a dimensionally reduced theory captures the leading effects in the near extremal limit. To make this argument, we study a closely related two-dimensional theory of Jackiw-Teitelboim gravity with dynamical defects. We show that this theory is equivalent to a matrix integral. 

 \medskip
  \noindent
  \end{abstract}
\bigskip \bigskip \bigskip

  \end{titlepage}

   \tableofcontents


\newpage
\section{Introduction}

In the quest to understand quantum mechanical theories of gravity, one central problem is to distinguish low energy theories with consistent ultraviolet completions from those without. For theories with asymptotically anti de Sitter (AdS) boundary conditions, this question becomes particularly sharp, since a UV complete theory can be rigorously defined through the AdS/CFT correspondence \cite{Maldacena:1997re}, and we by now have a detailed understanding of the  conditions that a CFT must satisfy to be well-described by a local theory of gravity at low energies \cite{Heemskerk:2009pn, *Heemskerk:2010ty, ElShowk:2011ag, Fitzpatrick:2012cg, Camanho:2014apa}. Perhaps the simplest question to ask is whether a consistent quantum theory of \emph{pure} gravity exists.

While this question is simple to ask, it is much harder to make meaningful progress.  A natural way to attempt to construct a quantum theory of pure gravity is by quantizing the classical theory, for example using a path integral over metrics, but it is very challenging to make sense of this in most cases. We can ameliorate these difficulties by studying low-dimensional models, in which gravity simplifies enough that we can hope for a more complete understanding.

 We will begin with three dimensional gravity, for which an apparent obstruction to the existence of a theory of pure gravity was pointed out in \cite{Benjamin:2019stq}. Our first major result is to find new contributions to the path integral of pure 3D gravity which overcome this obstacle. Our three-dimensional considerations will lead us naturally to consider a class of two-dimensional models which generalize Jackiw-Teitelboim (JT) gravity by inclusion of additional `defect' degrees of freedom. Our second major result relates to this class of 2D models, which we study in their own right independent of their 3D origin. The path integral of JT gravity was recently solved exactly by Saad, Shenker and Stanford \cite{Saad:2019lba}, and we draw heavily from their work. Their main result was that JT gravity has a dual quantum mechanical description as a double-scaled matrix integral. We find that such a description continues to hold when we include a gas of defects in JT gravity. We show that a matrix integral dual to this gravitational theory exists and generalizes \cite{Saad:2019lba}. We also argue that this theory is equivalent to another 2D dilaton gravity theory with a modified dilaton potential, appearing from integrating out the gas of defects. 
 
 We now review the context of the paper and summarize our main results.
 
\subsection{The spectrum of 3D pure gravity}

Gravity in three dimensions has several simplifying features. In particular, all classical solutions are locally isometric to empty AdS$_3$, and perturbative graviton excitations are determined by the Virasoro symmetry of the theory. Using these simplifications, Maloney and Witten \cite{Maloney:2007ud} and Keller and Maloney \cite{Keller:2014xba} constructed a candidate partition function from a pure 3D gravity path integral and found the corresponding spectrum; we will refer to these as the `MWK' partition function and spectrum. They took a very natural approach to the path integral, by classifying all classical solutions with appropriate boundary conditions, computing their classical action and all perturbative corrections (which happen to be exact at one loop) and finally summing their contributions.\footnote{The sum is not convergent and must be regularised. The ambiguity arising from the choice of regulator does not appear to be relevant for the features discussed here \cite{Alday:2019vdr}.} The solutions in question are the $SL(2,\ZZ)$ Euclidean black holes \cite{Maldacena:1998bw}, generalizations of the BTZ black hole \cite{Banados:1992wn}. Besides the vacuum and its Virasoro descendants describing empty AdS$_3$ and perturbative graviton excitations, the resulting spectrum is supported on precisely the energies and spins for which BTZ black holes exist. However, the MWK density of states has several problematic features. First, the corresponding spectrum is continuous, which we will comment on later.\footnote{A noncompact theory has a continuous spectrum, but with density of states proportional to a formally infinite factor of the volume of target space. Here the spectrum is continuous but finite, which is not compatible with a quantum mechanical Hamiltonian, even for such a noncompact theory.} More seriously, it was recently observed by \cite{Benjamin:2019stq} that the density of states is also negative in a particular `near-extremal' regime.

To explain this feature, we write the most important contributions to the MWK density of states for energies close to the edge of the spectrum. For a given value of (integer quantized) angular momentum $J$, the density of states is nonzero for energies $E>|J|$ in appropriate units (in terms of CFT quantities, $E=h+\bar{h}-\frac{c-1}{12}$ and $J=\bar{h}-h$), corresponding to the extremality bound of BTZ black holes. For large $|J|$ and energies very close to this bound, the density of states behaves like 
\begin{equation}\label{eq:rhoMWK}
		\rho^{(MWK)}_J(E) \sim a_0e^{S_0(J)} \sqrt{E-|J|} + a_1(-1)^J e^{\frac{S_0(J)}{2}} \frac{1}{\sqrt{E-|J|}},
\end{equation}
where $S_0(J) \gg 1$ is the Bekenstein-Hawking entropy given by the area of the extremal BTZ event horizon in Planck units, and $a_0,a_1$ are constants. The first term, coming from the ordinary BTZ black hole, is positive and dominates in almost all regimes. The second term, which comes from the simplest class of $SL(2,\ZZ)$ black holes, is exponentially suppressed by a relative factor of $e^{-\frac{S_0(J)}{2}}$ but enhanced very close to the edge. It becomes important when $E-|J|$ is exponentially small, and for odd spins it causes $\rho^{(MWK)}_J(E)$ to become negative in that regime. Terms from other classes of $SL(2,\ZZ)$ black holes are negligible, since they have the same $(\sqrt{E-|J|})^{-1/2}$ scaling with energy, but grow much slower with spin.

In order to resolve	this negativity, \cite{Benjamin:2019stq} and \cite{Benjamin:2020mfz} proposed that the dynamics of the theory must be modified by introducing particle degrees of freedom. The correction from a matter loop running round the horizon of the BTZ black hole gives an additional contribution to the density of states proportional to $e^{\alpha_m S_0(J)}(\sqrt{E-|J|})^{-1/2}$, where for a scalar of mass $m$ we have $\alpha_m=1-4mG_N$.\footnote{Here, the mass means the local mass, for example appearing as the coefficient of worldline length in a particle Lagrangian. The energy  $E\sim\frac{1}{8G_N}(1-\alpha^2)$ as measured from infinity is lower due to screening from gravitational backreaction.} For sufficiently light particles, with $\alpha_m\geq \frac{1}{2}$, the density grows fast enough to cure the negativity. The matter can be very heavy, consisting of particles with Planckian mass $m\sim \frac{1}{8G_N}$ and larger, but it would nonetheless be preferable to have a consistent theory with only metric degrees of freedom.

We propose an alternative resolution which does not require a modification of the theory, but rather the inclusion of additional contributions to the path integral over metrics. These contributions require spacetime topologies for which no classical solution exists, which explains their absence from the MWK partition function. We must therefore perform an `off-shell' path integral over an appropriate space of configurations, rather than a saddle-point approximation. Integrals over off-shell configurations were prominent in the recent analysis of the two-dimensional theory of Jackiw-Teitelboim (JT) gravity \cite{Saad:2019lba}. We will draw heavily on this work, which will turn out to have an extremely close relationship with the three-dimensional problem we are studying.

Before explaining this, let us describe how the additional topologies in the path integral result in a positive density of states. As opposed to the matter proposal of \cite{Benjamin:2019stq,Benjamin:2020mfz}, our new contributions grow more slowly with spin, but are more singular than $(E-|J|)^{-1/2}$ at the edge. Including some of the most important corrections, we find an expansion for the density behaving like
\begin{equation}\label{eq:rhoJexpansion}
	\begin{aligned}
	\rho_J(E) \sim& a_0\, e^{S_0(J)} \sqrt{E-|J|} +  a_1 \frac{(-1)^J e^{\frac{S_0(J)}{2}}}{\sqrt{E-|J|}} 
	\\&+ a_2 (E-|J|)^{-3/2} + a_3(-1)^J e^{-\frac{S_0(J)}{2}} (E-|J|)^{-5/2} + \cdots \ .
	\end{aligned}
\end{equation} 
The $SL(2,\ZZ)$ black hole now appears as the first term in an infinite series of nonperturbative corrections. In the regime where $\rho^{(MWK)}$ became negative, the later terms in the expansion are no longer suppressed so all terms must be considered together. Fortunately, we are able to compute all terms including the $a_\#$ coefficients. Once we sum over them, the density of states to leading order in the expansion near the edge takes the simple form
\begin{gather}
		\rho_J(E) \sim a_0 e^{S_0(J)} \sqrt{E-E_0(J)}, \qquad E>E_0(J), \\
		E_0(J)-|J| \propto -(-1)^J e^{-\frac{S_0(J)}{2}}.
\end{gather}
The interpretation is that the black hole extremality bound $E_0(J)$ has received a nonperturbative, spin-dependent correction to its classical value $|J|$. The negativity in \eqref{eq:rhoMWK} is an artifact of truncating the binomial series for the square root and then evaluating outside its radius of convergence.

The other families of $SL(2,\ZZ)$ black holes, which we have so far disregarded, are themselves the first terms in a multinomial series contributing further nonperturbative shifts. These shifts take the form
\begin{equation}\label{eq:pqShift}
	\delta E_0(J) \sim \cos\left(2\pi\tfrac{p}{q} J\right) e^{-\frac{S_0(J)}{q}},
\end{equation}
where $\frac{p}{q}$ is a rational number labelling the $SL(2,\ZZ)$ black hole in question, and we have ignored $J$-independent prefactors; we have so far been discussing the leading order shift with $q=2$. 

 An extremely similar phenomenon arises from the perturbative one-loop contribution of matter \cite{Maxfield:2019hdt}, but here we see a nonperturbative version from pure gravity alone. This was mostly studied from a dual CFT perspective in \cite{Maxfield:2019hdt}, where it was shown to be generic phenomenon for chaotic CFTs from crossing symmetry and modular invariance. We will revisit pure gravity from this bootstrap perspective in \cite{MTwip}, explaining how our proposal is consistent with modular invariance without introducing states far below the BTZ threshold, and how the arguments of \cite{Benjamin:2019stq} are evaded.

A different negativity of the MWK density of states, taking the form of $-6$ states at the massless, spinless BTZ  threshold, was observed in the original work \cite{Maloney:2007ud}. Allowing additional topologies, we seem to lose semiclassical control entirely for these small black holes: all the topologies we study (and likely many more) seem to contribute at the same order, and the near-extremal limit we use to study them is no longer a good approximation. Some new insight to control these untamed fluctuations of topology is required to say anything definitive in this regime.

\subsection{3D gravity in the near-extremal limit}
 
We now describe the new contributions to the path integral that give rise to the additional terms in \eqref{eq:rhoJexpansion}. The effect we are studying is important only for black holes very close to extremality. In this regime, the black hole develops a near-horizon region with the geometry of a circle fibered over AdS$_2$. We can usefully describe the physics using a two-dimensional theory, obtained by reducing on the circle. In an ensemble of fixed spin $J$, the resulting theory is perturbatively equivalent to Jackiw-Teitelboim gravity \cite{Jackiw:1984je,*Teitelboim:1983ux, Almheiri:2014cka}, a theory which has received much recent attention \cite{Jensen:2016pah,*Maldacena:2016upp,*Engelsoy:2016xyb, Grumiller:2016dbn, Cvetic:2016eiv, Mertens:2017mtv,*Lam:2018pvp, Harlow:2018tqv, Blommaert:2018oro, Kitaev:2018wpr, *Yang:2018gdb, Iliesiu:2019xuh, Saad:2019lba}. The precise map between perturbative physics in JT gravity and in near-extremal rotating BTZ black holes was described in \cite{Ghosh:2019rcj}, and was rigorously shown to hold for a generic CFT with large central charge using bootstrap methods.

The new contributions to the 3D gravity path integral we find in this paper come from extending the study of this two-dimensional theory to include new nonperturbative effects. The crucial new ingredient is that the theory contains nonperturbative objects, which we call `Kaluza-Klein instantons'. These correspond to configurations which appear singular from the two-dimensional geometry, with a conical defect localized at a point in the 2D spacetime, but are smooth in three dimensions (see also \cite{Dabholkar:2014ema}). They are somewhat analogous to Kaluza-Klein monopoles \cite{Sorkin:1983ns}, which are smooth solutions to a five-dimensional gravity theory which become singular Dirac monopoles when viewed from four dimensions (or 6-branes in type IIA supergravity, which arise as KK monopoles from reduction of a smooth solution to 11-dimensional supergravity \cite{Townsend:1995kk}). The KK instantons correspond to locations in AdS$_2$ where the circle fibration degenerates, and are labeled by a rational number $\frac{p}{q}\in (0,1)\cap \QQ$. An instanton creates a conical defect with opening angle $\frac{2\pi}{q}$ in the two-dimensional geometry, but gives a smooth three-dimensional metric since a closed loop around the defect is accompanied by a $2\pi \frac{p}{q}$ rotation in the circle fiber. The corresponding three-dimensional topologies are known as Seifert manifolds.

The KK instantons tend to attract, so the only classical solutions in this class are those with a single instanton, and these correspond to the $SL(2,\ZZ)$ black holes. The new contributions to the path integral we are studying have multiple instantons. In particular, the terms with coefficient $a_k$ in \eqref{eq:rhoJexpansion} come from $k$ instantons labeled by $\frac{p}{q}=\frac{1}{2}$, and the shift in \eqref{eq:pqShift} comes from including the instantons labeled by $\frac{p}{q}$.

We compute the relevant path integrals in the two-dimensional theory, and argue that this gives a good approximation to the three-dimensional pure gravity result in the near-extremal regime. We concentrate on the leading order KK instantons with $q=2$. Contributions from $q>2$ include energy- and spin-independent prefactors that do not seem to be captured fully by the two-dimensional theory alone. Nevertheless, their contribution is subleading in the near extremal limit. A full understanding requires a full 3D calculation, which we leave for future work. 

 We now turn to a summary of the two-dimensional calculations.

 \subsection{JT gravity with defects} 
 
Our three-dimensional considerations have thus led us naturally to consider a modification of JT gravity, by including dynamical instantons which source conical defects. In section \ref{sec:JT}, we study this theory in its own right.

Such a defect is labeled by two parameters, $\alpha \in (0,1)$ which specifies the conical angle $2\pi(1-\alpha)$ it sources, and the amplitude $\lambda$ for a defect to appear in the path integral, defined more precisely in section \ref{sec:JT}. We may have several `flavors' of defect, each with their own value of $\alpha_i$ and $\lambda_i$, so for $N_{\rm F}$ flavors we have a $2N_{\rm F}$-parameter generalization of JT gravity. A single such defect on the disc was studied by \cite{Mertens:2019tcm}, but here we sum over contributions with any number of defects and any topology.
 
 The nonperturbative path integral of JT gravity was solved recently by Saad, Shenker and Stanford \cite{Saad:2019lba} (see also \cite{Stanford:2019vob}) in a genus expansion, and we directly generalize their approach to an expansion in both genus $g$ and number of defects $k$. Their solution requires the Weil-Petersson (WP) volumes of the moduli space of hyperbolic surfaces with geodesic boundaries of specified lengths $b$, computed in \cite{Mirzakhani:2006fta, *Mirzakhani:2006eta}. Including defects requires us to generalize this to a moduli space of surfaces with cone points. Fortunately, the relevant volumes have been studied already \cite{tan2004generalizations,do2006weilpetersson, do2011moduli}, and they are obtained simply from the usual WP volumes by inserting a boundary with $b=2\pi i \alpha$ for each defect (proven for $\alpha\leq \tfrac{1}{2}$, but conjectured more generally). This fact was recently used in a slightly different context by \cite{Cotler:2019nbi}.
 
We first compute the defect contributions to the spectrum at low energy, and show that they give rise to an expansion of the form \eqref{eq:rhoJexpansion}, which shifts the edge of the spectrum. This gives a 2D toy model of the Maloney-Witten partition function, whereby we include only classical solutions and loop corrections; for JT gravity, this means only the disc and the disc with the insertion of a single defect. This can produce negative contributions to $\rho(E)$, inconsistent with a unitary theory. This problem in 2D is resolved in the manner described above when we sum over an arbitrary number of defects.

We then explicitly compute the sum over any number of defects at genus zero, using an explicit formula for WP volumes derived in \cite{mtms}. For one boundary, this gives a closed form expression for the density of states at leading order in the genus expansion \eqref{eq:ExactDOS}. For multiple boundaries, the result matches precisely what is required for a matrix integral dual generalizing \cite{Saad:2019lba}. Using a theorem of \cite{Eynard:2007kz}, we prove that this extends to all orders in the genus expansion (as well as providing several explicit checks). 

We also argue this theory is also equivalent to a 2D dilaton gravity with a modified dilaton potential, obtained by summing the defects as an instanton gas. 
 

 \subsection{Outline of paper}
 
 The paper is organized as follows. In section \ref{Sec:PI3Dgrav} we study the path integral of 3D gravity near extremality from the perspective of a dimensionally reduced theory. This leads us to suggest a new class of topologies to include in the path integral. In section \ref{sec:JT} we study a 2D theory of JT gravity with a gas of defects, motivated by the 3D considerations. We solve the theory in an expansion in genus and number of defects, and construct a matrix integral dual. In section \ref{sec:backto3D} we apply what we learnt back to 3D gravity, and we argue that the two-dimensional theory is a good approximation in the most interesting regime. We conclude in section \ref{sec:discussions} with discussion and open problems.

\paragraph{Note:} A related analysis of JT gravity with defects and matrix models was carried out independently by E. Witten and appeared in \cite{Witten:2020wvy}.

\section{The path integral of 3D gravity} \label{Sec:PI3Dgrav}

Our main target of study will be the spectrum of three-dimensional pure gravity with negative cosmological constant. This is a very simple theory in several ways, giving hope that we might understand it completely, if it indeed exists as a consistent quantum theory. In particular, all classical solutions are locally isometric to empty AdS$_3$, due to the absence of local propagating degrees of freedom, so can be classified in some cases. In addition, it enjoys an infinite-dimensional Virasoro symmetry group, which determines the perturbative excitations around the vacuum. 

Nonetheless, we will find it profitable to simplify the theory yet further, by studying more closely the `S-wave' path integral over metrics with a $U(1)$ symmetry. After dimensional reduction on the symmetry direction, we will find a two-dimensional theory that we can solve. Fortunately, the issues with the path integral of 3D gravity appear in the near extremal limit, for which this reduction in a good approximation (up to a subtlety discussed in section \ref{sec:backto3D}).

\subsection{Dimensional reduction}

We begin by describing the two-dimensional action that results from a dimensional reduction of pure 3D gravity, closely following \cite{Ghosh:2019rcj} to which we refer for more details.

We start from the Einstein-Hilbert action in three dimensions:
\begin{equation}\label{eq:EH}
	I_\text{EH} = -\frac{1}{16\pi G_N} \left[\int d^3 x\sqrt{g_3}\left(R_3+\tfrac{2}{\ell_3^2}\right) +2\int_\partial d^2x \sqrt{\gamma}(\kappa_3-\tfrac{1}{\ell_3}) \right].
\end{equation}
The subscripts on the metric $g_3$, curvature $R_3$, AdS length $\ell_3$, and extrinsic curvature $\kappa_3$ in the boundary term distinguish them from the two-dimensional quantities we encounter later. We impose boundary conditions that fix the induced metric $\gamma$ on the boundary to be a flat torus parameterized by spatial angle $\varphi$ and Euclidean time $t_E$,
\begin{equation}
	\gamma = \epsilon^{-2} dt_E d\varphi, \qquad (t_E,\varphi)\sim (t_E,\varphi+2\pi)\sim (t_E+\beta,\varphi+\theta), 
\end{equation}
where $\epsilon$ is a holographic renormalisation parameter which is taken to zero. We choose dimensionless coordinate $\varphi,t_E$, so that $\epsilon^{-1}$ has units of length, while $\beta$ is dimensionless. With these boundary conditions, the resulting path integral is interpreted as the partition function of the theory,
\begin{equation}
	Z(\beta,\theta) = \Tr(e^{-\beta H-i\theta J}) = \Tr\left[ e^{2\pi i \tau(L_0-\tfrac{c}{24})-2\pi i \bar{\tau}(\bar{L}_0-\tfrac{c}{24})}\right]
\end{equation}
where $H$ is the Hamiltonian and $J$ the angular momentum. We have also written this in terms of the left- and right-moving conformal generators $L_0$, $\bar{L}_0$ and modular parameter $\tau = \tfrac{\theta+i\beta}{2\pi}$, $\bar{\tau} = \tfrac{\theta-i\beta}{2\pi}$.

The simplification of the S-wave path integral is to consider not completely general metrics, but only those which respect a $U(1)$ symmetry, which we will take to act as translation in the $\varphi$ direction on the boundary. This means that we can write the three-dimensional metric as
\begin{equation}\label{eq:ansatz}
	g_3 = g_2 + \Phi^2 (d\varphi+A)^2,
\end{equation}
in terms of a two-dimensional metric $g_2$, dilaton $\Phi$ and Kaluza-Klein gauge field $A$, which can be taken to be independent of the coordinate $\varphi$. The three-dimensional diffeomorphisms that leave the metric in this form consist of two-dimensional diffeomorphisms, as well as shifts in $\varphi$ generated by $\lambda \partial_\varphi$ for any function $\lambda$ of the two-dimensional coordinates. The latter acts as gauge transformations $A\mapsto A+d\lambda$, so that $A$ is a $U(1)$ gauge field, with compact gauge group due to the $2\pi$ periodicity of $\varphi$.

Given this ansatz, we can now write the action \eqref{eq:EH} in terms of the two-dimensional fields and integrate over $\varphi$, giving an Einstein-Maxwell-dilaton theory \cite{Achucarro:1993fd}
\begin{equation}\label{eq:EMD}
	I_\text{EH} = -\frac{2\pi}{16\pi G_N}\left[\int d^2 x\sqrt{g_2}\Phi\left(R_2 - \tfrac{1}{4}\Phi^2 F_{ab}F^{ab}+\tfrac{2}{\ell_3^2}\right) +2\int_\partial ds\Phi(\kappa_2-\tfrac{1}{\ell_3}) \right].
\end{equation}
The indices $a,b$ run over the two remaining coordinates, $F=dA$ is the field strength of the Kaluza-Klein gauge field, and $s$ a proper length parameter on the boundary. Expressing the three-dimensional boundary conditions in terms of two-dimensional fields, we find that we must have constant dilaton $\Phi=\epsilon^{-1}$ at the cutoff and the proper length of the boundary is $\epsilon^{-1}\beta$. Finally, we must fix the holonomy of the gauge field $\int A=\theta$ (this being the only gauge invariant information contained in the one-dimensional gauge field pulled back to the boundary).

This will not be the most convenient form in which to study the theory, due to the presence of $A$. Fortunately, the simplicity of two-dimensional gauge fields allows us to integrate it out (see \cite{Ghosh:2019rcj} and references therein). Furthermore, by changing boundary conditions the resulting effective action will in fact be local in terms of $\Phi$ and $g_2$. To see this, we rewrite the Maxwell piece of the action by introducing an auxiliary scalar field $\mathcal{J}$:
\begin{gather}
	I_\text{Maxwell} = \frac{1}{32G_N}\int d^2x \sqrt{g_2} \Phi^3 F_{ab}F^{ab} =  \int \left[i\mathcal{J} F +\tfrac{1}{2} \mu \mathcal{J}^2\right], \\
	\text{where }\mu = 8G_N \Phi^{-3} d^2x \sqrt{g_2} .
\end{gather}
If we integrate out $\mathcal{J}$, we get back to the original action, in the meantime finding that it is related to the field strength via
\begin{equation}\label{eq:FJ}
	F = i \mathcal{J} \mu.
\end{equation}
In particular, note that for real Euclidean three-dimensional geometries $\mathcal{J}$ will be imaginary.

Now, we may instead integrate out the gauge field, which imposes the constraint that $\mathcal{J}$ is constant, and the Maxwell action becomes
\begin{equation}\label{eq:MaxwellRewrite}
	I_\text{Maxwell} = i\mathcal{J}\theta + \tfrac{1}{2}\mathcal{J}^2 \int \mu.
\end{equation}
In fact, we see that gauge-invariance requires $\mathcal{J}$ to be not only constant but an integer, since $\theta$ is a $2\pi$-periodic variable.

This still leaves us to sum over possible values of $\mathcal{J}$. But we can in fact identify $\mathcal{J}$ with the angular momentum $J$, which suggests that it is nicer to work in an ensemble of fixed spin, rather than with a chemical potential $\theta$ for spin. This means that we would like to study the spin-$J$ partition function
\begin{equation}\label{eq:ZJ}
	Z_J(\beta) = \frac{1}{2\pi}\int_0^{2\pi}d\theta \,e^{i J \theta} Z(\beta,\theta),
\end{equation}
which requires us to introduce the boundary term
\begin{equation}\label{eq:bdryTerm}
	I_\partial=-\int_\partial i\mathcal{J} A = -i\theta J
\end{equation}
before integrating over $\theta$. This precisely imposes the constraint that $\mathcal{J}=J$ (a Neumann boundary condition fixing the asymptotic field strength $F$), leaving behind the local effective action $\tfrac{1}{2}J^2\int \mu$ for the Maxwell field.

We thus find a theory of metric and dilaton only, governed by the two-dimensional action
\begin{equation}\label{eq:einsteinDilaton}
	I_\text{EH} = -\frac{2\pi}{16\pi G_N}\left[\int d^2 x\sqrt{g_2}\left(\Phi R_2 - \tfrac{(8G_N J)^2}{2}\Phi^{-3}+\tfrac{2}{\ell_3^2} \Phi\right) +2\int_\partial ds\Phi(\kappa_2-\tfrac{1}{\ell_3}) \right].
\end{equation}

However, this two-dimensional action does not quite capture the whole theory. We will find that we must include additional nonperturbative dynamical objects in the two-dimensional theory. To see why, we next examine the classical solutions of the three-dimensional theory.

\subsection{The classical solutions: $SL(2,\ZZ)$ black holes.}

The finite action classical solutions of three-dimensional gravity with torus boundary admit a complete classification \cite{Maloney:2007ud}. They fall into a discrete set of topological classes, with one solution in each class respecting the full $U(1)\times U(1)$ symmetry of the boundary torus. These solutions are usually called the `$SL(2,\ZZ)$ black holes', for reasons that will become clear. The other solutions add perturbative excitations of boundary gravitons, obtained by acting with an asymptotic symmetry generator, a diffeomorphism that does not fall off sufficiently rapidly to be regarded as part of the gauge symmetries.

Topologically, the $SL(2,\ZZ)$ black holes are simply solid tori; there are many such solutions, since we have a choice of which boundary cycle to fill in. Different choices are related by the $SL(2,\ZZ)$ mapping class group of the torus. To see this explicitly, we first use complex coordinates to write the torus as the complex plane identified by translations on a lattice,
\begin{equation}
	z= \frac{\varphi+i t_E}{2\pi},\qquad z\sim z+1\sim z+\tau,\qquad \tau=\frac{\theta+i\beta}{2\pi}.
\end{equation}
These coordinates pick out the spatial circle as a line running from the origin to $z=1$. But we can choose a different coordinate $w$ to pick out any simple closed curve in this way, running from the origin to $c\tau+d$ for coprime integers $c,d$ (and by choice of orientation we may set $c>0$):
\begin{equation}
	\tilde{z}= \frac{z}{c\tau+d},\qquad \tilde{z}\sim \tilde{z}+1\sim \tilde{z}+\tilde{\tau},\qquad \tilde{\tau}=\frac{a\tau+b}{c\tau+d}.
\end{equation}
Having picked the first generator of the lattice to be $c\tau+d$\footnote{The factor $c$ here parametrizes the $SL(2,\ZZ)$ transformation, not to be confused with the central charge.}, we can choose the second generator as $a\tau+b$, where in order that these form a basis we must choose $a,b$ such that $a d-b c=1$. We thus have $\left(\begin{smallmatrix}a&b\\c&d
\end{smallmatrix}\right)\in SL(2,\ZZ)$.

Now, to construct a Euclidean $SL(2,\ZZ)$ black hole, we simply define new coordinates as real and imaginary parts of $\tilde{z}$  through $\tilde{z}=\frac{1}{2\pi}(\tilde{\varphi}+i \tilde{t}_E)$, write the real and imaginary parts of the $\tilde{z}$ lattice identification as $\tilde{\tau} = \frac{1}{2\pi}(\tilde{\theta}+i\tilde{\beta})$, and write the metric of AdS$_3$ with $\tilde{\varphi}$ chosen as the spatial circle:
\begin{equation}\label{eq:SL2Z1}
	\ell_3^{-2} ds^2 = (1+\tilde{r}^2)d\tilde{t}_E^2 + \frac{d\tilde{r}^2}{1+\tilde{r}^2}+ \tilde{r}^2d\tilde{\varphi}^2, \qquad (\tilde{\varphi},\tilde{t}_E) \sim (\tilde{\varphi}+2\pi,\tilde{t}_E)\sim (\tilde{\varphi}+\tilde{\theta},\tilde{t}_E+\tilde{\beta}).
\end{equation}
Note that the choice of $a$ and $b$ (which is ambiguous up to the addition of a multiple of $c$ and $d$ respectively) does not affect this metric, since it acts only by shifting $\tilde{\theta}$ by a multiple of $2\pi$. The solutions are thus classified not by $PSL(2,\ZZ)$ elements, but by the coset $PSL(2,\ZZ)/\ZZ$. Equivalently, they are classified by the rational number $\rat=-\tfrac{d}{c}$ which gets mapped to infinity by $\rat\mapsto \frac{a\rat+b}{c\rat+d}$. The solutions are thus in one-to-one correspondence with rational numbers $\rat\in\QQ\cup\{\infty\}$, where infinity accounts for AdS$_3$ itself, corresponding to $c=0$. 

Now, to understand this solution in terms of the dimensionally reduced theory we would like to express the metric \eqref{eq:SL2Z1} in terms of the original untilded coordinates $\varphi,t_E$. To do this, we simply make the relevant substitutions and define a new radial coordinate $r$ as the coefficient of $d\varphi^2$. It is convenient to write the solution in terms of new parameters $r_\pm$, related to $\beta,\theta$ or $\tau,\bar{\tau}$ by
\begin{gather}\label{eq:rpm}
	\beta = \frac{2\pi}{c}\frac{r_+}{r_+^2-r_-^2},\qquad \theta =-2\pi\frac{d}{c}+ \frac{2\pi i}{c} \frac{r_-}{r_+^2-r_-^2}, \\
	\text{or}\quad \tau = \frac{i}{c(r_+-r_-)}-\frac{d}{c},\quad \bar{\tau} = -\frac{i}{c(r_++r_-)}-\frac{d}{c},
\end{gather}
where  $r_+$ is positive, and $r_-$ is imaginary for the Euclidean solutions of \eqref{eq:SL2Z1} with real $\theta$. We then have
\begin{gather}\label{eq:SL2Z2}
	\ell_3^{-2}ds^2 = \frac{dr^2}{f(r)}+f(r)dt_E^2 + r^2\left(d\varphi-i\frac{r_+r_-}{r^2}dt_E\right), \\
	\text{where } f(r)=\frac{(r^2-r_+^2)(r^2-r_-^2)}{r^2}, \qquad (t_E,\varphi)\sim (t_E,\varphi+2\pi)\sim (t_E+\beta,\varphi+\theta).
\end{gather}
This result takes the familiar form of the metric of the BTZ black hole \cite{Banados:1992wn}. Indeed, for $c=1,d=0$ this is the BTZ black hole, in Euclidean signature for imaginary $r_-$, and Lorentzian signature for real $r_-$ when we Wick rotate the time coordinate. However, it is not quite the BTZ black hole for $c\neq 1$, since we have slightly different identifications.

This metric is manifestly in the form of our ansatz \eqref{eq:ansatz}, so we may immediately write it down in terms of two-dimensional fields. Before we do this, we first address the subtlety that in the current form we have a nontrivial gauge bundle at infinity: due to the nontrivial identification $(t_E,\varphi)\sim (t_E+\beta,\varphi+\theta)$, when we go around the spatial circle we must also perform a gauge transformation. We can remove this by using a new angular coordinate $\varphi-\frac{\theta}{\beta}t_E$ which does not have such a shift when we go around the Euclidean time circle. From the two-dimensional perspective, this acts as a gauge transformation with gauge parameter $\lambda = \frac{\theta}{\beta}t_E$. With this understood, we can write our solution as follows:
\begin{align}
	g_2 &= \ell_3\left[\frac{dr^2}{f(r)}+f(r)dt_E^2\right],\\
	\Phi &= \ell_3 r,  \\
	A &= \left(\frac{\theta}{\beta}-i\frac{r_+r_-}{r^2}\right)dt_E.
\end{align}
The gauge field $A$ can be reconstructed (at least locally; we will see later how the holonomy arises) from the relation between the angular momentum $J$ and the field strength $F$ in \eqref{eq:FJ}. The volume form is given by $dr\wedge dt_E$, so we can write
\begin{equation}
	F = dA = 2i\frac{r_+r_-}{r^3}  dr\wedge dt_E = i\frac{\ell_3}{4G_N} r_+r_-\mu \implies J = \frac{\ell_3}{4G_N}r_+r_- \,.
\end{equation}
This is the usual expression for the angular momentum of the BTZ black hole, which can also be read off from the asymptotics of the three-dimensional metric. Once again, $J$ is imaginary for the Euclidean solution.

All seems well so far. But if we look more closely, we will see that this solution is apparently singular! To see this, we examine the metric near the origin $r=r_+$, where $f$ has its largest zero. Using a new coordinate $\rho$ defined as $r-r_+\sim \frac{r_+^2-r_-^2}{2r_+}\rho^2$ and the expression \eqref{eq:rpm} for $\beta$ in terms of $r_\pm$, we find
\begin{equation}
	\ell_3^{-2}g_2 \sim d\rho^2 + \frac{1}{c^2}\rho^2\left(\frac{2\pi}{\beta}dt_E\right)^2.
\end{equation}
Recalling that $t_E$ has period $\beta$, so that $\frac{2\pi}{\beta}t_E$ is an angular variable with period $2\pi$, we see that $g_2$ is smooth only if $c=1$. For larger $c$, the metric close to $\rho=0$ looks like a cone with opening angle $\frac{2\pi}{c}$.

To see how this is possible, given that we started with a smooth three-dimensional solution, we must also look at the gauge field near the conical defect. We have
\begin{equation}
	A \sim -2\pi\frac{d}{c}\,\frac{dt_E}{\beta},
\end{equation}
so that a small closed curve surrounding this point has nontrivial holonomy $\int A=-2\pi \frac{d}{c}$. Equivalently, the field strength $F$ has a delta-function source of this strength at $\rho=0$. This means that traversing a small loop around the origin in the two dimensional metric does not give rise to a closed curve in the three-dimensional space; rather, there is a nontrivial translation in the angular $\varphi$ direction by the holonomy. A closed curve only arises after going round $c$ times. A nice description of this is also given in \cite{Dabholkar:2014ema}.

Before giving a complete two-dimensional description of these defects, we note one important feature of the two-dimensional solution. By making large gauge transformations that wind around the circle, $A\mapsto A+\frac{2\pi n}{\beta}dt_E$ for integer $n$, we can shift $d$ by multiples of $c$. This moves us between isometric but inequivalent gravitational solutions, with $\theta\mapsto \theta+2n\pi$; they are not gauge equivalent because the identification with the boundary metric is different. We can simply sum over these solutions (or, more generally, any configurations related in the same way), which can be accounted for simply by quantizing spin. Given some contribution $\mathfrak{z}(\beta,\theta)$ to the path integral, we can rewrite a sum over its images $\theta\mapsto \theta+2n\pi$ using the Poisson summation formula,
\begin{equation}
	\sum_n \mathfrak{z}(\beta,\theta+2n\pi) = \sum_J e^{iJ\theta}\frac{1}{2\pi}\int d\vartheta e^{-i\vartheta J}\mathfrak{z}(\beta,\vartheta),
\end{equation}
and we see that the Poisson dual variable is precisely the angular momentum.

We thus see that in the fixed spin ensemble \eqref{eq:ZJ} with integer spin, it is sufficient to restrict to rational $\rat=-\frac{d}{c}$  in the interval $[0,1)$ (with $\rat=0$ corresponding to the usual BTZ black hole). Under the classification which identifies geometries only up to identification by such shifts $\theta \mapsto \theta+2n\pi$, the classes correspond to elements of the double coset $\ZZ\backslash PSL(2,\ZZ) /\ZZ$, where the groups of integers on the left and right both act by $\tau\mapsto \tau+1$. This is also the classification of solutions of a given angular momentum and either mass or temperature, which is more pertinent for us since we work an ensemble of fixed spin.

We note that the solutions we have been led naturally to consider, using the ensemble of fixed integer $J$, have complex Kaluza-Klein gauge field. These are real Euclidean solutions for the two-dimensional metric and gauge field, but have complex three-dimensional Euclidean metric. The off-shell configurations we generalize to later will be of the same nature. The $SL(2,\ZZ)$ black holes (though not the later generalizations) have real three-dimensional Lorentzian sections, which are nothing but the usual BTZ metric as can be seen from \eqref{eq:SL2Z2}.

To complete our review of the $SL(2,\ZZ)$ black hole solutions, we record their contribution to the density of states. Since we will be studying the theory exclusively in the near-extremal regime, we will give the answer only in that limit, taking
\begin{equation}
	|J| \gg \frac{\ell_3}{G_N}, \qquad 0<E-|J| \ll \frac{\ell_3}{G_N}.
\end{equation}
This means that the temperature is low and the horizon is large when measured in AdS units. For the exact result, see \cite{Benjamin:2019stq,Benjamin:2020mfz}.

First, we define $S_0(J)$ as the Bekenstein-Hawking entropy given by the area of the classical extremal BTZ black hole:\footnote{Since $G_N$ has a renormalisation ambiguity from gravitational quantum corrections, we define it more precisely via the dual CFT central charge $c$: we choose $c-1 = \frac{3\ell_3}{2G_N}$, which is the Brown-Hennaux relation with a natural one-loop shift.}
\begin{equation}
	S_0(J) = 2\pi \sqrt{\tfrac{\ell_3}{4G_N}|J|} \ .
\end{equation}
Now, we must treat the usual BTZ black hole solution as a special case in this limit. It contributes to the density of Virasoro primary states of spin $J$ as
\begin{equation}\label{eq:rhobtw1}
	\rho_J(E) \supset 2\pi \sqrt{\frac{4G_N}{\ell_3 |J|}}\ e^{S_0(J)} \ \sinh\left(2\pi \sqrt{\tfrac{\ell_3 }{8G_N}(E-|J|)}\right). \qquad \text{(BTZ)} 
\end{equation}
The other families of $SL(2,\ZZ)$ black holes, coming in families labeled by the rational parameter $\mathcal{Q}=\frac{p}{q} \in (0,1)\cap\QQ$, contribute as
\begin{equation}\label{rhobtw1b}
	\rho_J(E) \supset  \frac{\ell_3}{8G_N q} e^{-2\pi i \frac{p}{q} J}\left(1-e^{2\pi i \frac{(p^{-1})_q}{q}}\right) \sqrt{\frac{4G_N}{\ell_3 |J|}}  e^{\frac{1}{q}S_0(J)} \frac{\cosh\left(\frac{2\pi}{q} \sqrt{\frac{\ell_3 }{8G_N}(E-|J|)}\right)}{\sqrt{\tfrac{\ell_3 }{8G_N}(E-|J|)}}\ ,
\end{equation}
where $(p^{-1})_q$ denotes the inverse of $p$ modulo $q$, that is the solution to $p(p^{-1})_q -1 \in q\ZZ$. In particular, for $\rat = \tfrac{1}{2}$ we have
\begin{equation}\label{eq:rhobtw2}
	\rho_J(E) \supset  \frac{\ell_3}{8G_N} (-1)^J \sqrt{\frac{4G_N}{\ell_3 |J|}}  e^{\frac{1}{2}S_0(J)} \frac{\cosh\left(\pi \sqrt{\frac{\ell_3 }{8G_N}(E-|J|)}\right)}{\sqrt{\tfrac{\ell_3 }{8G_N}(E-|J|)}}\ .
\end{equation}

\subsection{Kaluza-Klein instantons}

From the $SL(2,\ZZ)$ black holes, we have seen that smooth configurations in the three-dimensional theory give rise to particular singular objects in the dimensionally reduced theory, which are localized in spacetime. We call these objects Kaluza-Klein instantons, in analogy to the Kaluza-Klein monopoles which arise from compactification from five to four spacetime dimensions \cite{Sorkin:1983ns}. We discuss here their incorporation into the two-dimensional theory.

We have a countably infinite set of species of instantons labeled by rational numbers strictly between zero and one, $\rat\in (0,1)\cap \QQ$; in the previous section we had $\rat=-\frac{d}{c}$, but here we will write $\rat=\frac{p}{q}$ ($1\leq p <q$, with $p$ and $q$ coprime). They provide a delta-function source for $F$ of strength $2\pi \frac{p}{q}$, and create a conical defect of angle $\frac{2\pi}{q}$. Since they are smooth in the three-dimensional geometry, they should not provide any singular contribution to the action once these boundary conditions are obeyed.

We will take these boundary conditions at the defect as a definition of a KK instanton, but it can also be helpful to understand them from an action formulation. To impose the boundary conditions for a KK instanton, we can add an action
 \begin{equation}\label{eq:instantonAction}
 	I_{p/q} = \left(1-\tfrac{1}{q}\right)\frac{2\pi\Phi}{4G_N} - 2\pi i \mathcal{J}\tfrac{p}{q}
 \end{equation}
to the Einstein-Maxwell-dilaton action \eqref{eq:EMD} after rewriting the Maxwell theory in terms of the auxiliary field $\mathcal{J}$ using \eqref{eq:MaxwellRewrite}. Here, the dilaton $\Phi$ is evaluated at the location of the instanton. In concert with the $\Phi R_2$ term in the dilaton action, this sources a conical defect of the desired strength at the instanton's location. This conical defect gives a delta-function contribution to the curvature $R_2$, which can be computed by the Gauss-Bonnet theorem applied to a small disc containing the defect. The resulting singular piece of the action \eqref{eq:einsteinDilaton} precisely cancels the dilaton term in the instanton action \eqref{eq:instantonAction}. Due to this cancellation, the overall contribution to the action from this dilaton piece is zero as desired.

The term $-2\pi i \mathcal{J} \frac{p}{q}$ accounts for the Maxwell piece of the action and induces the correct source for the field strength. With this addition, the relation \eqref{eq:FJ} is modified to
\begin{equation}
	F = i \mathcal{J} \mu + 2\pi \frac{p}{q} \delta,
\end{equation}
including a source (with $\delta$ a two-form delta-function defined so that $\int \delta f$ for a scalar function $f$ gives the value of $f$ at the location of the instanton). We still have that $\mathcal{J}$ is a constant, which our boundary conditions set equal to the desired spin $J$. After integrating out the gauge field we can therefore simply replace the auxiliary field $\mathcal{J}$ by the integer $J$ chosen by boundary conditions. We can think of the resulting action $-2\pi i J \frac{p}{q}$ as a contribution to the boundary term \eqref{eq:bdryTerm} $I_\partial = -i\theta J$ inserted to change to the fixed spin ensemble, coming from the additional holonomy sourced by the instanton.
 
Since the action \eqref{eq:instantonAction} is of order $\frac{1}{G_N}$, the instantons are nonperturbative, solitonic `D-instanton'-like objects. An analogous phenomenon occurs in type IIA supergravity, with solitonic D6 branes arising as the Kaluza-Klein monopole upon compactification of a smooth solution of 11-dimensional supergravity \cite{Townsend:1995kk}.
 
The advantage of formulating this two-dimensional theory with KK instantons is that we can now consider the effect of including multiple instantons in the path integral. We do not obtain new classical solutions in this way, since the instantons tend to attract to minimize the action. Nonetheless, we can in several circumstances perform the path integral over off-shell configurations. The resulting three-dimensional topologies are known as Seifert manifolds.

The action \eqref{eq:instantonAction} defines the KK instantons classically, but a full quantum mechanical treatment requires more input. In particular, at one loop we must specify an amplitude for the insertion of each instanton, or a measure for integrating over the locations at which we may insert it (see section \ref{sec:instGas}). This amplitude or measure may depend nontrivially on background fields, in particular the dilaton. Properly determining this requires a full three-dimensional analysis. However, in the near-extremal limit we describe in a moment, the geometry approaches a rigid AdS$_2$ background with the dilaton approximately constant, so any natural measure appears to reduce to a simple preferred choice. Only a normalization constant (which can depend on $J$) then remains unfixed, and this can be determined by matching to the $SL(2,\ZZ)$ black holes. A more careful understanding of this is an important problem for the future.

\subsection{Near-extremal limit}\label{sec:nearExt}

One regime in which we are able to study the Kaluza-Klein instantons in detail is the near-extremal limit of rapidly rotating black holes at low-temperature. We will see that, as well as being tractable, this is in fact the regime where they give rise to important physical effects.

As an example of a more generic phenomenon, our Einstein-dilaton theory \eqref{eq:einsteinDilaton} reduces in the near-extremal limit to Jackiw–Teitelboim gravity. We first very briefly review this, referring the reader to \cite{Ghosh:2019rcj} and references therein for details.

We can begin by looking for AdS$_2$ solutions of the theory. These must have constant dilaton at a zero of the dilaton potential, which for us sets
\begin{equation}
	\Phi_0 = \sqrt{4 G_N \ell_3 J}.
\end{equation}
The two-dimensional AdS radius is then set by the gradient of the potential at $\Phi=\Phi_0$, giving the AdS$_2$ radius $\ell_2 = \tfrac{1}{2}\ell_3$. This AdS$_2$ space is the near-horizon geometry of a near-extremal rotating BTZ black hole (completed to the three-dimensional `self-dual orbifold' geometry \cite{Coussaert:1994tu}, which is the analogue of near horizon extreme Kerr). To incorporate fluctuations away from this AdS$_2$ geometry, we write $\Phi = \Phi_0+4G_N \phi$ and expand to linear order in $\phi$. This leads us to the action of JT gravity,
\begin{equation}
	I_\text{JT}= -S_0 \chi - \frac{1}{2}\int d^2 x \sqrt{g_2} \phi\left( R_2 +\frac{2}{\ell_2^2}\right)-\int_\partial ds\phi(\kappa_2-\tfrac{1}{\ell_2}).
\end{equation}
The first term here is the two-dimensional Einstein-Hilbert action, which is topological, proportional to the Euler characteristic $\chi$. Its coefficient $S_0$ is the extremal entropy of BTZ,
\begin{equation}
	S_0 = 2\pi \sqrt{\frac{\ell_3 J}{4G_N}},
\end{equation}
which we take to be very large. The final boundary term is introduced at an artificial location where the value of the dilaton is set at a constant $\phi_\partial$ satisfying $1\ll \phi_\partial \ll S_0$, demarking the edge of the `near-horizon' region. Inside this near-horizon region, the dilaton is sufficiently small that the linearized approximation leading to JT gravity is valid. Outside this region, quantum corrections are suppressed by the large parameter $S_0$, so we may solve the theory classically.

With our fixed $J$ boundary conditions, this classical solution from the region far from the black hole has two important effects. First, it contributes an action $\beta J$, which shifts the energy by the value of extremal BTZ, $M=J$. Second, it induces boundary conditions at our artificial boundary $\partial$, setting the ratio of the proper length $L_\partial$ of the boundary to the dilaton value $\phi_\partial$ (with both taken very large):
\begin{equation}
	\frac{L_\partial}{\phi_\partial} = \frac{\ell_2}{\gamma}\beta, \qquad \text{where}\quad \gamma = \frac{\ell_3}{16G_N}.
\end{equation}
The parameter $\gamma$ sets the energy or temperature $\gamma^{-1}$ below which quantum fluctuations become strong. In the Schwarzian description of the theory \cite{Jensen:2016pah}, $\gamma$ appears as the coefficient of the Schwarzian action.

Now we simply need to incorporate the Kaluza-Klein instantons into this near-extremal theory. We can do this by simply rewriting the instanton action \ref{eq:instantonAction} in terms of the JT variables:
 \begin{equation}\label{eq:3DJTinstanton}
 	I_{p/q} = \left(1-\tfrac{1}{q}\right)S_0 + 2\pi\left(1-\tfrac{1}{q}\right)\phi - 2\pi i J\tfrac{p}{q}.
 \end{equation}
We will study the theory of JT gravity with such defects in detail in section \ref{sec:JT}.

\subsection{Instanton gas}\label{sec:instGas}

Before turning to the analysis of JT gravity with defects outlined above, we mention a possible alternative approach, summing the KK instantons as an instanton gas. This incorporates the effect of the instantons into a shift of the dilaton potential. This method is usually introduced as an approximation, taking a sum over instantons in a limit that they are well-separated and do not interact. Here, however, if we think of our instantons as fundamental pointlike objects in the two-dimensional theory, there is no such approximation required in principle.

Note first that we can introduce a single instanton by an insertion into the path integral of
\begin{equation}
	\int d^2x \sqrt{g_2}\, \Omega(x) \exp\left[-I_{p/q}(x)\right],
\end{equation}
where $I_{p/q}(x)$ is the action \eqref{eq:instantonAction} for an instanton at the point $x$, and we integrate over all possible insertion points. We have included a measure $\Omega$, which is possibly a function of fields in the theory. This is either taken as part of the definition of the instanton, or must be determined if we use a definition independent of the action formulation, as is the case for our theory obtained from dimensional reduction. The measure is of subleading importance in the classical limit, and $\Omega$ should approach a constant in the near-extremal limit, so we will be content to leave $\Omega$ as an unknown function in this section's more qualitative discussion.

Now, we can include many instantons by simply inserting many such factors into the path integral. To avoid overcounting equivalent configurations from permuting instantons, we must divide the $k$-instanton contributions by $k!$. The sum over all sectors then exponentiates:
\begin{equation}
	\begin{aligned}
		\sum_{n=0}^\infty \frac{1}{k!} \prod_{l=1}^k &\int d^2x_l\sqrt{g_2(x_l)}\,\Omega(x_l) e^{-I_{p/q}(x_l)} \\
		&= \exp\left[\int d^2x\sqrt{g_2(x)}\; \Omega(x)\, e^{-I_{p/q}(x)} \right].
	\end{aligned}
\end{equation}
But now this amounts to an insertion into the path integral of the exponential of a local integral. In other words, it is equivalent to adding a new local term to the action,
\begin{equation}
	I \rightarrow I-\int d^2x\sqrt{g_2(x)}\; \Omega(x)\, e^{-I_{p/q}(x)}.
\end{equation}
Our instanton action \eqref{eq:instantonAction} is simply linear in the dilaton, so the instanton gas adds an exponential to the dilaton potential.

More explicitly, we can write the bulk part of our dilaton action \eqref{eq:einsteinDilaton} as
\begin{equation}
	I = -\frac{2\pi}{16\pi G_N}\int d^2 x\sqrt{g_2}\left(\Phi R_2 -U(\Phi)\right),
\end{equation}
where the dilaton potential is $U(\Phi)=\tfrac{1}{2}(8G_NJ)^2\Phi^{-3} - \frac{2}{\ell_3^2}\Phi$. The instanton gas from the action $I_{p/q}$ shifts $U$ by addition of a term proportional to
\begin{equation}
	U_{p/q}(\Phi) \approx e^{2\pi i J \frac{p}{q}} \exp\left[-(1-q^{-1})\frac{2\pi\Phi}{4G_N}\right],
\end{equation}
where we should remember that the prefactor is subject to determination of the measure $\Omega$.

First, these terms will be of leading importance if we have a regime where $\Phi$ is of order $G_N$. This occurs close to the threshold for the lightest black hole at small spins. There, higher topologies, including (but not limited to) the manifolds represented by the KK instantons, are no longer suppressed, and we appear to lose all hope of describing the physics semiclassically.

The second important regime, which we concentrate on in this work, occurs at large spin $J$ but very low temperature. This means that $\Phi$ is still large so the new exponential terms from instantons are small, but there are physical effects nonetheless, such as a shift of the ground state energy.

We now have two potential ways to compute the effects of instantons, either by computing the path integral with a fixed number $k$ of instanton insertions and summing over $k$, or by using the instanton gas potential from summing their effects. These approaches have their utility in complementary regimes. For circumstances where the typical instanton number is small, the first approach is useful; indeed, in the next section we will be able to explicitly calculate the $k$-instanton path integrals. Conversely, when we have a high density, the effects may be describable in semiclassical terms from the instanton gas potential, arising as the collective behavior of a condensate of instantons. We will discuss this in section \ref{Sec:InstantonGasSemiClassics}.

\section{JT gravity with defects}\label{sec:JT}

Our analysis of dimensionally reduced three-dimensional gravity above, in particular in the near-extremal limit of section \ref{sec:nearExt}, leads us naturally to a two-dimensional theory of Jackiw-Teitelboim gravity with the inclusion of fundamental defects. In this section, we study this family of  theories in their own right. 

\subsection{The JT path integral with defects}\label{sec:JTpathintwdef}

We here review the genus expansion of amplitudes in JT gravity following \cite{Saad:2019lba}, making the necessary extensions to incorporate defects. JT gravity, as encountered in the near extremal limit of BTZ in section \ref{sec:nearExt}, is a theory of a two-dimensional metric $g_2$ and a scalar dilaton $\phi$, with action
\begin{equation}\label{eq:JTaction}
	I_\text{JT}= -S_0 \chi - \frac{1}{2}\int d^2 x \sqrt{g_2} \phi\left( R_2 +2\right)-\int_\partial ds\phi(\kappa_2-1).
\end{equation}
We have here chosen units to set the two-dimensional AdS length $\ell_2$ to unity. The coefficient $S_0$ of the Euler characteristic $\chi$ parameterizes the suppression of higher topologies, which will organize into an asymptotic series in $e^{-S_0}$. We should therefore regard $S_0$ as a large parameter.

We impose boundary conditions on circular boundaries `at infinity', which means the we fix the proper length $L_\partial \gg 1$ of the boundary, and ultimately take $L_\partial$ to infinity. We take the dilaton to be a constant $\phi_\partial$ on the boundary, which we also take to be large, but fixing the ratio $\frac{L_\partial}{\phi_\partial}$.
\begin{equation}
	\text{Boundary condition:}\qquad L_\partial,\phi_\partial\to\infty,\qquad \frac{L_\partial}{\phi_\partial} = \frac{\beta}{\gamma} \quad\text{held fixed.}
\end{equation}
The dimensionless parameter $\gamma$ is fixed once and for all, and acts only to rescale the units in which $\beta$ is measured. In section \ref{sec:nearExt} we had $\gamma = \frac{\ell_3}{16G_N} =\frac{c}{24}$ using the Brown-Henneaux value of central charge \cite{Brown:1986nw}. This means that $\beta$ of order $\gamma$ corresponds to temperature of order one in Planck units, not in AdS units. We refer to $\beta$ as the renormalized length of the boundary, or an inverse temperature.

Our theory in addition contains dynamical `defects' or `instantons', parameterized in terms of a defect angle $\alpha$ and a weighting $\lambda$. As in \cite{Mertens:2019tcm}, we can include a defect by an insertion in the path integral proportional to
\begin{equation}\label{eq:JTinstaction}
	\lambda \int d^2x \sqrt{g_2} \, e^{-2\pi(1-\alpha) \phi(x)}.
\end{equation}
For our purposes, we will take this more as a heuristic guide, and use the expansion we develop below as our precise definition for the defects and the parameter $\lambda$; we explore this path integral definition a little more in the context of the instanton gas in section \ref{Sec:InstantonGasSemiClassics}. We may also have several species of defect labeled by an index $i=1,\ldots, N_{\rm F}$, with parameters $\lambda_i,\alpha_i$. The path integral is defined to sum over insertions of any number of each species of defect. We will take $0<\alpha<1$, though to be conservative, for reasons that will become clear later, we can restrict further to $0<\alpha\leq \frac{1}{2}$.

 For the $p/q$ KK instanton in our 3D analysis, the action \eqref{eq:3DJTinstanton} gives us a defect
\begin{equation}
	\alpha = \frac{1}{q},\qquad \lambda_{p/q} \propto \exp\left[-\left(1-\tfrac{1}{q}\right)S_0+2\pi i J \tfrac{p}{q}\right],
\end{equation}
but in this section we take arbitrary parameters.  

\begin{figure}
\begin{center}
 \hspace{-0.3cm} \begin{tikzpicture}[scale=0.9]
\pgftext{\includegraphics[scale=0.5]{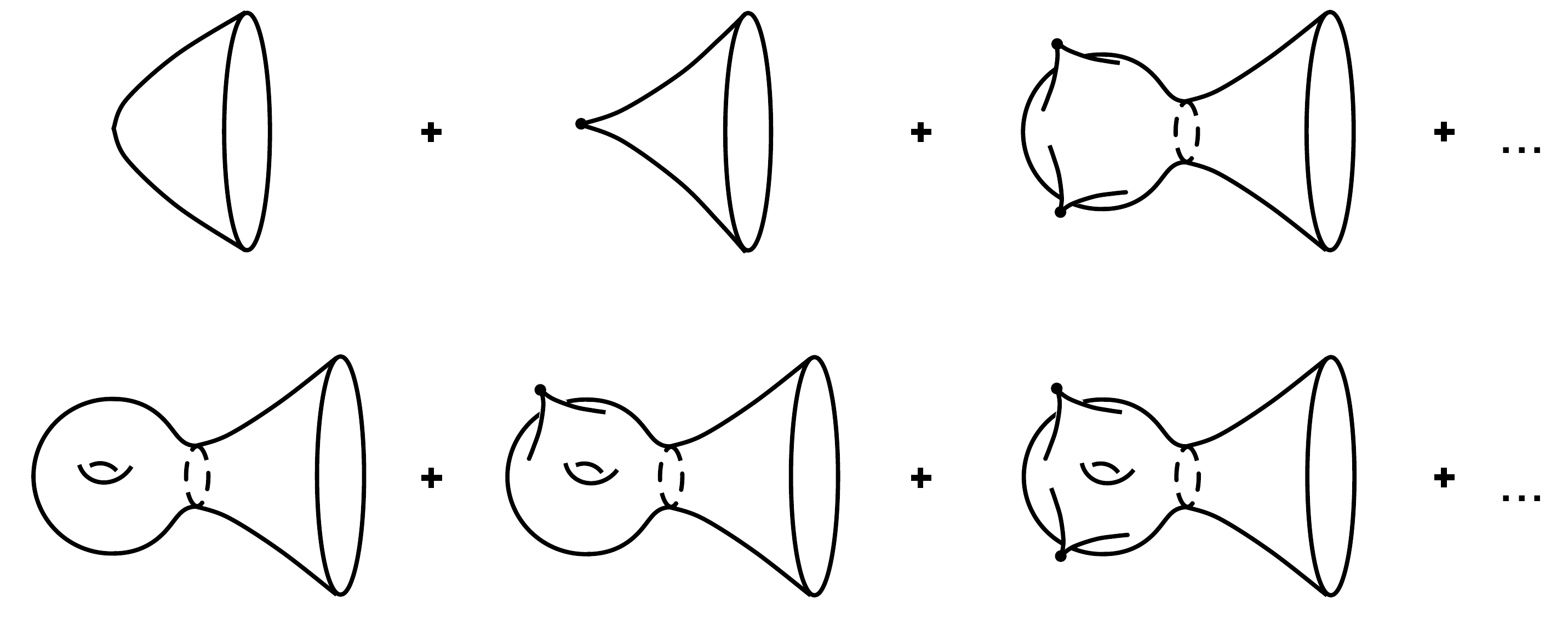}} at (2,0);
\draw (-8.5,-1.5) node  {$g=1:$};
\draw (-8.5,1.6) node  {$g=0:$};
  \end{tikzpicture}
 \caption{The first few topologies contributing to the expansion of  $\big\langle Z(\beta)\big\rangle$, as in \eqref{eq:Zexpansion} with $n=1$. The top row shows the topologies for the disc $n=1,g=0$, with some number of defects $k=0,1,2,\ldots$, of order $e^{S_0}\lambda^k$. The second row shows the $n=1,g=1$ contributions for $k=0,1,\ldots$, of order $e^{-S_0}\lambda^k$.
 \label{fig:sumoverdef1}}
\end{center}
\end{figure}

The general amplitude we are interested in is computed by the path integral with $n$ boundaries, each with boundary condition parameterized by its own renormalized length $\beta$. Without loss of generality, it is sufficient to compute the connected amplitude, and following \cite{Saad:2019lba} we denote this by $\big\langle Z(\beta_1)\cdots Z(\beta_n)\big\rangle_C$. This amplitude can be expanded as a sum over sectors give by different topologies, labeled by the genus $g$ of the connected spacetime, as well as the number of defects $k_i$ of each species $i$, with $k=\sum_i k_i$ in total:
\begin{equation}\label{eq:Zexpansion}
	\big\langle Z(\beta_1)\cdots Z(\beta_n)\big\rangle_C = \sum_{g,k_1,k_2,\ldots=0}^\infty e^{-(2g+n-2)S_0} \left(\prod_i \frac{\lambda_i^{k_i}}{k_i!}\right) Z_{g,n,k}(\beta_1,\ldots,\beta_n;\alpha_1,\ldots,\alpha_k)
\end{equation}
The symmetry factors $k_i!$ in this definition of $Z_{g,n,k}$ are included to account for overcounting by permutations of identical defects. With this definition, the $k$ defects in $Z_{g,n,k}$ can be regarded as distinguishable, labeled only by their defect parameters $\alpha$, and it becomes irrelevant whether two defects with the same $\alpha$ are of the same or different species. We sketch the first few terms in this expansion for a single boundary $n=1$ in figure \ref{fig:sumoverdef1}.

The remarkable result of \cite{Saad:2019lba} was that the amplitudes $Z_{g,n}$ (without defects) can be computed from the Weil-Petersson (WP) volumes of the moduli spaces of constant-curvature surfaces with boundary. To understand this result, we note first that the dilaton appears linearly in the action \eqref{eq:JTaction}, so acts as a Lagrange multiplier constraining the curvature $R_2=-2$. The resulting metrics are locally unique, so the path integral reduces to an integral over locations of the boundary (governed by the Schwarzian theory), and over the finite-dimensional moduli space of surface with a given topology. More explicitly, with the exceptions of the $n=1$, $g=0$ amplitude with one boundary and disc topology, every constant curvature surface with $n$ asymptotic boundaries has a unique geodesic homotopic to each boundary. We can cut the surface along these geodesics, which we take to have lengths $b_1,\ldots,b_n$. We then have cut the spacetime into a genus $g$ surface with $n$ geodesic boundaries of lengths $b_1,\ldots,b_n$ (the `convex core'), and $n$ `trumpets' with one geodesic boundary and one asymptotic boundary. This split is shown for the $g=1,n=1$ case in the bottom diagram of figure \ref{fig:sumoverdef1}, where the dotted line corresponds to the geodesic of length $b$. On the asymptotic boundaries, we must integrate over all ways in which the boundary conditions of renormalized length $\beta$ can be obeyed on the trumpet geometry, with the result
\begin{equation}
	Z_{\text{trumpet}}(\beta,b) = \sqrt{\frac{\gamma}{2\pi\beta}}  e^{-\frac{\gamma}{2}\frac{b^2}{\beta}}.
\end{equation}
Next, we must integrate over all constant curvature surfaces of genus $g$ with $n$ geodesic boundaries of the specified lengths. This is a compact space of dimension $2n+6(g-1)$, and it turns out that the correct measure on this space is provided by the WP symplectic form. The result of the integral is thus the WP volume, denoted $V_{g,n}(b_1,\ldots,b_n)$. Finally, we must integrate over the lengths $b$ of the geodesics separating the trumpets from the convex core with the correct measure $b db$. Putting these pieces together, we have
\begin{equation}\label{eq:Vexp1}
	Z_{g,n,k=0}(\beta_1,\ldots,\beta_n) = \int_0^\infty b_1 db_1 Z_{\text{trumpet}}(\beta_1,b_1) \cdots \int_0^\infty b_n db_n Z_{\text{trumpet}}(\beta_n,b_n) \, V_{g,n}(b_1,\ldots,b_n).
\end{equation}

There are two exceptional cases for which this formula does not apply. The first is the $n=1,g=0$ disc, for which we have
\begin{equation}\label{eq:JTdisc}
	Z_{g=0,n=1,k=0}(\beta) = \sqrt{\frac{\gamma^3}{2\pi\beta^3}} \ e^{2\pi^2\frac{\gamma}{\beta}}.
\end{equation}
The second is the $n=2,g=0$ double trumpet, which follows from the above if we set $V_{0,2}(b_1,b_2)=\tfrac{1}{2}\delta(b_1^2-b_2^2) =\tfrac{1}{b_1}\delta(b_1-b_2) $, with the factor of $\frac{1}{b}$ informally understood as coming from the residual gauge symmetry of rotating both boundaries along the length $b$ of the separating geodesic:
\begin{equation}
	Z_{g=0,n=2,k=0}(\beta_1,\beta_2)= \int_0^\infty b db\ Z_{\text{trumpet}}(\beta_1,b) Z_{\text{trumpet}}(\beta_2,b) = \frac{\sqrt{\beta_1\beta_2}}{2\pi(\beta_1+\beta_2)}\,.
\end{equation}

The expression \eqref{eq:Vexp1} is only as useful as our knowledge of the volumes $V_{g,n}$. Fortunately, they can be efficiently computed due to a recursion relation found by Mirzakhani \cite{Mirzakhani:2006fta}. In particular, the volumes are even polynomials in $b_1,\ldots,b_n$, with degree equal to the dimension of moduli space $2n+6(g-1)$.

This concludes our brief review of \cite{Saad:2019lba}, giving the expansion of the amplitudes without defects. It is now a rather simple matter to include the defects, using a result on the volumes of moduli spaces of hyperbolic surfaces with conical defects, also used by \cite{Cotler:2019nbi} in a somewhat different context. First, let us understand the effect of the defects when we integrate out the dilaton. Since the defect insertion \eqref{eq:JTinstaction} is exponential in the dilaton, it can be thought of as another term in the action which is linear in the dilaton. With this inclusion, the bulk JT action becomes $-\tfrac{1}{2}\int \phi(R_2+2-4\pi(1-\alpha)\delta_{\text{inst}})$, where $\delta_{\text{inst}}$ is a delta-function at the location of the defect. When we integrate out $\phi$, we impose constant curvature everywhere except at the locations of defects, where we have delta-function sources of positive curvature. These are conical defects, with defect angle $2\pi(1-\alpha)$.

Now, just as before, we are left with an integral over surfaces of constant curvature, now with conical points. And just as before, excepting some special cases, every such surface has a unique geodesic homotopic to each asymptotic boundary, where we do not allow homotopies to pass through any defect. This split is made along the dotted line for all but the first two diagrams in figure \ref{fig:sumoverdef1}; the first is the disc which we have already addressed, and the second will be discussed in a moment. We thus have the same formula as before, except that we must use volumes of moduli spaces with conical points. The WP measure has a natural generalization to this setting. The fact that this is the correct measure for our integrals was derived in \cite{Cotler:2019nbi}. We thus can simply generalize \eqref{eq:Vexp1} to
 \begin{equation}\label{eq:Vexp2}
 \begin{aligned}
 	Z_{g,n,k}&(\beta_1,\ldots,\beta_n;\alpha_1,\ldots,\alpha_k) =\\
 	& \int_0^\infty b_1 db_1 Z_{\text{trumpet}}(\beta_1,b_1) \cdots \int_0^\infty b_n db_n Z_{\text{trumpet}}(\beta_n,b_n) \, V_{g,n,k}(b_1,\ldots,b_n;\alpha_1,\ldots,\alpha_k),
 \end{aligned}
\end{equation}
 where $V_{g,n,k}$ are the WP volumes of genus $g$, with $n$ geodesic boundaries with the specified lengths and $k$ cone points with the specified defect angles. We take these amplitudes as a constructive definition of what we mean by JT gravity with defects \footnote{This definition of the theory explicitly excludes the possibility of a defect hitting the Schwarzian boundary curve. As we will see later this process is suppressed anyways for the Schwarzian boundary conditions, as long as $\alpha<1$.}.

Before discussing these new volumes with cone points, we address the special cases with boundaries which are not homotopic to geodesics, so we cannot split the surface into trumpets and surfaces with geodesic boundary, and formula \eqref{eq:Vexp2} does not obviously apply. This occurs only for disc topology, with genus $g=0$  and a single boundary $n=1$, when the total defect angle is less than $2\pi$, $\sum (1-\alpha)\leq 1$. To see that this occurs, we can use the Gauss-Bonnet theorem to compute the area of a negatively curved disc ($\chi=1$) with geodesic boundary and defects, finding $A=-2\pi +2\pi \sum (1-\alpha)$; this area is negative if the total defect angle is too small. This always occurs for a single defect $k=1$, which we treat as a special case below. We can avoid meeting any other special cases by restricting our attention to sufficiently strong defects, $\alpha\leq \frac{1}{2}$.

We now turn to the volumes $V_{g,n,k}$ including cone points. Remarkably, this does not require any new ingredient! These volumes can be computed from the volume polynomials $V_{g,n+k}$ without defects but with $n+k$ boundaries, simply be evaluating $k$ of the boundary lengths at imaginary values $b=2\pi i \alpha$:
\begin{equation}\label{eq:coneVols}
	V_{g,n,k}(b_1,\ldots b_n; \alpha_1,\ldots, \alpha_k) = V_{g,n+k}(b_1,\ldots b_n, b_{n+1}=2\pi i \alpha_1,\ldots, b_{n+k}=2\pi i\alpha_k).
\end{equation}
This has been proven for $\alpha\leq \frac{1}{2}$ \cite{tan2004generalizations} (see also \cite{do2006weilpetersson, do2011moduli}), with the restriction essentially due to the special cases discussed above. This formula was recently applied to de Sitter JT gravity in \cite{Cotler:2019nbi}.

We now address the special case $k=1$. The relevant single-defect path integral was computed in \cite{Mertens:2019tcm}, with the result
\begin{equation}\label{eq:1defect}
 	Z_{g=0,n=1,k=1}(\beta;\alpha) =\sqrt{\frac{\gamma}{2\pi\beta}}  \ e^{2\pi^2\frac{ \gamma}{\beta}\alpha^2}.
\end{equation}
In fact, rather remarkably, this result can be obtained from \eqref{eq:Vexp2}, by rather formal use of the two-boundary volume $V_{0,2}(b_1,b_2) =\tfrac{1}{b_1}\delta(b_1-b_2) $ which we applied earlier to the double trumpet:
\begin{equation}
 	Z_{g=0,n=1,k=1}(\beta;\alpha) = Z_{\text{trumpet}}(\beta,b = 2\pi i \alpha).
\end{equation}
We take this as evidence that the formula \eqref{eq:Vexp2} for the amplitudes, along with \eqref{eq:coneVols} for the volume polynomials $V_{g,n,k}$, is correct even when the argument that led there does not apply. We would therefore be extremely surprised if the results we will obtain fail to apply for general $\alpha\in (0,1)$, though strictly speaking they are proven only for $\alpha\leq \tfrac{1}{2}$.

Before beginning our analysis of the amplitudes from \eqref{eq:Vexp2}, we recall the interpretation of the $n$-boundary path integral $\langle Z(\beta_1)\cdots Z(\beta_n)\rangle$ in JT gravity without defects. As observed by Eynard and Orantin \cite{Eynard:2004mh,Eynard:2007kz}, Mirzakhani's recursion relation \cite{Mirzakhani:2006fta} is closely related to the topological recursion obeyed by matrix integrals. The matrix over which we are integrating is to be interpreted as the Hamiltonian $H$ of a dual quantum mechanical theory, and $Z(\beta)=\Tr(e^{-\beta H})$  is the corresponding partition function. However, since we are integrating over $H$ this Hamiltonian is not unique, but rather selected from a random distribution determined by the measure in the matrix integral, so that for each $\beta$, $Z(\beta)$ is a random variable. We then interpret the amplitudes $\langle Z(\beta_1)\cdots Z(\beta_n)\rangle$ as moments of this distribution,
\begin{equation}
	\langle Z(\beta_1)\cdots Z(\beta_n)\rangle = \int dH\; \mu(H) \, \Tr(e^{-\beta_1 H})\cdots \Tr(e^{-\beta_n H}),
\end{equation}
and learn about the measure $\mu$ from the JT path integral. From the agreement between Mirzakhani's recursion relations and topological recursion, we learn that at any order in the genus expansion, the moments agree with a particularly special type of ensemble, namely a matrix integral with measure
\begin{equation}
	\mu(H) \propto e^{-L \Tr V(H)}
\end{equation}
for some potential $V$, taking $H$ to be an $L\times L$ Hermitian matrix (and $dH$ is the flat measure on independent components). More precisely, we must take a `double-scaled' limit of such integrals, where we take $L\to\infty$ and tune the potential $V$ such that the resulting average density of states at leading order in the large $L$ expansion agrees with the disc amplitude \eqref{eq:JTdisc}. The structure of such matrix integrals is extremely rigid, since the leading order density of states (or equivalently the disc amplitude) completely determines $Z_{g,n}$ for all $g,n$.

In the following we will find that such an interpretation exists for our modified theory of JT gravity with defects. We will show analytically that the amplitudes $\langle Z(\beta_1)\cdots Z(\beta_n)\rangle$ are consistent with the moments of some ensemble of Hamiltonians at any genus. This will allow us to find an explicit solution for the matrix integral measure in the double scaling limit. 

\subsection{The disc with defects: $n=1,g=0$ amplitudes}\label{sec:disc}

We begin our analysis of JT gravity with defects by looking at the disc with any number of defect insertions, the amplitudes $Z_{g=0,n=1,k}$ for any $k$. We will mostly restrict to a single species, commenting later on the extension to multiple types of defect. 

First, we note that most contributions to the path integral discussed above do not come from classical solutions. One way to see this is to use the equation of motion coming from variation of the metric in the action \eqref{eq:JTaction}. This equation is equivalent to the statement that $\epsilon^{ab}\partial_b\phi$ is a Killing vector for $g_2$ \cite{Banks:1990mk}, implying that the level sets of $\phi$ are the integral curves of the Killing vector, coinciding with orbits of the symmetry it generates. For a given surface with constant curvature metric $g_2$, a classical solution to JT therefore exists if and only if the metric has a globally defined continuous symmetry. The only surfaces with such symmetries are the disc and double trumpet (for which the corresponding solution does not have the desired boundary conditions, since the dilaton must go to $-\infty$ at one end). Once we include defects, the dilaton must be stationary at their locations, so they lie at fixed points of the symmetry. The only additional classical solution is therefore the disc with all defects localized at the origin. For $\alpha< \tfrac{1}{2}$, two defects on a constant curvature metric are always separated by a finite minimal distance at which the surface develops a cusp, so we can never have more than one defect in a single location (and likewise for the marginal case $\alpha=\tfrac{1}{2}$, for which we can bring the defects as close as desired, but they recede to infinite distance, limiting to a cusp geometry).

As a result, if we were to take the approach (analogous to Maloney and Witten in the case of three-dimensional pure gravity) of classifying all classical solutions, computing the perturbative expansion to all orders, and summing the results, we would only have two contributions,
\begin{align}\label{eq:onedefpf}
	\langle Z(\beta)\rangle_{\text{naive}} &= e^{S_0}Z_{g=0,n=1,k=0}(\beta) + e^{S_0}\lambda Z_{g=0,n=1,k=1}(\beta) \\
	&= e^{S_0}\sqrt{\frac{\gamma^3}{2\pi\beta^3}} \ e^{2\pi^2\frac{\gamma}{\beta}} + e^{S_0}\lambda \sqrt{\frac{\gamma}{2\pi\beta}}  \ e^{2\pi^2\frac{ \gamma}{\beta}\alpha^2}
\end{align}
where we used \eqref{eq:JTdisc} and \eqref{eq:1defect}. We can then extract the corresponding density of states by taking an inverse Laplace transform, finding
\begin{equation}\label{eq:rhonaive}
	\rho_\text{naive}(E) =  e^{S_0} \frac{\gamma }{2 \pi ^2} \left[\sinh \left(2 \pi  \sqrt{2\gamma  E}\right) + 2\pi\lambda  \frac{\cosh \left(2 \pi \alpha \sqrt{2\gamma  E}\right)}{\sqrt{2\gamma E}}\right] , \quad E>0.
\end{equation}
For $|\lambda|\ll 1$, the first term is almost always dominant, except at very small energies, where we have
\begin{equation}\label{eq:naiverho}
	\rho_\text{naive}(E) \sim   e^{S_0} \frac{\gamma }{\pi} \left( \sqrt{2\gamma  E} +  \frac{\lambda}{\sqrt{2 \gamma E}}\right),
\end{equation}
so the second term is suppressed by a factor of $\lambda$, but enhanced by a factor of $E^{-1}$. Now, any consistent unitary dual (even one involving an average over Hamiltonians) must have positive density of states. For $\lambda<0$, our naive density of states violates this for $E<\frac{|\lambda|}{2\gamma}$.

In the context of three-dimensional gravity, this is precisely the problematic feature of the Maloney-Witten-Keller partition function pointed out by  \cite{Benjamin:2019stq}. The defect in question in that case is the $\rat=\tfrac{1}{2}$ KK instanton described in the previous section, whose one-defect solution corresponds to an $SL(2,\ZZ)$ black hole \cite{Maldacena:1998bw} in three dimensions. Indeed, the sum of the BTZ density \eqref{eq:rhobtw1} and $\rat=\tfrac{1}{2}$ $SL(2,\ZZ)$ density \eqref{eq:rhobtw2}  takes precisely the form of \eqref{eq:rhonaive} for an appropriate $\lambda\propto (-1)^J e^{-S_0(J)/2}$.

If we include only classical solutions and fluctuations around them, we must conclude that this theory does not have a unitary dual. To solve the negativity problem we would be forced to alter the dynamics of the theory, for example by introducing an additional species of defect with $\lambda' \geq |\lambda|$. In the 3D gravity context, this can be achieved by introducing a particle of mass $m\leq1/(16G_N)$, as discussed in the introduction and later in section \ref{sec:discMatter}. This is the two-dimensional version of the proposal of \cite{Benjamin:2019stq, Benjamin:2020mfz}.

However, it turns out that the negativity is automatically resolved by including the path integral over off-shell configurations including multiple defects on the disc. To see how this might occur, we look first at the two-defect result, which is simple to obtain from \eqref{eq:Vexp2} and \eqref{eq:coneVols} using $V_{0,3}=1$. This volume is clear from the observation that there is a unique hyperbolic surface with three boundaries of fixed length (i.e.~a pair of pants), so the moduli space in question is a single point. We find the result
\begin{equation}\label{eq:Z012}
	Z_{g=0,n=1,k=2}(\beta;\alpha,\alpha) = \sqrt{\frac{\beta}{2\pi\gamma}},
\end{equation}
which we can include in our expansion of the density of states (recalling the symmetry factor of $2$ in the definition \eqref{eq:Zexpansion}):
\begin{equation}
	\rho_\text{disc}(E) =  e^{S_0} \frac{\gamma }{\pi} \left( \sqrt{2\gamma  E} +  \frac{\lambda}{\sqrt{2 \gamma E}} -\frac{\lambda^2}{2(2\gamma E)^{3/2}} +\cdots \right) 
\end{equation}
The density $E^{-3/2}$ (supported at positive $E$) is not an  integrable function, so strictly speaking we must regard the last term as a distribution which regulates the divergence at $E\to 0$ in an appropriate way\footnote{Specifically, it is the principal value distribution $\mathcal{P}E^{-3/2}$, defined through its pairing with a smooth test function $\psi$ which decays rapidly as $E\to +\infty$ by $\int (\mathcal{P}E^{-3/2}) \psi(E) = \int_0^\infty dE\, E^{-3/2}(\psi(E)-\psi(0))$. Roughly speaking, we have subtracted an `infinite delta function' at zero energy to cancel the divergence of the integral. Choosing $\psi(E)=e^{-\beta E}$ gives us the Laplace transform \eqref{eq:Z012}.}. The new term is once again suppressed by a factor of $\lambda$, but more singular at small energy. This suggests that the expansion in powers of $k$ is breaking down at $E$ of order $\lambda$, so we cannot trust the perturbative series in the regime where the negativity occurs.

In fact we can do better, since we can compute the leading order term in $Z_{g=0,n=1,k}(\beta)$ at low temperature explicitly and sum the resulting series of leading order corrections. The limit when our asymptotic boundary becomes very long corresponds taking the corresponding geodesic length $b\to\infty$ with all other lengths or defect angles fixed. Since $V_{g,n}$ is an even polynomial in the lengths of order $2n+6g-6$, we are picking out the monomial $b^{2n+6g-g}$, and the result is independent of all other lengths or defect angles. We compute the coefficient in appendix \ref{app:wp} using results from \cite{Mirzakhani:2006fta,ITZYKSON_1992}, finding
\begin{equation}
	V_{g,n}(b,\ldots) \sim \frac{1}{(24)^g g!} \frac{1}{(3g-3+n)!} \left( \frac{b^2}{2}\right)^{3g-3+n}.
\end{equation}
Specializing to the case of multiple defects on the disc $g=0$ using \eqref{eq:coneVols}, we have
\begin{equation}
	V_{g=0,n=1,k} (b;\alpha_1,\ldots,\alpha_k) = \frac{1}{(k-2)!} \left( \frac{b^2}{2} \right)^{k-2} +O(b^{2k-6}),
\end{equation}
and integrating this against the trumpet gives simply
\begin{equation}
	Z_{g=0,n=1,k}(\beta) = \frac{1}{\sqrt{2\pi}} \left(\frac{\beta}{\gamma}\right)^{k-\tfrac{3}{2}} (1+O(\beta^{-1})).
\end{equation}
The derivation applies for $k\geq 2$, but the result also is valid for the special cases $k=0,1$.

Now we simply sum over $k$ as in \eqref{eq:Zexpansion} including the required symmetry factors, simply finding that the defects give an exponential correction to leading order at low temperature:
\begin{align}
	Z_\text{disc}(\beta) &= e^{S_0}\sum_{k=0}^\infty \frac{\lambda^k}{k!}Z_{0,1,k}(\beta) \\
	&= \frac{1}{\sqrt{2\pi}} \left(\frac{\gamma}{\beta}\right)^{\tfrac{3}{2}} e^{\gamma^{-1}\lambda\beta}(1+O(\beta^{-1})).
\end{align}
In term of the density of states, this corresponds simply to a shift in the ground state energy,
\begin{equation}
	\rho_\text{disc}(E) \sim e^{S_0} \frac{\gamma }{\pi}  \sqrt{2\gamma  (E-E_0)}, \qquad E>E_0 \sim -\gamma^{-1}\lambda.
\end{equation}
The naive density of states \eqref{eq:naiverho} arises from taking a binomial expansion and truncating after the first two terms. The negative density of states appears in a regime where that series fails to converge. It was this an artifact of the truncation, and has been resolved in a rather elegant manner by including the off-shell configurations of multiple defects.

At this leading order in the low-temperature limit, it is straightforward to generalize to include several species of defect, and their contributions to $E_0$ simply add,
\begin{equation}
	E_0 \sim -\gamma^{-1} \sum \lambda_i \ .
\end{equation}

The contributions from $k\geq 2$ defects include new terms decaying as $\beta^{-1/2}$ and growing as odd positive powers of $\beta$, which themselves have the potential of causing a negative density of states or other pathological low-energy behavior. Experimentally, by explicit calculations (we compute up to $k=6$ defects using the list of volume polynomials in appendix B of \cite{do2011moduli}), we find that all higher corrections fit into the same pattern, shifting the edge of the spectrum at higher orders in $\lambda$. The higher order corrections are no longer independent of $\alpha$. Explicitly, the first few orders for a single defect are given by
\begin{equation}\label{eq:E01}
\gamma E_0(\lambda) = -\lambda - \lambda^2 \pi^2(1-2\alpha^2) - \frac{\lambda^3}{3} \pi^4(5-18\alpha^2+15\alpha^4) +\mathcal{O}(\lambda^4),
\end{equation}
and for two species of defect we find
\begin{equation}\label{eq:E02}
	\gamma E_0(\lambda) = -\lambda_1-\lambda_2-\pi^2 (\lambda_1+\lambda_2)(\lambda_1(1-2\alpha_1^2)+\lambda_2(1-2\alpha_2^2))+\cdots \ .
\end{equation}
We emphasize that this  structure places extremely stringent constraints relating the volume polynomials. The amplitude $Z_{0,1,k}$ contains terms proportional to $\beta^{\frac{1}{2}+p}$ for $p=0,1,\cdots, k-2$; for $p\geq 2$ there is a unique coefficient determined by lower orders in $k$ that is compatible with the interpretation as a shift of $E_0$. The next order in $E_0$ can then be read off from the $p=1$ term, and finally the $p=0$ term describes a change in the functional form of the density of states, so $\rho_\text{disc}$ is not simply a uniform translation of the Schwarzian density of states.

In the next section, we will perform these calculations exactly and explain more directly the origin of the structure that emerges from the perturbative calculation done above.

\subsection{The matrix integral dual}\label{sec:JTdefMM}

In the previous section we studied JT gravity with a gas of defect perturbatively in the weighting parameter $\lambda$. In this section we will obtain exact expressions at genus zero from resuming over an arbitrary number of defects and show that the result for $\lb Z(\beta_1) \ldots Z(\beta_n) \rb_C$ matches precisely with a matrix integral. In section \ref{sec:highergenus} we give a more formal, but less explicitly, proof that is valid also for higher genus contributions. 

\subsubsection{Two boundary amplitude}\label{sec:twoBoundary}
We will begin by considering the two boundary correlator $\lb Z(\beta_1) Z(\beta_2)\rb_C$ at genus zero, including defects. This correlator is universal for all double-scaled (GUE) matrix models. Its independent of the spectral curve which defines the theory and depends only on the position of the edge of the spectrum $E_0$,  
\beq\label{eq:universal2loopMM}
\big\lb \Tr(e^{-\beta_1 H}) \, \Tr(e^{-\beta_2 H}) \big\rb_{C,g=0,\mathrm{MM}} = \frac{1}{2\pi} \frac{\sqrt{\beta_1 \beta_2}}{\beta_1+\beta_2} e^{- E_0 (\beta_1+\beta_2)},
\eeq
where we include the annotation ${\rm MM}$ to emphasize that this result comes from a matrix model. If our model defects gives this result, it is a `smoking gun' giving us strong evidence that there is a dual matrix integral.

To compute this cylinder amplitude, we write the answer in the defect expansion
\beq
\lb Z(\beta_1) Z(\beta_2)\rb_{C,g=0} = \sum_{k=0}^\infty \frac{\lambda^k}{k!}  \lb Z(\beta_1) Z(\beta_2)\rb_{0,2,k}, 
\eeq
where the sum is over the number of defects present in the gravitational path integral, and
\beq
\lb Z(\beta_1) Z(\beta_2)\rb_{0,2,k} \equiv \int_0^\infty \left( \prod_{i=1,2} b_i db_i Z_{\rm trumpet} (\beta_i,b_i) \right) V_{0,2+k}(b_1,b_2,2\pi i\alpha, \ldots, 2\pi i \alpha).
\eeq
The quantity in the right hand side is the WP volume with $2+k$ boundaries. We used the fact stated above that the volume with defects is a simple analytic continuation. To simplify the expression, the case without defects $k=0$ is included in the sum with the understanding that $V_{0,2}(b_1,b_2) =\tfrac{1}{b_1}\delta(b_1-b_2)$.

We now make use of the explicit formula derived in \cite{mtms} for the genus zero WP volumes:
\begin{equation}\label{eq:genus0V}
	V_{0,n}(b_1,\ldots,b_n) = \frac{1}{2} \left.\left(\frac{\partial}{\partial x}\right)^{n-3} \left[I_0\left(b_1 \sqrt{u_{\rm JT}(x)}\right)\cdots I_0\left(b_n \sqrt{u_{\rm JT}(x)}\right) u_{\rm JT}'(x) \right]\right|_{x=0},
\end{equation}
valid for $n\geq 2$. $I_0$ is the modified Bessel function of the first kind. The function $u_{\rm JT}(x)$ is defined implicitly by
\begin{equation}
	\sqrt{u_{\rm JT}(x)}\hspace{0.1cm}J_1\left(2\pi\sqrt{u_{\rm JT}(x)}\right)=2\pi x \,,
\end{equation}
and primes denote derivatives with respect to $x$. In a moment we will briefly review the role played by $u_{\rm JT}$ in the matrix model, but for now we will simply take the formula as given and apply it to our problem.

 Using this explicit formula for WP volumes we can find the contribution with $k$ defects to the double trumpet 
\begin{equation}
\begin{gathered}
\lb Z(\beta_1) Z(\beta_2)\rb_{0,2,k} =\frac{\sqrt{\beta_1 \beta_2} }{4\pi \gamma}  \left.\Big(\frac{\partial}{\partial x} \Big)^{k-1} \left[ u'_{\rm JT}(x) e^{\frac{1}{2\gamma}u_{\rm JT}(x)(\beta_1 + \beta_2)} \left( J_0 (2 \pi \alpha \sqrt{u_{\rm JT}(x)})\right)^k \right] \right|_{x=0} \\
 =\frac{\sqrt{\beta_1 \beta_2} }{2\pi (\beta_1 +\beta_2)}    \left. \Big(\frac{\partial}{\partial x} \Big)^{k-1} \left[ \left( e^{\frac{1}{2\gamma}u_{\rm JT}(x)(\beta_1 + \beta_2)}\right)' \left( J_0 (2 \pi \alpha \sqrt{u_{\rm JT}(x)})\right)^k\right] \right|_{x=0}.
 \end{gathered}
\end{equation}
To derive this formula we integrated over $b_1$ and $b_2$ which produces the prefactor and a simple exponential of $\beta_1$ and $\beta_2$. Putting everything together we obtain 
\bea
\lb Z(\beta_1) Z(\beta_2)\rb_{C,g=0} &=&\frac{\sqrt{\beta_1 \beta_2} }{2\pi (\beta_1 +\beta_2)} \nonumber\\
&&\hspace{-2.5cm}\times \left. \left( 1 +  \sum_{k=1}^\infty \frac{\lambda^k}{k!}\left(\frac{\partial}{\partial x} \right)^{k-1} \left[ \left( e^{\frac{1}{2\gamma} u_{\rm JT}(x)(\beta_1 + \beta_2)}\right)' \left( J_0 (2 \pi \alpha \sqrt{u_{\rm JT}(x)})\right)^k\right] \right) \right|_{x=0}, 
\eea

So far, we have just been plugging the formula \eqref{eq:genus0V} for the Weil-Petersson volumes into the expression for the amplitudes with $k$ defects. At this point, we can explicitly perform the sum over all defects using the Lagrange reversion theorem (see Appendix \ref{App:LRT} for a statement of the theorem). We obtain precisely the answer expected from the double-scaled matrix integral, given by
\beq\label{two-loopmmjt}
\lb Z(\beta_1) Z(\beta_2)\rb_{C,g=0} = \frac{1}{2\pi} \frac{\sqrt{\beta_1 \beta_2} }{\beta_1 +\beta_2} e^{-\frac{1}{2\gamma} u(x) (\beta_1+\beta_2)} \Big|_{x=0},
\eeq
where we defined the function $u(x)$ implicitly through 
\beq
u(x) \equiv - u_{\rm JT}(x_\star (x)) ,~~~{\rm with}~~ x_{\star} = x + \lambda J_0 \left(2\pi \alpha \sqrt{u_{\rm JT}(x_\star)}\right).
\eeq
For later convenience, we have chosen to define $u(x)$ with a minus sign relative to the conventional definition used for $u_{\rm JT}(x)$. We can eliminate the variable $x_\star$ from these expressions to obtain a more direct implicit definition for $u(x)$,
\beq\label{JTdefSE}
\frac{\sqrt{u(x)}}{2\pi} \hspace{0.1cm} I_1 \left( 2\pi \sqrt{u(x)} \right) + \lambda \hspace{0.1cm}I_0 \left(2 \pi \alpha \sqrt{u(x)} \right) = x,
\eeq
where we used that $I_n(x) = i^{-n} J_n (ix)$ to simplify the expression. In the old matrix model literature \cite{Brezin:1990rb, *Douglas:1989ve, *Gross:1989vs, Banks:1989df} this equation is called the (genus zero) string equation, $u(x)$ is called the heat capacity and $x$ is related to the leading KdV parameter. This function characterizes the theory completely in the double scaling limit and is equivalent to giving the genus zero density of states (or equivalently the spectral curve), which we will compute later. It is evident that taking $\lambda \to0$ sends $u(x) \to - u_{\rm JT}(x)$, recovering the JT gravity string equation. 

Taking the $x\to0$ limit of the expression above we get the final answer for the two-loop amplitude corresponding to JT gravity with a gas of defects 
\bea\label{eq:Zg=0}
\lb Z(\beta_1) Z(\beta_2) \rb_{C,g=0} &=&  \frac{1}{2\pi} \frac{\sqrt{\beta_1 \beta_2} }{\beta_1 +\beta_2} e^{- E_0 (\beta_1+\beta_2)}  \\
&=&\big\lb \Tr(e^{-\beta_1 H}) ~\Tr(e^{-\beta_2 H}) \big\rb_{\mathrm{MM},g=0},
\eea 
where the exact edge of the spectrum is given by $E_0 =(2\gamma)^{-1}u(x=0)$. The zero-point energy can be written more explicitly, using the string equation, as the largest solution $E_0$ of the equation
\beq
\sqrt{2\gamma E_0}\hspace{0.1cm}I_1\left(2\pi \sqrt{2\gamma E_0}\right) +2 \pi \lambda \hspace{0.1cm}I_0 \left(2\pi \alpha \sqrt{2\gamma E_0}\right)=0.
\eeq
The gravitational result \eqref{eq:Zg=0} from summing over defects thus takes the precise functional form \eqref{eq:universal2loopMM} required for a double-scaled matrix integral.

This result can be straightforwardly generalized to allow a number of flavors $N_{\rm F}$ of defects with weighting $\lambda_i$ and angles $\alpha_i$ for $i=1,\ldots , N_{\rm F}$. Using the second version of the Lagrange reversion theorem described in Appendix \ref{App:LRT} we can show the two-loop amplitude takes the same form \eqref{two-loopmmjt} but with a string equation given by
\beq\label{eqJTdefMSE}
\frac{\sqrt{u}}{2\pi} I_1 \left( 2\pi \sqrt{u}\right) + \sum_{i=1}^{N_{\rm F}} \lambda_i \hspace{0.1cm}I_0 \left(2 \pi \alpha_i \sqrt{u} \right) = x,
\eeq
where we leave implicit that $u$ should be thought of as a function of $x$. Taking $x\to0$ gives the equation that the zero-point energy satisfies 
\beq\label{eqEdgeJTdefM}
\sqrt{2\gamma E_0} \hspace{0.1cm} I_1\left(2\pi \sqrt{2\gamma E_0}\right) + 2 \pi  \sum_{i=1}^{N_{\rm F}} \lambda_i \hspace{0.1cm} I_0 \left(2\pi \alpha_i \sqrt{2\gamma E_0} \right)=0.
\eeq
These expressions can be verified by Taylor expanding in all the $\lambda$'s, comparing with the perturbative results \eqref{eq:E01}, \eqref{eq:E02}, and we checked that they agree up to seventh order in $\lambda$.

We find it remarkable that the defects appear linearly at the level of the string equation \eqref{eqJTdefMSE}. This is completely obscure from the expansion in terms of Weil-Petersson volumes. As we will see, all the nonlinearities can be traced back to the shift $E_0$ of the edge of the spectrum.

\subsubsection{All genus-zero amplitudes}

The procedure in the previous section can be generalized to the genus zero amplitude with any number of boundaries (including the one-boundary disc amplitude studied perturbatively in section \ref{sec:disc}, which we will treat as a special case in the next subsection). We will find that these all match the results from the matrix integral we met above. 

The gravitational calculation for the connected amplitude $\lb Z(\beta_1) \cdots Z(\beta_n) \rb_{C,g=0}$ with $n$ boundaries involves a sum over defects for a genus zero topology, adding $Z_{g=0,n,k}(\beta_1,\ldots,\beta_n)$ for all $k$ (and summing over species if $N_{\rm F}>1$). Using the explicit formula for the WP volumes, for the case of one specie of defects, the $n$-loop amplitude becomes
\bea
\lb Z(\beta_1) \cdots Z(\beta_n) \rb_{C,g=0} &=& \frac{e^{(2-n)S_0}}{2\pi^{n/2} (2\gamma)^{(n-2)/2}} \frac{ \sqrt{\beta_1 \cdots \beta_n}}{\beta_1 + \cdots + \beta_n} \nonumber\\
&&\hspace{-3cm}\times   \sum_{k=0}^\infty \frac{\lambda^k}{k!} \left.\left(\frac{\partial}{\partial x} \right)^{n+k-3} \left[ \left( e^{\frac{1}{2\gamma} u_{\rm JT}(x)(\beta_1 + \cdots + \beta_n)}\right)' \left( J_0 (2 \pi \alpha \sqrt{u_{\rm JT}(x)})\right)^k\right] \right|_{x=0}.
\eea
We can again evaluate this sum using the Lagrange reversion theorem. Since the derivation is almost identical to the case with $n=2$ we will just quote the final result:
\beq\label{eq:multiloopg0JT}
\lb Z(\beta_1) \cdots Z(\beta_n) \rb_{C,g=0} = e^{(2-n)S_0}\frac{\sqrt{\beta_1 \cdots \beta_n}}{2\pi^{n/2} (2\gamma)^{(n-2)/2}}\left.\left(\frac{\partial}{\partial x} \right)^{n-2} \frac{e^{-\frac{1}{2\gamma}u(x)(\beta_1 + \cdots + \beta_n)}}{\beta_1 + \cdots + \beta_n}  \right|_{x=0}.
\eeq
The function $u(x)$ appearing here is exactly the same one that appeared above for the special case $n=2$, defined thorugh \eqref{JTdefSE}. The matrix model multi-loop amplitude at genus zero was derived some time ago in \cite{Ambjorn:1990ji, Moore:1991ir}; for a recent discussion see \cite{mtms}. The expression derived in those references is precisely equivalent to \eqref{eq:multiloopg0JT}. Therefore we have shown here that at genus zero, or equivalently to leading order in $S_0$, JT gravity with a gas of defects is equivalent to a matrix integral
\beq
\lb Z(\beta_1) \ldots Z(\beta_n) \rb_{\rm JT+def} =\big\lb \Tr(e^{-\beta_1 H}) \ldots \Tr(e^{-\beta_n H}) \big\rb_{\rm MM} \ .
\eeq
When multiple defect flavors are present, we obtain \eqref{eq:multiloopg0JT}, where $u(x)$ is given by \eqref{eqJTdefMSE}. Again, this is easily derived using the second version of the reversion theorem in Appendix \ref{App:LRT}. In the next section we will analyze the density of states one can derive from these formulas, before discussion higher genus amplitudes.

\subsection{Density of States}\label{sec:DOSJTDEF}
We show that JT gravity with defects has a matrix integral dual, interpreted as an ensemble of random Hamiltonians \cite{Saad:2019lba}. From this perspective the spectral curve plays an important role and is related to the averaged density of states of the dual theory. In this section we will analyze this quantity. 

 The partition function with a single boundary can be derived using the same method as before, combining the exact formula for the genus zero WP volumes and the reversion theorem. We obtain the result
\beq
\lb Z(\beta) \rb_{g=0} =e^{S_0} \sqrt{\frac{\gamma}{2\pi \beta}} \int_0^\infty dx ~e^{-\frac{1}{2\gamma}u(x) \beta},
\eeq
where $u(x)$ is defined by \eqref{eqJTdefMSE}. Once again, this matches with the answer for a matrix integral with heat capacity equal to $u(x)$, see for example \cite{Okuyama:2019xbv, *Okuyama:2020ncd, Johnson:2019eik}. Using \eqref{eqJTdefMSE}, this gives us an exact expression in the defect weighting parameter $\lambda$. 

The density of states extracted from this formula can be obtained by an inverse Laplace transform. Changing the integration variable to $u$, we have the explicit expression
\beq\label{eq:ExactDOS}
\langle \rho(E)\rangle_{g=0} = \frac{\gamma \hspace{0.1cm} e^{S_0} }{2\pi } \int_{2\gamma E_0}^{2\gamma E} \frac{du}{\sqrt{2\gamma E-u}}\left( I_0\left(2\pi \sqrt{u}\right) + \sum_i \lambda_i \frac{2\pi \alpha_i}{\sqrt{u}} I_1\left(2\pi \alpha_i \sqrt{u} \right)\right) .
\eeq
The integrand is basically the derivative with respect to $u$ of the string equation. In the right hand side the density of states depends on the coupling $\lambda$ not only through the linear term in the integrand, but also through the implicit $\lambda$ dependence of the zero-point energy $E_0$. This is defined as the largest root of an equation that we repeat here for convenience:
\beq\label{eqEdgeJTdefM2}
\sqrt{2\gamma E_0} \hspace{0.1cm} I_1\left(2\pi \sqrt{2\gamma E_0}\right) + 2 \pi  \sum_i \lambda_i \hspace{0.1cm} I_0 \left(2\pi \alpha_i \sqrt{2\gamma E_0} \right)=0.
\eeq
Unless the solution to this equation is degenerate, this immediately implies that we have a smooth density with a square-root edge
\begin{equation}\label{eq:exactsqrtedge}
	\rho_\text{disc}(E) \approx \sqrt{E-E_0}\, 
\end{equation}
as we found perturbatively in section \ref{sec:disc}. The prefactor is proportional to the integrand in the expression above evaluated at $u\to 2\gamma E_0$. 
 
Now, we should check whether the model we have arrived at makes sense as a double-scaled matrix model (at least perturbatively in the genus expansion; we leave aside possible nonperturbative instabilities for now which are also present in JT gravity). For this, we require that the density of states $\rho_\text{disc}$ is positive for all $E>E_0$. Now that we have an expression for this density at finite $\lambda$, we can explore this for a range of parameters $\alpha$ and $\lambda$. For simplicity we will restrict to the case of a single species of defect.

 First we can look at large energies. In this regime, it is possible to approximate the density of states by $\rho(E) \sim e^{2 \pi \sqrt{2\gamma E}} $ for $\alpha<1$ and $\rho(E) \sim \lambda e^{2 \pi \alpha \sqrt{2\gamma E}}$ for $\alpha >1$. This change of the ultraviolet behavior for $\alpha>1$ is not unexpected from the gravitational perspective, since such defects would be favored to proliferate and destroy (or at least substantially modify) the asymptotic region of the geometry. As before, we will now focus on $0<\alpha<1$.

\begin{figure} 
\begin{center}
\hspace{-0.3cm}
\begin{tikzpicture}[scale=1]
\pgftext{\includegraphics[scale=0.4]{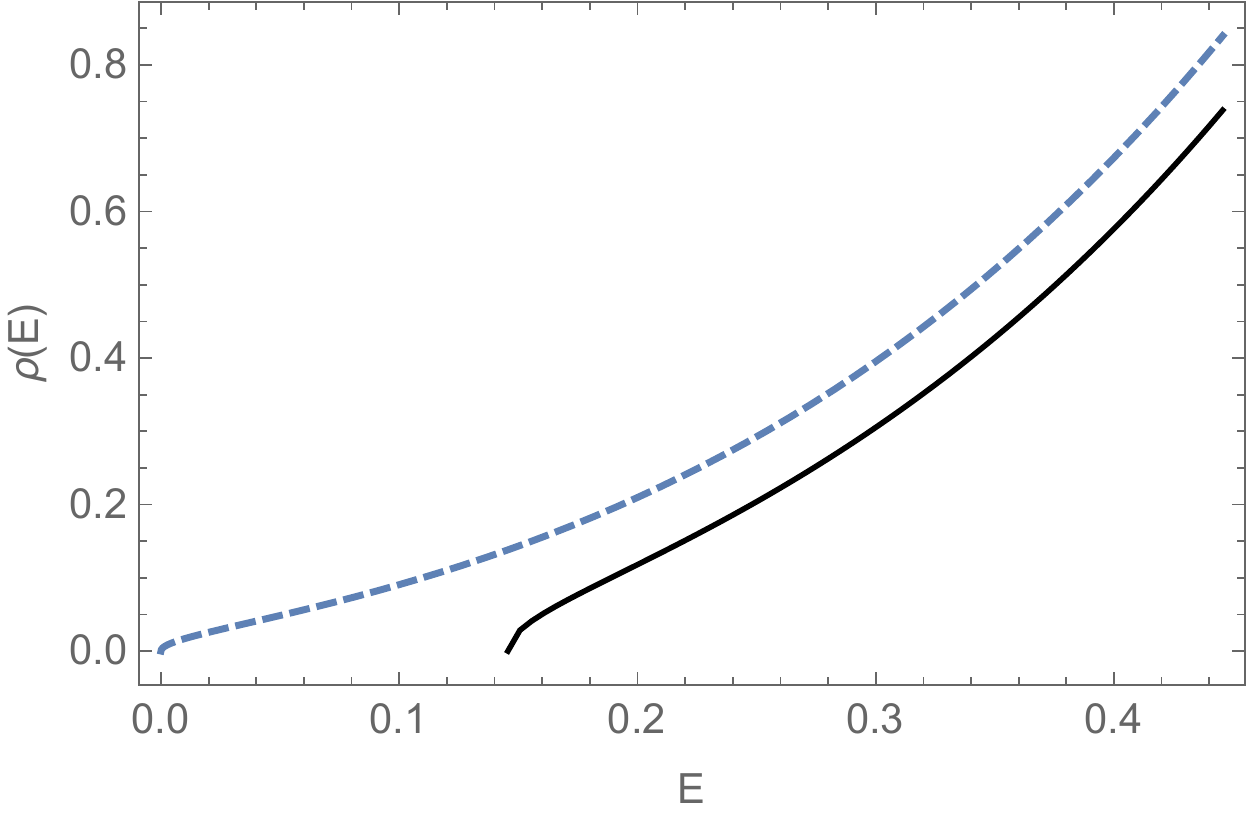}} at (0,0);
      \draw (0,-2) node {\small (a) $\alpha=1/2$, $\lambda=-0.1$};
    \end{tikzpicture}
    ~~~~~\begin{tikzpicture}[scale=1]
\pgftext{\includegraphics[scale=0.4]{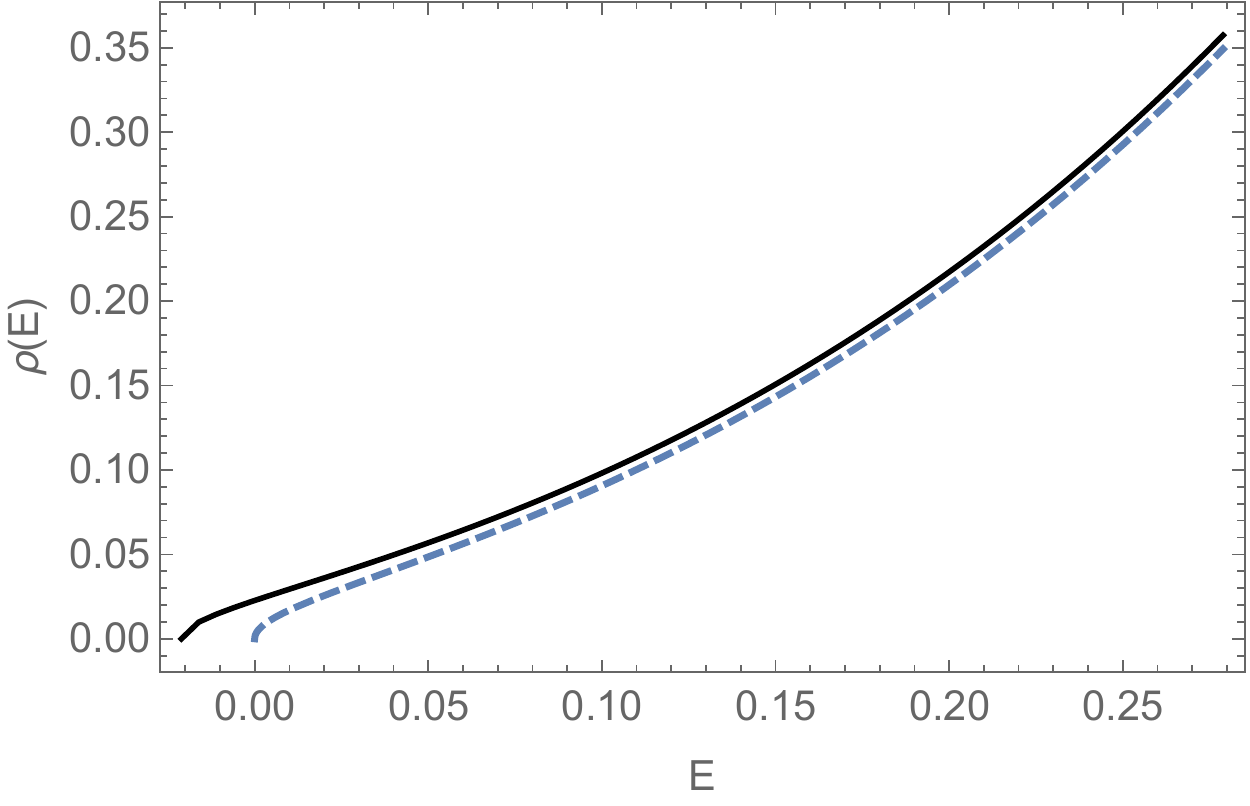}} at (0,0);
      \draw (0,-2) node {\small (b) $\alpha=1/2$, $\lambda=0.01$};
    \end{tikzpicture}
 ~~~~~   \begin{tikzpicture}[scale=1]
 \pgftext{\includegraphics[scale=0.4]{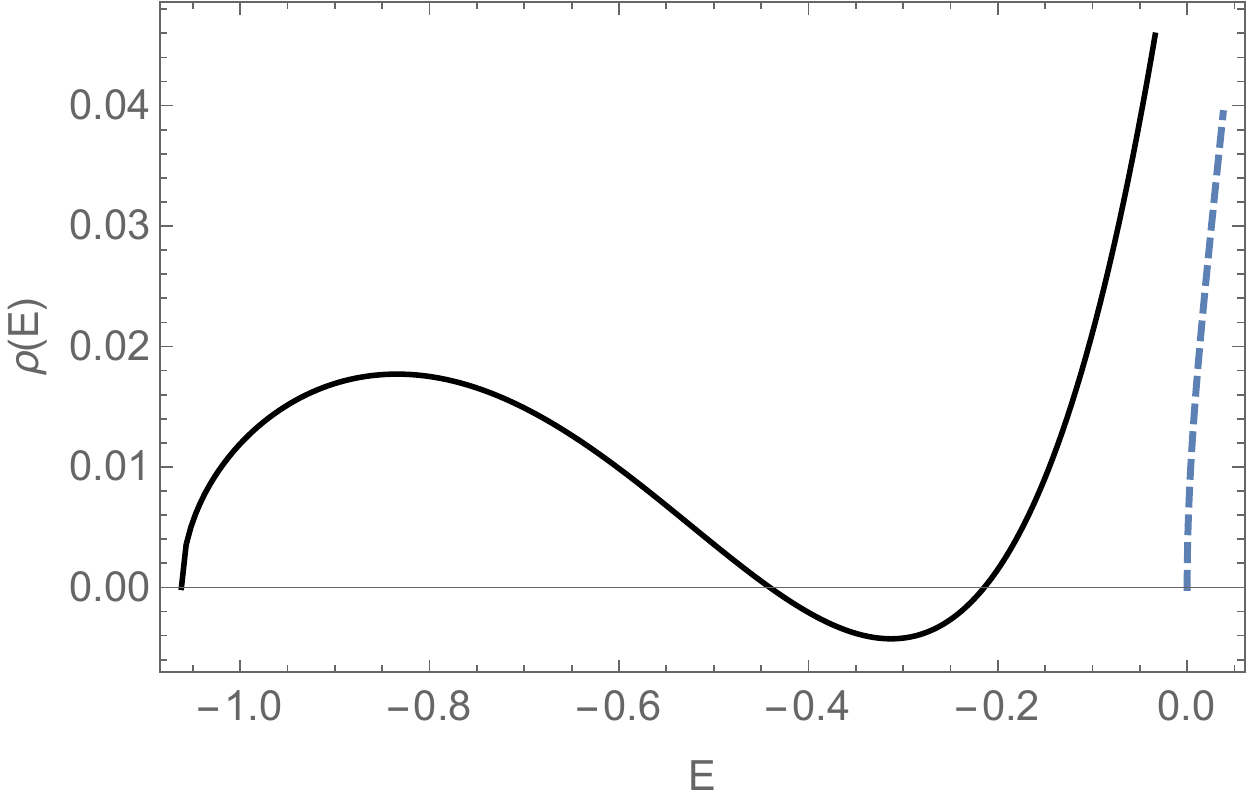}} at (0,0);
      \draw (0,-2) node {\small (c) $\alpha=1/2$, $\lambda=0.08$};
    \end{tikzpicture}
    \end{center}
    \vspace{-0.4cm}
    \caption{\small \label{fig:DOSnumerical} Density of states of JT gravity with (black) and without (dashed blue) defects, for $2\gamma=1$. (a) For $\alpha=1/2<\alpha_c$ and $\lambda=-0.1<0$ we see $E_0>0$ as expected and the theory is fine (b) For $\lambda = 0.01$ smaller than $\lambda_c(\alpha=1/2)\approx 0.06$ we get $E_0<0$ and the theory is fine (c) For $\lambda=0.08>\lambda_c(1/2)$ the density of states becomes negative in a finite range of energies.}
\end{figure}

We find that for all such $\alpha$, there is a range of $\lambda$ for which $\rho_\text{disc}$ is positive. However, this can fail for sufficiently large $\lambda$. There are several cases to consider; we show some representative examples from numerical integration of \eqref{eq:ExactDOS} in figure \ref{fig:DOSnumerical}.  
\begin{itemize}
\item For $\alpha_c < \alpha <1$, where the critical value $\alpha_c \approx 0.627$ is the ratio between the first zero of $J_0$ and the first zero of $J_1$, the density of states is positive for all energies, for any choice of $\lambda$.
\item For $0<\alpha<\alpha_c$ and $\lambda<0$ the density of states is positive. 
	 An example is shown in figure \ref{fig:DOSnumerical}(a).
\item For $0<\alpha<\alpha_c$ and $\lambda>0$, the theory is well behaved for couplings smaller than a critical value $\lambda_c(\alpha)$. This critical coupling is given implicitly by solving the following two equations simultaneously
\begin{gather}
	\sqrt{2\gamma E_c}\hspace{0.1cm} I_1\left(2\pi \sqrt{2\gamma E_c} \right) + 2\pi \lambda_c \hspace{0.1cm}I_0 \left( 2\pi \alpha \sqrt{2\gamma E_c}\right)=0,\label{eq:Ec} \\
\sqrt{2\gamma E_c}\hspace{0.1cm}I_0\left(2\pi \sqrt{2\gamma E_c} \right)+ 2\pi \alpha \lambda_c \hspace{0.1cm}I_1 \left(2\pi \alpha\sqrt{2\gamma E_c}\right) =0 .
\end{gather}
The first equation gives the zero-point energy at the critical coupling $\lambda_c$. The second equation determines the point where the solution to the first equation becomes degenerate, which means that the suppressed prefactor in \eqref{eq:exactsqrtedge} vanishes. Eliminating the critical zero-point energy $E_c$ gives the desired relation $\lambda_c(\alpha)$. Solving this numerically, we find that $\lambda_c(\alpha)$ goes smoothly between values $\lambda_c(0)\approx 0.03$ and $\lambda_c(\alpha_c)\approx 0.12$. For small enough $\lambda \lesssim 0.03$, the theory has a positive density of states for any $\alpha<1$. An example is shown in figure \ref{fig:DOSnumerical}(b). For the KK instantons, the coupling $\lambda$ is exponentially small in $S_0$, so this is the relevant case.
\item For $0<\alpha<\alpha_c$ and $\lambda>\lambda_c(\alpha)$, the density of states is no longer positive for all $E$, so we cannot interpret our model as a double-scaled matrix integral. The solution to \eqref{eq:Ec} determining the edge of the spectrum $E_0$ jumps to a smaller value, and we then find a finite range of energies $E_1<E<E_2$ for which $\rho_{\text{disc}}(E)<0$. A representative example is shown in figure \ref{fig:DOSnumerical}(c).
\end{itemize}

\subsection{Higher genus}\label{sec:highergenus}

We have now defined a matrix integral which reproduces all genus zero amplitudes of JT gravity with defects by explicit computation. We do not have a closed form expression for the volumes of moduli space at higher genus, so we must take a slightly more indirect approach. In this section, we give a formal proof that the amplitudes with defects \eqref{eq:Vexp2}  satisfy the topological recursion of a matrix integral, using a theorem of Eynard and Orantin. We thus have a duality to all orders in the genus expansion, generalizing \cite{Saad:2019lba}. To simplify we will set $\gamma=\frac{1}{2}$ in this section. Since the proof is rather formal, we also present explicit checks at higher genus in Appendix \ref{app:checks}, where we also write down the recursion explicitly. 

We begin by introducing some notation following \cite{Saad:2019lba}. First define the coordinate $z$ in terms of the matrix eigenvalue $E$ as $z^2=-E$. The spectral curve is defined through the disc density of states as $y(z)=- i \pi e^{-S_0} \rho_{\rm disc}(E)$. For example pure JT gravity has $y_{\rm JT}(z) = \sin(2\pi z)/(4\pi)$. Then define the functions 
\beq
W_{g,n}(z_1,\ldots, z_n) = (-1)^n 2^n z_1 \ldots z_n R_{g,n}(-z_1^2,\ldots, -z_n^2),
\eeq
proportional to the genus $g$ contribution to the connected correlator of $n$ resolvents,
\beq
R_n(E_1,\ldots, E_n) \equiv \left\lb {\rm Tr} \frac{1}{E_1 - H} \ldots {\rm Tr} \frac{1}{E_n - H} \right\rb_{\rm conn.}.
\eeq
The expectation value is taken within the matrix model ensemble. This is related through a Laplace transform to the connected partition function ${\rm Tr}(e^{-\beta H})$ correlator which we denote by $Z_{g,n}(\beta_1,\ldots, \beta_n)$.  For the exceptional case $g=0,n=1$ the resolvent of the double-scaled integral is not convergent, but we can instead define $W_{0,1}(z)=-2z y(z)$.

The main result we will use is the deformation theorem derived by Eynard and Orantin, Theorem 5.1 of \cite{Eynard:2007kz} (see also section 4.3.2 of \cite{Eynard:2015aea} and for a summary of other interesting properties of the topological recursion \cite{Eynard:2014zxa}). This starts with a given solution of the topological recursion and considers a one-parameter deformation, parameterized by $\lambda$ (which for us will be the defect amplitude). We denote various quantities after the deformation by $y(z;\lambda)$, $W_{g,n}(z_1,\ldots, z_n;\lambda)$ and $Z_{g,n}(\beta_1,\ldots, \beta_n;\lambda)$. The undeformed case corresponds to $\lambda \to 0$, returning us to $y(z)$, $W_{g,n}(z_1,\ldots, z_n)$ and $Z_{g,n}(\beta_1,\ldots, \beta_n)$.

\paragraph{Deformation Theorem:} The statement of Eynard and Orantin's theorem requires the choice of a closed curve $\gamma$ in the complex plane and a function $f(z)$ (which for simplicity we take to be independent of $\lambda$). We require that the spectral curve and the cylinder amplitude behave as follows under the deformation:
\begin{align}
	\frac{\partial}{\partial \lambda} y(z_1;\lambda)  &= -\frac{1}{2z_1}\oint_{\gamma} \frac{dz}{2\pi i} f(z)  W_{0,2}(z,z_1;\lambda) ,\label{eq:cond1} \\
 \frac{\partial}{\partial \lambda}  W_{0,2}(z_1,z_2;\lambda) &= \oint_\gamma \frac{dz}{2\pi i} f(z) W_{0,3}(z,z_1,z_2;\lambda). \label{eq:cond2}
\end{align}
Then, applying topological recursion with these deformed amplitudes as input, the solution satisfies
\beq\label{eq:EOsol}
\frac{\partial}{\partial \lambda}  W_{g,n}(z_1,\ldots, z_n;\lambda)  = \oint_\gamma  \frac{dz}{2\pi i}  f(z) W_{g,n+1}(z,z_1,\ldots, z_n;\lambda). 
\eeq
The deformed amplitudes are thus completely determined at all genus by the closed curve $\gamma$ and function $f(z)$ in a simple way.  Since the resolvent correlators are invariant under permutations of its arguments the choice of the variable being integrated in the right hand side is irrelevant.

\paragraph{Solution at finite $\lambda$:}
From the above, it is simple to construct a solution to topological recursion to all orders in $\lambda$:
\begin{equation}\label{eq:Wexpansion}
	W_{g,n}(z_1,\ldots, z_n;\lambda) = \sum_{k=0}^\infty \frac{\lambda^k}{k!} W_{g,n,k}(z_1,\ldots, z_n).
\end{equation}
For our model, the amplitudes $W_{g,n,k}(z_1,\ldots, z_n)$ will be equivalent to the expansion coefficients $Z_{g,n,k}(\beta_1,\ldots, \beta_n)$ in \eqref{eq:Zexpansion}, up to an integral transform to convert between resolvents and partition functions. By repeated use of the relation \eqref{eq:EOsol}, the $W_{g,n,k}$ are given by multiple integrals, as
\begin{equation}\label{eq:Wcoefficients}
W_{g,n,k}(z_1,\ldots, z_n) = \oint_\gamma \frac{d\tilde{z}_1}{2\pi i} f(\tilde{z}_1) \cdots \oint_\gamma \frac{d\tilde{z}_k}{2\pi i} f(\tilde{z}_k) W_{g,n+k}(z_1,\ldots,z_n, \tilde{z}_1, \ldots, \tilde{z}_k).
\end{equation}
The same equation applies for the cylinder \eqref{eq:cond2}, and from \eqref{eq:cond1} we have a similar expansion for the spectral curve,
\begin{equation}
	y(z;\lambda) = y(z) -\frac{1}{2z}\sum_{k=1}^\infty \frac{\lambda^k}{k!} W_{0,1,k}(z).
\end{equation}
If the family of solutions to topological recursion is analytic in $\lambda$ in a neighborhood of the origin, this series converges to the solution. In fact, it is sufficient for the input to the topological recursion (the deformed spectral curve $ y(z;\lambda)$ and $W_{0,2}(z_1,z_2;\lambda$)) to be analytic, since the topological recursion preserves analyticity. This is the case for our example of the sum over defects. 

\paragraph{JT gravity with defects:} We will now apply the general result above to our specific problem. We will find an appropriate choice of contour $\gamma$ and function $f(z)$ to implement the deformation induced by inclusion of defects, before showing that the coefficients $W_{g,n,k}$ in \eqref{eq:Wcoefficients} are precisely those given by the path integral with $k$ defects \eqref{eq:Vexp2}.

From equation \eqref{eq:Wexpansion}, it is roughly clear what we need to do: the extra $\tilde{z}$ variables give us extra boundaries in JT gravity, and we need to choose $f$ and $\gamma$ so that the integrals act to replace those boundaries with defects. Let us make this more explicit. The Weil-Petersson volumes are simply related to the expansion of the resolvent in JT gravity by a Laplace transform \cite{Eynard:2007fi},
\begin{equation}\label{eq:WJT}
	W_{g,n}(z_1,\ldots,z_n) = \int_0^\infty b_1 db_1 e^{-b_1 z_1} \int_0^\infty b_n db_n e^{-b_n z_n} V_{g,n}(b_1,\ldots,b_n).
\end{equation}
This is equivalent to the JT path integral \eqref{eq:Vexp1}, with the trumpet factor replaced by the Laplace transform so that we get the resolvent rather than the partition function. Our defect expansion of the resolvent \eqref{eq:Vexp2}, using the WP volumes with cone points \eqref{eq:coneVols}, is then
\begin{equation}\label{eq:Wgnk}
	W_{g,n,k}(z_1,\ldots,z_n) = \int_0^\infty b_1 db_1 e^{-b_1 z_1} \int_0^\infty b_n db_n e^{-b_n z_n} V_{g,n}(b_1,\ldots,b_n,\underbrace{2\pi i\alpha,\ldots,2\pi i\alpha}_k).
\end{equation}
 This can be derived by combining the relation between resolvents and partition function correlators 
\beq\label{baba}
W_{g,n,k}(z_1,\ldots,z_n) = 2^n z_1 \ldots z_n \int_0^\infty d\beta_1 \ldots d\beta_n e^{-(\beta_1 z_1^2 + \ldots + \beta_n z_n^2)} Z_{g,n,k}(\beta_1,\ldots, \beta_n). 
\eeq
with the identity 
\beq\label{abab}
e^{-b z} = 2 z \int_0^\infty d\beta e^{-\beta z^2 } Z_{\rm trumpet}(\beta,b),~~~~~Z_{\rm trumpet}(\beta,b) = \frac{e^{-\frac{b^2}{4\beta}}}{2\sqrt{\pi \beta}}.
\eeq

Comparing the gravity answer \eqref{eq:Wgnk} to the deformation of the topological recursion \eqref{eq:Wcoefficients}, we should choose the integral $\oint \frac{d\tilde{z}}{2\pi i} f(\tilde{z}) \cdots$ to implement the inverse Laplace transform, stripping off some of the integrals in \eqref{eq:WJT} and setting lengths in the Weil-Petersson volumes to $b=2\pi i \alpha$. Because the volumes are all even polynomials in $b$, this is equivalent to choosing
\begin{equation}
	f(\tilde{z}) = \frac{\sin(2\pi\alpha \tilde{z})}{2\pi \alpha},
\end{equation}
and taking $\gamma$  to be any curve encircling the origin, as we will check in detail now.

Each term in the WP volume will depend on the lengths $b$ as a monomial $b^{2m}$. Taking the Laplace transform, this contributes to the JT resolvent as $(2m+1)! \tilde{z}^{-2(m+1)}$. Now we insert this into \eqref{eq:Wcoefficients}, where the integral picks out the residue of $(2m+1)! \tilde{z}^{-2(m+1)} \frac{\sin(2\pi\alpha \tilde{z})}{2\pi \alpha}$ at $\tilde{z}=0$, which is precisely $(2\pi i \alpha)^{2m}$. The overall result is therefore to evaluate the WP volume with $b$ replaced by $2\pi i \alpha$, as required by \eqref{eq:Wgnk}.

This analysis applies for $n\geq 2$ and all genus. The $n=1,g=0$ case, which is the disc with defects giving the deformed spectral curve $y(z;\lambda)$, can be treated similarly. There is one exception, coming from the $g=0,n=1,k=1$ disc amplitude with a single defect, where the corresponding Weil-Petersson volume is not defined. This contributes to leading order in the deformation of the spectral curve, which from section \eqref{sec:disc} is given by
\begin{equation}
	y(z;\lambda) = \frac{\sin(2\pi z)}{4\pi} - \lambda \frac{\cos(2\pi \alpha z)}{2z}  + \mathcal{O}(\lambda^2).
\end{equation}
This can be checked separately to follow from the same deformation, though the curve $\gamma$ must be large enough to enclose the point $z$ at which we are evaluating the spectral curve.\footnote{The theorem still applies, since we can perform the deformation before taking the double-scaling limit, introducing a cutoff $a$ as in section 2.4 of \cite{Saad:2019lba}. Then for any $a$ we can choose a fixed curve $\gamma$, which we take to infinity as we take the double-scaled limit $a\to \infty$.}

This establishes that the addition of defects is a special case of Eynard and Orantin's deformation theorem, and hence the resulting genus expansion obeys the topological recursion. The argument generalizes straightforwardly to multiple defects, by treating each species of defect as a separate deformation. 

\section{Back to 3D gravity}\label{sec:backto3D}

Our analysis of 3D gravity near extremality in section \ref{Sec:PI3Dgrav} motivated us to study a simplified model for the path integral on Seifert manifolds. This simplified model is an example of JT gravity with defects, as studied in section \ref{sec:JT}. When we included only classical solutions, this model had a potential problem with a negative density of states, similarly to the Maloney-Witten-Keller partition function \eqref{eq:rhoMWK}. We saw that this problem was naturally resolved by including the path integral with multiple defects, and instead we found a shift in the edge of the spectrum of energies. Our proposal is that the same mechanism applies in the full three-dimensional problem: a sum over Seifert manifolds cures the negativity arising from the $SL(2,\mathbb{Z})$ black holes, and we instead have a nonperturbative shift of the extremality bound.

To  establish this mechanism conclusively, and for a more detailed quantitative study, it is important to solve the complete three-dimensional problem of computing the path integral on Seifert manifolds. Nonetheless, in the regime of large spin and very low temperature where the effect is relevant, we have principled reasons to believe that the dimensionally reduced theory is a good approximation. To that end, in this section we will also discuss and quantify various sources of corrections to this approximation.

Here, we will use a shifted and rescaled energy which is natural from the near-extremal point of view, measured in units of the characteristic scale of quantum fluctuations:
\beq
E_{\rm JT} \equiv \frac{\ell_3}{8G_N} (E-|J|).
\eeq
The leading contribution to the density of states near the extremal limit is given by the BTZ black hole
\beq
\rho_0(E_{\rm JT}) \sim e^{S_0(J)} \sinh\left( 2 \pi \sqrt{E_{\rm JT}}\right)
\eeq
where we ignore a (spin dependent) multiplicative constant that can in any case be absorbed as a logarithmic correction to $S_0(J)$. From the 2D JT perspective this is the contribution to the partition function from the disk with no defects. The leading correction to this contribution comes from the $SL(2,\mathbb{Z})$ black hole with $\rat=\tfrac{1}{2}$, given near extremality by 
\beq
\rho_{1} (E_{\rm JT}) \sim 4\pi^2 e^{S_0(J)} \lambda_{\frac{1}{2}} \frac{\cosh\left(\pi \sqrt{E_{\rm JT}}\right)}{\sqrt{E_{\rm JT}}},~~~~~\lambda_{\frac{1}{2}} \equiv \frac{1}{2} \frac{\ell_3}{32 \pi^2 G_N} (-1)^J e^{-\frac{1}{2} S_0(J)}.
\eeq
From the JT perspective this is precisely the one-defect contribution with $\alpha=1/2$ and $\lambda = \lambda_{1/2}$ defined above. For the exponential classical dependence on parameters, including the important factor of $(-1)^J$, this agrees with the action \eqref{eq:3DJTinstanton} obtained earlier.


The potential pathologies in the spectrum occur at exponentially low temperatures, so we will focus on the following `double scaling' limit  \cite{Benjamin:2019stq} at very low energies above extremality:
\beq\label{eq:doubleLimit}
J\to\infty, \quad E_{\rm JT} \to 0,\qquad\text{with}\quad \frac{E_{\rm JT}}{|\lambda_{\frac{1}{2}}|} \sim E_{\rm JT} \, e^{\frac{1}{2} S_0(J)} \quad\text{fixed}.
\eeq
In this regime the $\rat=\tfrac{1}{2}$ family of $SL(2,\mathbb{Z})$ black holes is relevant at leading order, but all others are negligible. Then the density of states from classical solutions is approximately 
\beq
\rho(E_{\rm JT}) \sim 4 \pi^2 e^{S_0(J)} \left( \sqrt{E_{\rm JT}} + \frac{\lambda_{1/2}}{\sqrt{E_{\rm JT}}} \right).
\eeq 
In our double-scaling limit, these two terms are of the same order. If $\lambda_{\frac{1}{2}}<0$ (odd $J$), then for $E_{\rm JT} < |\lambda_{\frac{1}{2}}|$ the density of states can become negative. 

But now, we can apply the lesson of section \ref{sec:disc} and include the path integral over multiple defects. Here, this defects are KK instantons, which lift to smooth Seifert three-manifolds with $U(1)$ symmetry. In the near-extremal limit we are studying, we approximate the full path integral by the dimensional reduction, so we can directly apply the results of the previous section. The resulting density of states is now free of negativities, and instead we have a shift of the extremal energy:
\beq
\rho \left(E_{\rm JT} \right) \to e^{S_0(J)} \sqrt{E_{\rm JT} - E_{\rm JT}^{(0)}} ,~~~~E_{\rm JT}^{(0)} = - \frac{1}{2} \frac{\ell_3}{32 \pi^2 G_N} (-1)^J e^{-\frac{1}{2} S_0(J)}.
\eeq
Here, we wrote the shift in the extremal energy is written in terms of the variable $E_{\rm JT}$. Going back to the original energy $E$, the extremal value $E_0(J)$ is shifted by a non-perturbative correction anticipated in the introduction
\beq\label{eq:extremalityshiftS4}
E_{0}(J) = |J| -  \frac{1}{(2 \pi)^2 }(-1)^J  e^{-\frac{1}{2} S_0(J)}.
\eeq

Now, there is one important contribution to the three-dimensional path integral which is not captured by the dimensional reduction in this limit, namely the one-loop effect of Kaluza-Klein modes of the graviton. In the large spin limit, the transverse circle direction is very large so these modes are not heavy (though their interactions are suppressed \cite{Ghosh:2019rcj}, see also \cite{higherdim}), and they should be included. Intuitively, this loop determinant accounts for boundary gravitons on top of our solutions, which correspond to Virasoro descendants in the dual. We therefore expect our results to capture the density of Virasoro primary states, not of all states; these are not greatly different, since at large central charge, most states carry a very small amount of their energy in descendants. This intuition is supported by matching the JT density of states, as well as the density with a single defect, to the energy dependence of the low-energy spectrum of primaries from BTZ and $SL(2,\ZZ)$ black holes.  For a single KK instanton, we can simply match $\lambda$ with the coefficient in the exact $SL(2,\ZZ)$ contribution from three dimensions, since we have the correct energy dependence. Our underlying assumption is that the graviton KK modes do not introduce new effects for multiple KK instantons, at least at leading order. This is not unreasonable, since the gravitons propagate only at the boundary, but demands more careful justification from a full three-dimensional calculation \footnote{We can explicitly check this intuition in a similar situation using the results of \cite{CotlerJensenDoubleTrumpet}. Their partition function for primary states reduces to the JT gravity answer for the double trumpet. This is not unreasonable since the WP measure used in the JT gravity calculation can be deduced from $SL(2,\mathbb{R})$ BF theory, which in turn is a reduction of Chern-Simons. We expect the same to be true for the geometries considered here.}.

If we were to go to lower temperatures still, the contributions from other families of $SL(2,\ZZ)$ black holes become important. These should also give rise to shifts of the extremality bound, with the $\rat=\frac{p}{q}$ contribution proportional to
\begin{equation}
	\delta E_0(J) \propto e^{-2\pi i \frac{p}{q}J}e^{(1-q^{-1})S_0(J)}
\end{equation}
at leading order. The proportionality constant includes some additional spin-independent phase originating with the one-loop effects discussed above (in particular, from the fact that the vacuum has a null descendant); this can be seen from the $SL(2,\ZZ)$ black hole contribution in \eqref{eq:rhobtw1}. The imaginary part cancels between terms labelled by $\rat$ and $1-\rat$. When we add the contributions from all $p$ coprime to a given $q$, these phases give rise to Kloosterman sums, which appear in the MWK density \cite{Maloney:2007ud,Keller:2014xba,Benjamin:2019stq}. From this, the BTZ extremality bound appears to depend on spin in a complicated number-theoretic way.

Finally, we analyze the order of magnitude of corrections to this result and see that they are negligible in the double scaling limit \eqref{eq:doubleLimit}. The corrections can be suppressed either due to the large spin limit or due to the small energies above extremality. 

\paragraph{Other saddles:} First, we look at contributions from KK instantons with $q>2$. At small energies their contribution to the density of states is of order $\rho_q(E) \sim e^{-(1-\frac{1}{q}) S_0}/\sqrt{E_{\rm JT}}$, with $q>2$. Since we are focusing on energies of order $E_{\rm JT} \sim e^{-\frac{1}{2} S_0}$ all these contributions are exponentially suppressed in the large spin limit by a factor of $e^{-(\frac{1}{2}-\frac{1}{q})S_0} \ll 1$ and therefore can be neglected. It is crucial for this conclusion to hold that contributions from  $q>2$ do not grow any faster at low energies than those with $q=2$.

\paragraph{Corrections to dilaton potential from dimensional reduction:} A second source of corrections is the fact that in the throat, the 2D dilaton gravity which appears from the reduction of 3D gravity has non-linear corrections to the dilaton potential. The leading correction to the dilaton potential in the near extremal limit is $U(\phi) = -2\phi + \varepsilon \phi^2$, where $\varepsilon$ is a fixed coupling of order $1/S_0$ which can be read off from equation \eqref{eq:einsteinDilaton}. This produces a shift in the free energy $\delta \log Z \sim \varepsilon \beta^{-2}$, computed in detail by Kitaev and Suh \cite{Kitaev:2017awl}, which translates in a relative shift of the density of states $\delta \rho/\rho_0 \sim \varepsilon E_{\rm JT}^{2}$. In the limit we are interested in this is of order $\delta \rho/\rho_0 \sim e^{-S_0}/S_0$ which is also negligible: from our very low energy perspective it is a UV effect, and we are interested in the deep IR.

\paragraph{Other topologies:} Another source of corrections are higher topologies appearing in the near-horizon region. We can study a class of these which arise in JT gravity, by adding handles to AdS$_2$. The relative shift in the density of states from adding a handle on the disc is $\delta \rho/\rho \sim e^{-2S_0}/E_{\rm JT}^3$. This grows at low energies, but not sufficiently quickly to overcome the exponential suppression: for $E_{\rm JT}\sim e^{-\frac{1}{2}S_0}$ this is of order $\delta \rho/\rho \sim e^{-\frac{1}{2}S_0}$.

Interestingly, these higher genus corrections become important at energies $E_\text{JT}\sim e^{-\frac{2}{3}S_0}$, which is the same scale at which the $q=3$ KK instantons become relevant. At very low energies, we expect the handles to resum into the universal form for a matrix integral with a square-root density at genus zero, namely
\beq
\rho(E)= e^{\frac{2S_0}{3}}\Big[ {\rm Ai}'(\xi)^2-\xi {\rm Ai}(\xi)^2 \Big],~~~~\xi = - e^{\frac{2S_0}{3}} (E-E_0).
\eeq
We do not then have a sharp edge of the spectrum, since it is smoothed out on this scale, and has an exponentially small tail for $E<E_0$. The shifts in $E_0$ from the $SL(2,\ZZ)$ black holes at $q>2$ are therefore not particularly meaningful, since the spectrum is expected to be smooth at the relevant scale.

\section{Discussion}\label{sec:discussions}

Previous attempts to define the Euclidean path integral of pure 3D gravity \cite{Maloney:2007ud, Keller:2014xba} have considered only the quantum fluctuations around saddle point geometries, and obtained results in tension with an interpretation as a unitary quantum mechanical theory \cite{Benjamin:2019stq}. We have proposed a mechanism to resolve this tension, by including certain topologies in the path integral for which there is no classical solution. We have not computed the path integral over such topologies in the full theory, but instead reduced to a more tractable problem by concentrating on the near extremal limit, where their effects are most important. In this regime the physics is approximated by a dimensionally reduced two dimensional theory of JT gravity, with the addition of new instanton-like objects inherited from the reduction from three dimensions. These objects correspond to points where the circle fibration over which we are reducing becomes degenerate, so the two-dimensional geometry is singular while the full three-dimensional geometry remains smooth. The corresponding three-geometries --- Seifert manifolds --- are the new topologies we include in the path integral, and we argue that they cure the negative density of states found in \cite{Benjamin:2019stq}, replacing it with a nonperturbative shift in the energy of extremal rotating BTZ black holes.

In the process we needed to understand the path integral of JT gravity with a sum over defects, which we studied in its own right in section \ref{sec:JT}. We discovered that this family of theories is dual to a matrix integral, generalizing the pure JT result of \cite{Saad:2019lba}.   

We conclude with some comments on open questions and future directions. 

\subsection{An ensemble dual for three-dimensional pure gravity?}

Our cure for the pathological negative density of states revives the hope that there may be a consistent theory of pure 3D quantum gravity, with only metric degrees of freedom. This relied on two-dimensional calculations closely related to those in JT gravity \cite{Saad:2019lba}, so we are led naturally to the idea that a dual CFT description should be of the same form: that is, we do not have a single dual CFT, but rather an ensemble of duals. Such an interpretation would explain the second pathology of the MWK density of states, namely that it is continuous: this is not consistent for a single theory, but is the expected result from an average over a family of theories.

The ensemble interpretation offers an explanation for the path integral over geometries that connect several disconnected Euclidean boundaries, such as those discussed in section \eqref{sec:twoBoundary}. The simplest of these  corresponds to  a three-dimensional geometry with the topology of a torus times an interval, with two torus boundaries living at the ends of the interval; work computing the path integral on this topology \cite{CotlerJensenDoubleTrumpet} appeared while this paper was in preparation. Similar `spacetime wormhole' geometries connecting disjoint boundaries have also made a recent appearance in discussion of the information paradox \cite{Almheiri:2019qdq,Penington:2019kki}, and put the `baby universe' discussions of \cite{Coleman:1988cy,Giddings:1988cx,Giddings:1988wv} in an asymptotically AdS context \cite{Marolf:2020xie}. At least naively, these wormholes give rise to `connected correlators' between boundaries, so the partition function on two disjoint tori is different from the product of the separate partition functions on each torus. This is incompatible for single local dual theory, but can be interpreted as a result of statistical correlations in an ensemble of theories.

For two-dimensional gravity, a dual description is a one-dimensional quantum mechanics, for which matrix integrals provide a natural class of ensembles. For 3D gravity, we instead require a two-dimensional CFT dual, and there is no such obvious candidate ensemble to select from, at least for irrational theories (see \cite{Maloney:2020nni, Afkhami-Jeddi:2020ezh} for discussion of an ensemble of free theories and the possibility of an exotic gravitational dual). Our findings strongly suggest that the spectrum of Virasoro primaries at a given spin is well-described by a matrix integral near extremality. However, this cannot be the whole story, since it contains no sign of spatial locality. An ensemble of consistent local CFTs certainly requires correlations between sectors of different spins (for example to impose modular invariance), and seems likely to also require deviations from random matrix statistics even within a sector of fixed spin. In particular, without marginal operators we do not expect a continuous moduli space of theories, but rather a set of isolated CFTs; it is hard to imagine what an appropriate ensemble may look like in this case. Perhaps there is a large set of CFTs that resemble pure gravity, which mimic a continuum in a large central charge limit.

We also have more dynamical data and constraints coming from boundaries of higher genus. For example, we can interpret the results of \cite{Cardy:2017qhl,Collier:2019weq} as an ensemble average (rather than a microcanonical average) for the OPE coefficients of pure state black holes, arising from a gravitational calculation of genus 2 partition functions. More gravitational calculations with different boundary conditions and topologies will be helpful to elucidate the properties of a putative ensemble (see \cite{CotlerJensenDoubleTrumpet,Belin:2020hea} for first steps in this direction). Such results provide us with the opportunity of a window into the statistics of chaotic quantum systems \cite{Saad:2019pqd,Pollack:2020gfa,Collier:2018exn,Belin:2020hea}. Another possible source of inspiration is provided by the two-dimensional analog of the SYK model studied in \cite{Murugan:2017eto}; while not directly related to gravity, this model is a tractable example of an ensemble of irrational CFTs (other relevant models are \cite{Berkooz:2016cvq, Turiaci:2017zwd}).

\subsection{The full three-dimensional path integral}

To make our off-shell path integrals tractable, we truncated the full three-dimensional theory to a two-dimensional sector. We have argued that this truncation is a good approximation to the physics of interest near extremality, but it is nonetheless important to upgrade this to a full three-dimensional calculation. Besides verifying that the two-dimensional model is indeed a good approximation, such a calculation would be useful in its own right, to extend the range of validity away from extremality and study more detailed statistics (such as correlations between sectors of different spin) in a putative ensemble dual.

The geometries we studied are Seifert three-manifolds, which may be a tractable class of manifolds on which to study the gravitational path integral more completely, for example generalizing the methods of \cite{Cotler:2018zff} and \cite{CotlerJensenDoubleTrumpet}. At least in perturbation theory, 3D pure gravity is closely related to a noncompact Chern-Simons theory \cite{Witten:2007kt}, so we may be able to draw inspiration from previous studies of Chern-Simons theories on Seifert manifolds (some examples are \cite{Beasley:2005vf, Blau:2013oha}).

Due to the emergence of the near-horizon AdS$_2$ region where quantum fluctuations are important, it is plausible that the most important topologies in the near-extremal limit are those we study, respecting the symmetry of the large transverse circle direction. However, we are not guaranteed that this will be the case. While a systematic study of all three-manifolds seems beyond reach, we would like to find some organizing principle to guide us to the most relevant topologies.

Our results suggest that there is one regime where topological fluctuations become completely uncontrolled, namely for small non-rotating black holes. Our two-dimensional approximation is not useful there (for example, it becomes important that we integrate only over positive values of the dilaton), but at least formally we note that the genus expansion breaks down entirely, so the infinite set of topologies we consider (and likely may more) will be of equal importance at leading order.

\subsection{Consistency with modular invariance}

After observing the negativity of the MWK spectral density, \cite{Benjamin:2019stq} argued that a positive density of states required the existence of sufficiently light states, well below the black hole threshold, from modular invariance. On the other hand, our gravitational calculation indicates that there is a cure to the negativity without introducing any such light states. We will explain how this is consistent with modular invariance and the loophole in the argument of \cite{Benjamin:2019stq} in a companion paper \cite{MTwip}, which we briefly preview.

Modular invariance gives universal results for the density of states in an asymptotic limit of large spin $|J|\to\infty$ and fixed twist $E-|J|$ \cite{Kusuki:2018wpa,Collier:2018exn,Maxfield:2019hdt}. We will state a precise and rigorous version of this statement in \cite{MTwip}, as well as a generalization giving (for instance) the density of even spin states minus the density of odd spin states, following ideas of \cite{Benjamin:2019stq}. The most important terms in this large spin expansion arise as the modular transforms of the operators of lowest twist.

However, this is not quite the limit in which the negative density of states appears. Instead, to probe this regime we require a combined limit of large spin and exponentially low twist, which is more subtle to study. Since our proposed cure for the negativity does not involve new terms which grow rapidly with spin, but instead terms that become more important at very low twist, it does not require the existence of light operators. Instead, it implies a small correction to the density of states relative to the Cardy formula. The required correction to the degeneracy of black hole states grows polynomially with spin, giving an exponentially small correction to the entropy.

\subsection{Including matter}\label{sec:discMatter}

If we consider a three-dimensional theory of gravity with matter, we can ask how it affects our near-extremal two-dimensional description. We obtain a useful perspective by including the matter as `first-quantized' particles, using a point-particle action rather than a field description. An interesting effect arises from particles which run round the $S^1$ on which we reduce. Taking the worldlines of such particles to be independent of the AdS$_2$ directions, they appear as pointlike objects in the two-dimensional theory. Indeed, they give new examples of the defects we studied in section \ref{sec:JT}. For a particle of mass $m$, the resulting defect is parameterized by $\alpha = 1-4G_N m$ and $\lambda \sim e^{-4 m G_N S_0(J)}$.

This gives a two-dimensional perspective on the proposal of \cite{Benjamin:2019stq,Benjamin:2020mfz} to cure the negative density of states with matter, as discussed after equation \eqref{eq:naiverho}. To cancel the contribution of a KK instanton defect without considering off-shell geometries, one can add a matter defect with $\lambda$ growing sufficiently rapidly with spin, which means $m<\frac{1}{8G_N}$.

Of course, we can now include multiple matter defects, and as a result find a shift in the black hole extremality bound. Precisely this effect was obtained from a three-dimensional analysis of one-loop corrections to the extremal BTZ geometry in \cite{Maxfield:2019hdt}, for light matter (excluding backreaction). Such an effect was also demonstrated for a generic irrational CFT using conformal bootstrap methods, which gave a result that includes backreaction when applied to a gravitational dual. The consistency between this full three-dimensional analysis and the two-dimensional reduced theory gives us extra confidence  in the approximations we used in this paper.

%
%
%
%

\subsection{The instanton gas for JT with defects}

\label{Sec:InstantonGasSemiClassics}

We gave a solution to JT gravity with defects by constructing the path integral with a definite number of defects, and then summing their results. A possible alternative approach to pursue is provided by the instanton gas described in section \ref{sec:instGas}, where we instead perform the sum before doing the path integral. This leads us to propose that this theory is also equivalent to a theory of dilaton gravity.

As discussed in \cite{Mertens:2019tcm} and sections \ref{sec:instGas} and \ref{sec:JTpathintwdef}, a defect insertion in JT gravity is equivalent to inserting an operator exponential in the dilaton. Including an integral over the location of the defect, we can include defects by insertions of an operator proportional to
\begin{equation}\label{eq:instInsertion}
	\int d^2x\sqrt{g_2}\, e^{-2\pi(1-\alpha)\phi(x)}
\end{equation}
in the path integral, as in \eqref{eq:JTinstaction}. At this point, we could carefully relate this to our earlier calculations by gauge-fixing diffeomorphisms and integrating out $\phi$ to arrive at an integral over moduli space, checking that we land on the Weil-Petersson volume form. In particular, such a calculation would determine the normalization of a defect insertion in terms of $\lambda$. We will not follow this route here, instead fixing the normalization by comparing the two methods.

As described in section \ref{sec:instGas}, we now can insert any number of defects in the path integral and sum over them including symmetry factors, giving the exponential of \eqref{eq:instInsertion}. This can be incorporated as an additional local term added to the JT action \eqref{eq:JTaction}, giving
\begin{equation}
	I_\text{IG} = - \frac{1}{2}\int d^2 x \sqrt{g_2} \left( \phi R_2 +U(\phi)\right),
\end{equation}
with potential
\begin{equation}
	U(\phi) = 2\phi+4\pi(1-\alpha)\lambda \, e^{-2\pi(1-\alpha)\phi},
\end{equation}
with the normalisation chosen to match parameters (see appendix \ref{app:IG}). This is a model of dilaton gravity in the class studied by \cite{Almheiri:2014cka} (see also \cite{Grumiller:2007ju, Kyono:2017jtc, Witten:2020ert}), which approaches JT gravity at large dilaton (as long as $\alpha<1$, since otherwise defects will proliferate at large $\phi$, and destroy or at least modify the asymptotic region).

This argument leads us to conjecture that we have an equivalence between three theories: JT gravity with defects, the matrix integral of section \ref{sec:JTdefMM}, and now a theory of dilaton gravity. We make some very preliminary comments, and leave more detailed study to the future. See the independent work \cite{Witten:2020ert} for a discussion on the relationship between the dilaton gravity theory and the sum over defects.

\begin{figure} 
\begin{center}
\begin{tikzpicture}[scale=1]
      \draw[->] (-2,0) -- (2,0) node[right] {\small $\phi$};
      \draw[->] (0,-1.8) -- (0,2) node[above] {\small $U(\phi)$};
      \draw[scale=0.1,domain=-3:3.2,smooth,variable=\x,blue,thick] plot ({5*(\x+0.7)},{2*\x-2*0.4*exp(-\x)+1.6});
      \draw (0,-2.7) node {\small (a) $\lambda<0$};
    \end{tikzpicture}
    \hspace{1cm}
    \begin{tikzpicture}[scale=1]
      \draw[->] (-2,0) -- (2,0) node[right] {\small $\phi$};
      \draw[->] (0,-1.8) -- (0,2) node[above] {\small $U(\phi)$};
      \draw[scale=0.11,domain=-3.4:4.5,smooth,variable=\x,blue,thick] plot ({5*(\x-0.6)},{2*\x+2*0.3*exp(-\x)-1});
      \draw (0,-2.7) node {\small (b) $0<\lambda<\lambda_c$};
    \end{tikzpicture}
    \hspace{1cm}
    \begin{tikzpicture}[scale=1]
      \draw[->] (-2,0) -- (2,0) node[right] {\small $\phi$};
      \draw[->] (0,-1.8) -- (0,2) node[above] {\small $U(\phi)$};
      \draw[scale=0.1,domain=-3:4,smooth,variable=\x,blue,thick] plot ({5*\x},{2*\x+2*0.5*exp(-\x)+1});
      \draw (0,-2.7) node {\small (c) $\lambda_c<\lambda$};
    \end{tikzpicture}
    \end{center}
    \vspace{-0.4cm}
    \caption{\label{fig:shapedilpot} Shape of the dilaton potential for (a) $\lambda<0$ with only a single zero at a positive value of $\phi$ (b) $0<\lambda<\lambda_c$ with two zeros at negative values and $\phi_0$ the largest and (c) $\lambda_c < \lambda$ with no zeros.}
\end{figure}
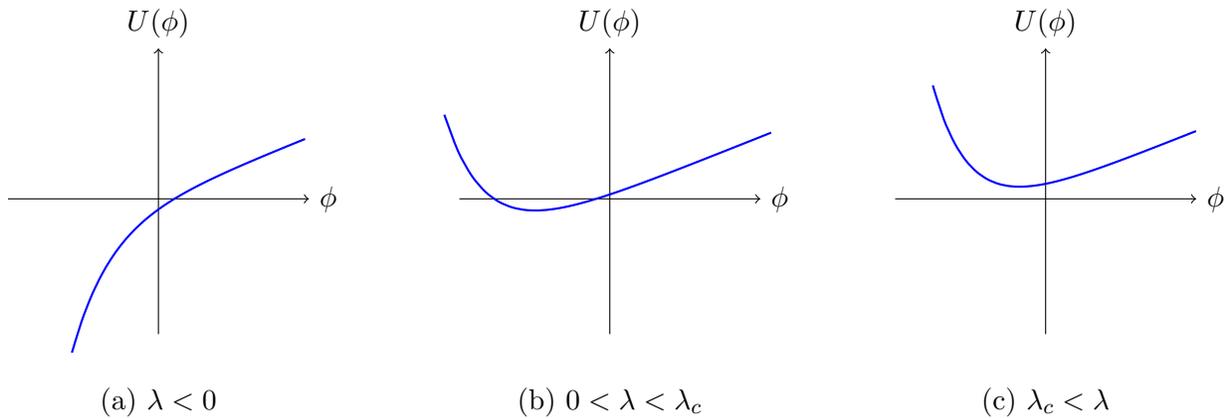

We discuss the semiclassical physics of this model in appendix \ref{app:IG}. Here, we will highlight one aspect, determining the classical stability of the model. This depends on qualitative features of the dilaton potential, for which we have three cases, shown in figure \ref{fig:shapedilpot}.
First, for $\lambda<0$ the potential $U$ is monotone with a single zero, which means that the model is classically stable. However, this is no longer true for $\lambda>0$, since the potential increases without bound as $\phi\to -\infty$; this leads to classical solutions with energy unbounded below. For coupling smaller than a critical value, $0<\lambda<\lambda_c$, the potential $U$ has two zeros. The consequence is that we have sensible black hole solutions for every temperature, and we can at least make sense of the model perturbatively. For $\lambda>\lambda_c$, the potential is positive for all $\phi$, and as a consequence we do not have classical solutions below a critical temperature: the model is not stable even perturbatively at low energies. As long as $\lambda$ is not too large, there is some hope that we are saved from this fate quantum mechanically. 
For example, perhaps the solutions that lead to the instability of the theory are themselves unstable to the spacetime pinching off near the minimum of the dilaton potential, where we have potentially large quantum corrections.

At a very qualitative level, this stability analysis chimes with our findings for the density of states in the matrix model at finite $\lambda$ in section \ref{sec:DOSJTDEF}. The model is well-behaved for any negative $\lambda$, but does not make sense for low temperatures when $\lambda$ is sufficiently large and positive. We potentially have a semiclassical interpretation of the regime in which the theory is well-defined, which deserves better understanding.

We can construct more general dilaton gravity models by including multiple species of defect, each of which should simply add another exponential term to the potential $U$. If our conjecture is correct, we thus have a large class of dilaton gravity theories with matrix integral duals. Through this dual, we have the opportunity to understand such models nonperturbatively. An interesting application is to studying flows in the bulk between geometries that are not necessarily AdS$_2$. This can be intuitively understood as a condensation of defects that modifies the geometry deep in the bulk. An important example is given by \cite{Anninos:2017hhn, Anninos:2018svg}. For a dilaton potential with a certain shape it is possible to produce flows where the geometry inside the bulk is dS$_2$ (see also \cite{Maldacena:2019cbz, Cotler:2019nbi}). From the semiclassical analysis we see this may be possible for $\lambda>0$ and $\alpha$ close to one. Another possible application is to understand the path integral over finite spaces following \cite{Iliesiu:2020zld} and using the exact density of states as a seed for the solution of the Wheeler-de Witt equation. We leave  such explorations for future work.

\paragraph{Acknowledgements} 

We thank N.\ Benjamin, L\ Iliesiu, A.\ Maloney, T.\ Mertens, D.\ Stanford and E.\ Witten for useful discussions. HM is funded by a Len DeBenedictis Postdoctoral Fellowship and under NSF grant PHY1801805, and receives additional support from the University of California. GJT is supported by a Fundamental Physics Fellowship. 
 
\appendix 

\section{Weil-Petersson volumes}\label{app:wp}
In this appendix we will give a simple derivation of the WP volume on a surface with genus $g$ and $n$ geodesic boundaries of length $\mathbf{b}=(b_1,\ldots,b_n)$ where one of them has a very large length $b_1\to \infty$. The leading term is this limit is given by
\beq\label{app:eq:idwpv}
V_{g,n}(\mathbf{b}) = \frac{1}{(24)^g g!} \frac{(b_1^2/2)^{3g-3+n}}{(3g-3+n)!}(1 +\mathcal{O}(b_1^{-1})).
\eeq
To show this we first write down the general WP volume as 
\beq
V_{g,n}(\mathbf{b}) = \sum_{\mathbf{\alpha}} \frac{(\alpha_1,\ldots, \alpha_n)_{g,n}}{2^{|\alpha|} \prod_{i=1}^n \alpha_i!}  b_1^{2\alpha_1} \ldots b_n^{2\alpha_n},
\eeq
where $\alpha_i$ are a set of positive integers, $(\alpha_1,\ldots, \alpha_n)_{g,n}$ are the coefficients multiplying each power of the boundary length and $|\alpha| = \sum_{i=1}^n \alpha_i$ (not to be confused with the defect angles which will not appear in this appendix). The coefficients of highest order on the geodesic length can be obtained with the string equation. This applies when, in this notation, we have $|\alpha|= 3g-3+n$. Then the string equation is  
\beq
(\alpha_1, \ldots, \alpha_{n-1}, 0)_{g,n} = \sum_{\alpha_i \neq 0 } (\alpha_1,\ldots, \alpha_i-1, \ldots, \alpha_{n-1})_{g,n-1},
\eeq
which Mirzhakani showed to be equivalent to her recursion relations in theorems 7.1 and 7.2 of \cite{Mirzakhani:2006fta}. For our purposes since we are interested in taking only one boundary length to be large, we will consider the simpler case with top power $\alpha_1=3g-3+n$ and $\alpha_{i>1}=0$. The string equation becomes very simply 
\beq
(3g-3+n, 0, \ldots, 0)_{g,n} = (3g-3+n-1, 0, \ldots, 0)_{g,n-1}.
\eeq
This equation shows that the coefficient $(3g-3+n, 0, \ldots, 0)_{g,n}$ is independent of $n$. Finally we need to compute this coefficient for some particular value of $n$. This calculation is easier to do for a single boundary so that $n=1$. This was computed by Itzykson and Zuber \cite{ITZYKSON_1992} obtaining 
\beq
(3g-3+n, 0, \ldots, 0)_{g,n} = \frac{1}{(24)^g g!}. 
\eeq
Putting everything together we get the leading power we needed 
\beq
V_{g,n}(\mathbf{b})  = \frac{1}{(24)^g g!}\frac{1}{2^{3g-3+n}(3g-3+n)!} b_1^{6g-6+2n} + \ldots, 
\eeq
where the dots indicate terms in the polynomial with lower powers of $b_1$. This is precisely the relation  \eqref{app:eq:idwpv} we wanted to prove. The dilaton and string equation prove more useful in order to derive this result, and also to derive subleading terms, than Mirzhakani recursion relation\footnote{We thank A.\ Maloney for discussions on this.}. 

This can be seen more directly for the case of genus zero $g=0$. In this case we can use the explicit formula for the WP volumes found in \cite{mtms}, given by 
\beq\label{eq:WPsphereMT}
  V_{0,n}(\mathbf{b}) =\frac{1}{2} \left. \Big( \frac{\partial}{\partial x} \Big)^{n-3} \left[ u'_{\rm JT}(x) \prod_{i=1}^n I_0 \left(b_i \sqrt{u_{\rm JT}(x)} \right) \right] \right|_{x=0},
  \eeq
where $u_{\rm JT}(x)$ is the JT gravity specific heat, defined implicitly through
\beq
\sqrt{u_{\rm JT}} \hspace{0.1cm}J_1\left(2\pi \sqrt{u_{\rm JT}}\right)=2\pi x.
\eeq
When $b_1\to\infty$ this is approximated by the leading term where the $n-3$ derivatives act on the first Bessel function $n-3$ times, giving 
\beq
  V_{0,n}(\mathbf{b})  \approx \frac{1}{2} \frac{(u_{\rm JT}'(0))^{n-2}}{(n-3)! 2^{2n-6}} b_1^{2n-6}.
\eeq
We can efficiently compute the derivatives of $u_{\rm JT}(x)$ with respect to $x$ at $x=0$ using the Lagrange inversion theorem. This gives $u_{\rm JT}'(0)=2$ and we reproduce \eqref{app:eq:idwpv} for genus zero.

\section{Lagrange reversion theorem} \label{App:LRT}
In section \ref{sec:JTdefMM} we showed that, at genus zero, all the multi-loop amplitudes of JT gravity with a gas of defects match precisely with the results from a matrix integral. In order to show that we used a theorem, which we explain in more detail here. The \href{https://en.wikipedia.org/wiki/Lagrange_reversion_theorem#:~:text=In%20mathematics%2C%20the%20Lagrange%20reversion,of%20compositions%20with%20such%20functions.&text=Lagrange's%20reversion%20theorem%20is%20used%20to%20obtain%20numerical%20solutions%20to%20Kepler's%20equation.}{Lagrange reversion theorem} 
can be used to resum series which have the following form 
\beq
g(x)+ \sum_{n=1}^\infty \frac{y^n}{n!} \Big(\frac{\partial}{\partial x}\Big)^{n-1}\Big[ f(x)^n g'(x)\Big],
\eeq
where $f(x)$ and $g(x)$ are arbitrary functions of $x$, $y$ is another variable, and $g'(x)$ indicates the derivative of $g(x)$ with respect to $x$. It is useful to introduce the variable $v$, defined implicitly as a function of $x$ (and the parameter $y$) through the equation 
\beq
v = x + y f(x). 
\eeq
Then the theorem states that the answer for the infinite sum written above is 
\beq\label{eq:LRT}
g(v) =g(x)+ \sum_{n=1}^\infty \frac{y^n}{n!} \Big(\frac{\partial}{\partial x}\Big)^{n-1}\Big[ f(x)^n g'(x)\Big].
\eeq
In the main text we applied this theorem with $f(x) \to J_0\big(2\pi \alpha \sqrt{u_{\rm JT}(x)} \big)$, $g(x)\to e^{\frac{1}{2\gamma} u_{\rm JT}(x) \beta}$, $y\to \lambda$, and $u_{\rm JT}(x)$ is a given function defined in section \ref{sec:JTdefMM}. 

When dealing with multiple types of defects we need a slight generalization of this theorem. If we allow a number $K$ of variables and functions $y_i$, $f_i(x)$ for $i=1,\ldots,K$, and we define the variable $v$ implicitly as a function of $x$ through
\beq
v = x + \sum_{i=1}^K y_i f_i(x),
\eeq
then we can use the previous theorem to show that 
\beq\label{eq:multiLRT}
g(v) = g(x)+\sum_{\{n_i\}} \prod_{i=1}^K \frac{y_i^{n_i}}{n_i!} \Big( \frac{\partial}{\partial x} \Big)^{\sum_{i=1}^K n -1} \left[ g'(x) \left( \prod_{i=1}^K f_i(x)^{n_i}\right) \right],
\eeq
for any function $g(x)$. The sum runs over a set of $K$ non-negative integers $n_i$ excluding the case where they are all zero. In the main text we applied this with $f_i(x) \to J_0\big(2\pi \alpha_i \sqrt{u_{\rm JT}(x)}\big)$, $y_i\to \lambda_i$, and $K$ is the total number of defect flavors. This theorem is a simple generalization of the previous one if we call $y \to y_1$ and $f(x) \to \sum_i \frac{y_i}{y_1} f_i(x)$, and apply the multinomial theorem to $f(x)^n$ in \eqref{eq:LRT}, giving  \eqref{eq:multiLRT}.

\section{Explicit higher genus checks}\label{app:checks}

In this appendix we include some explicit checks that JT gravity with defects is equivalent to a matrix integral at higher genus. To simplify the expressions, we will set $\gamma=1/2$ in this section.

The pure JT gravity has spectral curve $y_{\rm JT}(z) = \sin(2 \pi z)/(4\pi)$. The partition function connected correlator from the matrix model perspective ${\rm Tr}(e^{-\beta H})$ will be denoted by $Z^{{\scriptscriptstyle  {\rm MM}}}_{g,n}(\beta_1,\ldots, \beta_n)$. Define the special cases $W_{0,1}(z)=-2 z y(z)$ and $W_{0,2}(z_1,z_2) = (z_1-z_2)^{-2}$. The latter is equivalent to the universal answer presented in \eqref{eq:universal2loopMM} with the edge of the spectrum $E_0$ set to zero. The topological recursion relation is
\begin{align}
\label{eynard}
&W_{g,n}(z_1,J) = \\
&\text{Res}_{z\to 0}\left\{\frac{1}{(z_1^2-z^2)} \frac{1}{4y(z)}\left[W_{g-1,n-1}(z,-z,J) + \sum_{h,I,h',I'} W_{h,1+I}(z,I)W_{h',1+I'}(-z,I')\right]\right\}, \nonumber
\end{align}
where $h+h'=g$ and $I \cup I' =J$ denotes a subset of the labels $z_2, \hdots z_n$, and the sum excludes the cases $(h=g,I=J)$ and $(h'=g,I'=J)$. 

This form of the recursion applies for a density of states that starts at $E=0$ ($z=0$). Therefore, when we apply it to our disk density of states for JT gravity plus defects we need to shift $E\to E+ E_0$, before using \eqref{eynard}. After this shift, we already proved analytically that JT gravity plus defects exactly gives $W_{0,2}(z_1,z_2) = (z_1-z_2)^{-2}$, which is also necessary in order for the recursion to make sense. Moreover all $W_{0,n}(z_1,\ldots, z_n)$ can be easily computed exactly from \eqref{eq:multiloopg0JT}, and they satisfy the recursion. 

The calculation is tedious and we will give the main final results at each order, to show how surprising it is that the gravitational theory with defects satisfies the topological recursion. We introduce some notation first, expanding the following quantities as 
\bea
y(z) &=& y_{\rm JT}(z) \hspace{0.24cm}+ \lambda \hspace{0.1cm}y_1(z)\hspace{0.1cm}+ \lambda^2 y_2(z)\hspace{0.25cm} +\ldots , \\ 
W_{g,n}(\mathbf{z}) &=& W^{{\scriptscriptstyle  {\rm JT}}}_{g,n}(\mathbf{z}) +\lambda W_{g,n,1}(\mathbf{z}) + \lambda^2 W_{g,n,2}(\mathbf{z})+\ldots,\\
Z^{{\scriptscriptstyle  {\rm MM}}}_{g,n}(\bm{\beta}) &=& Z^{{\scriptscriptstyle  {\rm JT}}}_{g,n}(\bm{\beta}) \hspace{0.05cm}+\lambda Z^{{\scriptscriptstyle  {\rm MM}}}_{g,n,1}(\bm{\beta})\hspace{0.04cm} + \lambda^2 Z^{{\scriptscriptstyle  {\rm MM}}}_{g,n,2}(\bm{\beta})\hspace{0.05cm}+\ldots ,
\ea
where to ease notation we wrote $\mathbf{z} = (z_1,\ldots, z_n)$ and $\bm{\beta}=(\beta_1,\ldots, \beta_n) $. The leading contributions, which match with pure JT gravity, were computed in \cite{Saad:2019lba} and we will not repeat them here. While in this section $y$'s and $W$'s are defined with respect to the shifted ensembles with edge at $E_0\to0$, the quantity $Z^{{\scriptscriptstyle  {\rm MM}}}_{g,n}$ is defined by the ensemble with the edge at $E_0$. This is done to be able to directly compare with the gravitational answer \eqref{eq:Vexp2}.

The general procedure we apply below is the following. First, we compute the correction to the spectral curve from the defects, and shift the zero-point energy to zero. Second, we insert this result in the topological recursion to compute the correction to $W_{g,n}$ and from it obtain the resolvent correlator. Third, an inverse Laplace transform gives the expectation value of a product of ${\rm Tr}(e^{-\beta (H+E_0)})$. Finally, we multiply this by a factor of $e^{E_0(\beta_1+\ldots +\beta_n)}$ to correct the shift in energy, obtaining the quantity we call $Z^{{\scriptscriptstyle  {\rm MM}}}_{g,n}(\beta_1,\ldots, \beta_n)$ above, expand in $\lambda$, and compare with the answer from JT gravity with defects order by order in $\lambda$.

\paragraph{One boundary, one handle, one defect:} The first correction to the density of states from one defect is $\delta \rho = \lambda \cosh 2\pi \alpha \sqrt{E} /(2\pi \sqrt{E})$. Correcting by the zero-point energy, which to this order is $E_0 \sim - 2 \lambda$, and switching to $y(z)$ we obtain 
\beq
y_1(z) = - \frac{\cos 2\pi \alpha z - \cos 2 \pi z}{2z }.
\eeq
The first term in the numerator comes from the one-defect amplitude, while the second comes from the shift in the energy of $y_{\rm JT}(z)$. To compute $W_{1,1}$ from the topological recursion we only need $y(z)$ and $W_{0,2}$. The leading correction in $\lambda$ gives
\beq
W_{1,1,1}(z_1) = \frac{\pi^2 (1-\alpha^2)(3+(3-\alpha^2)\pi^2z_1^2)}{12 z_1^4}.
\eeq
A Laplace transform of the full resolvent to this order, multiplication by $e^{\beta E_0}$, and Taylor expansion in $\lambda$ gives the correction to the partition function 
\beq
Z^{{\scriptscriptstyle  {\rm MM}}}_{1,1,1}(\beta) =  \frac{\pi^{7/2}(3-4\alpha^2+\alpha^4)}{12} \sqrt{\beta} - \frac{\pi^{3/2}(\alpha^2-2)}{6} \beta^{3/2} + \frac{1}{6\sqrt{\pi}} \beta^{5/2} \ .
\eeq

This is equal to the gravitational JT gravity plus defect calculation, which at this order following \eqref{eq:Vexp2} is given by 
\bea
Z_{1,1,1}^{{\rm JT+def}}(\beta)&=& \int_0^\infty bdb Z_{\rm trumpet}(\beta,b) V_{1,2}(b,2\pi i \alpha) \nonumber\\
&=& \frac{\pi^{7/2}(3-4\alpha^2+\alpha^4)}{12} \sqrt{\beta} - \frac{\pi^{3/2}(\alpha^2-2)}{6} \beta^{3/2} + \frac{1}{6\sqrt{\pi}} \beta^{5/2},
\eea
where we used the explicit form of the WP volumes tabulated by Do in Appendix B of \cite{do2011moduli}. 

\paragraph{One boundary, one handle, two defects:} Now we expand the density of states and shift by $E_0$ to quadratic order in $\lambda$. We get the next shift to the spectral curve 
\beq
y_2(z)= \frac{(2(1-2\alpha^2)\pi^2 z^2-1)\cos 2 \pi z + 2 \cos 2 \pi \alpha z - 2 \pi z(\sin 2\pi z-2\alpha \sin 2\pi \alpha z)-1}{4 z^{3}}
\eeq 
The next correction to the resolvent from the topological recursion is given by
\beq
W_{1,1,2} (z_1) =  \frac{9(5-6\alpha^2(2-\alpha^2))\pi^4+4(2-\alpha^2)(7-16\alpha^2+7\alpha^4)\pi^6 z_1^2}{72 z_1^4}
\eeq
After Laplace transform, multiplication by $e^{\beta E_0}$, and Taylor expansion in $\lambda$, the next correction to the partition function is 
\beq
Z^{{\scriptscriptstyle  {\rm MM}}}_{1,1,2}(\beta) ={\scriptstyle \frac{(2-\alpha^2)(7-16\alpha^2+7\alpha^4)\pi^{11/2}\sqrt{\beta}}{18} + \frac{(13-8\alpha^2 (3-\alpha^2))\pi^{7/2}\beta^{3/2}}{12}  + \frac{2(1-\alpha^2)\pi^{3/2}\beta^{5/2}}{3}  + \frac{\beta^{7/2}}{6\sqrt{\pi}}}
\eeq
This is equal to the gravitational JT gravity plus defect calculation, which at this order is given by
\bea
Z_{1,1,2}^{{\rm JT+def}}(\beta)&=&\frac{1}{2!} \int_0^\infty bdb Z_{\rm trumpet}(\beta,b) V_{1,3}(b,2\pi i \alpha, 2\pi i \alpha) \nonumber\\
&=& {\scriptstyle \frac{(2-\alpha^2)(7-16\alpha^2+7\alpha^4)\pi^{11/2}\sqrt{\beta}}{18} + \frac{(13-8\alpha^2 (3-\alpha^2))\pi^{7/2}\beta^{3/2}}{12}  + \frac{2(1-\alpha^2)\pi^{3/2}\beta^{5/2}}{3}  + \frac{\beta^{7/2}}{6\sqrt{\pi}}},
\eea
using the explicit form of $V_{1,3}(b_1,b_2,b_3)$.

\paragraph{Two boundaries, one handle, one defect:}
In this case knowing $y(z)$ to linear order at $\lambda$ is enough. The recursion involves $W_{1,1}$ (which we already computed)  and $W_{0,3}$ (which we can compute exactly in the previous section).  We get 
\beq
W_{1,2,1} (z_1,z_2)= {\scriptstyle \frac{ (1- \alpha^2 ) \pi^2 (45 z_2^4 +9 z_1^2 z_2^2 (3+(5- \alpha^2 ) \pi^2 z_2^2) + z_1^4(45+9 ( 5- \alpha^2 ) \pi^2 z_2^2 +(28-11\alpha^2 + \alpha^4 ) z_2^4 ) ) }{ 18 z_1^6 z_2^6 }}
\eeq
And the correction to the partition function is 
\bea
Z^{{\scriptscriptstyle  {\rm MM}}}_{1,2,1}(\beta_1,\beta_2) &=& {\scriptstyle \frac{\sqrt{\beta_1 \beta_2} }{18\pi} (-12 \pi ^2 ((\alpha ^2-2) \beta_1^2+(\alpha ^2-3) \beta_1 \beta_2 +(\alpha ^2-2)\beta_2^2)+3 \pi ^4 (2 (\alpha ^2-6) \alpha ^2+13) (\beta_1+\beta_2)}\nonumber\\
&&  {\scriptstyle -\pi^6 (\alpha^6-12 \alpha ^4+39 \alpha ^2-28)+6 (\beta_1+\beta_2) (\beta_1^2+\beta_1\beta_2 +\beta_2^2)) }
\eea

This function is precisely equal to the gravitational calculation, which is 
\beq
Z^{{\rm JT+def}}_{1,2,1}(\beta_1,\beta_2) = \int_0^\infty \left( \prod_{i=1,2} b_i db_i Z_{\rm trumpet} (\beta_i,b_i) \right)V_{1,3}(b_1,b_2,2\pi i \alpha)
\eeq

\paragraph{One boundary, two handles, one defect:} With the ingredients computed so far we can continue the topological recursion and compute $W_{2,1}$ to linear order. We obtain
\bea
W_{2,1,1}(z_1) &=& {\scriptscriptstyle \frac{-210 \pi ^2 (\alpha ^2-1)(2 \pi ^8 (\alpha ^8-43 \alpha ^6+641 \alpha ^4-3767 \alpha ^2+558) z_1^8-3 \pi ^6 (15 \alpha ^6-449 \alpha ^4+3895 \alpha ^2+595) z_1^6}{1814400 z_1^{10}}} \\
&& \hspace{-2cm} {\scriptscriptstyle +\frac{30 \pi ^4 (29 \alpha ^4-475 \alpha ^2-388) z_1^4-9135 \pi ^2 (\alpha ^2+3) z_1^2-28350)-10 \pi ^2 (\alpha ^2-1) (1890 \pi ^2 (145 \pi ^6 z_1^6+338 \pi ^4 z_1^4+695 \pi ^2 z_1^2+1015) z_1^2+1488375)}{1814400 z_1^{10}}}\nonumber
\ea
The partition function is 
\bea
Z^{{\scriptscriptstyle  {\rm MM}}}_{2,1,1}(\beta)\hspace{-0.2cm} &=&\hspace{-0.2cm} {\scriptstyle \frac{\pi ^{15/2} (15 \alpha ^8-464 \alpha ^6+4344 \alpha ^4-13440 \alpha ^2+10850) \beta ^{3/2}}{4320}+\frac{\pi ^{11/2} (-29 \alpha ^6+504 \alpha ^4-2172 \alpha ^2+2204) \beta ^{5/2}}{1080}+\frac{\pi ^{3/2} (44-15 \alpha ^2) \beta ^{9/2}}{180} }\nonumber\\
&&{\scriptstyle +\frac{ \pi ^{7/2} (29 (\alpha ^2-8) \alpha ^2+342) \beta ^{7/2}}{360}+\frac{\pi ^{19/2} (1-\alpha^2) (\alpha ^8-43 \alpha ^6+641 \alpha ^4-3767 \alpha ^2+7083) \sqrt{\beta }}{4320}+\frac{\beta ^{11/2}}{36 \sqrt{\pi }}}
\eea

This is equivalent to the gravitational calculation
\beq
Z^{{\rm JT+def}}_{2,1,1}(\beta) = \int_0^\infty b db\hspace{0.1cm} Z_{\rm trumpet} (\beta,b) V_{2,2}(b, 2\pi i \alpha)
\eeq

\section{The instanton gas theory}\label{app:IG}
In this appendix we will present more details regarding the instanton gas theory, which we define as a 2D dilaton gravity with a specific potential originating from summing over defects. We will analyze the semiclassical thermodynamics and match it with the high energy limit of JT gravity with defects, as a basic check of the equivalence between the two theories. 

As explained in section \ref{Sec:InstantonGasSemiClassics}, by formally summing defects as insertions in the path integral, we obtain a dilaton gravity theory with action
\begin{equation}
	I_\text{IG} = - \frac{1}{2}\int d^2 x \sqrt{g_2} \left( \phi R_2 +U(\phi)\right),
\end{equation}
and potential
\begin{equation}
	U(\phi) = 2\phi+\sum_{i=1}^{N_{\rm F}} 4\pi(1-\alpha_i)\lambda_i e^{-2\pi(1-\alpha_i)\phi}.
\end{equation}
This theory emerges from integrating out the species of defects with angle $\alpha_i$ and coupling $\lambda_i$, with $i=1,\ldots, N_{\rm F}$. Here we will study the case of a single species with parameters $\alpha$ and $\lambda$. 

To analyze this theory in the semiclassical limit it is convenient to introduce the coupling $\Lambda \equiv (2\pi(1-\alpha))^2 \lambda$. The the potential can be written as 
\begin{equation}
	U(\phi) = 2\phi+2\Lambda \frac{e^{-2\pi(1-\alpha)\phi}}{2\pi (1-\alpha)}.
\end{equation}
The potential approached the JT gravity one for large $\phi$ as long as $\alpha<1$, which is expected since otherwise we get an instability driven by the proliferation of instantons towards the asymptotic boundary where $\phi\to\infty$.

We will here simply analyze the dilaton model at the classical level. The model then only depends nontrivially on $\Lambda$, as can be seen by rewriting the action with a rescaled dilaton $\tilde{\phi} = 2\pi(1-\alpha)\phi$:
\begin{equation}
	I_\mathrm{IG} = - \frac{1}{4\pi(1-\alpha)}\int d^2 x \sqrt{g_2} \left( \tilde{\phi} R_2 +\tilde{U}(\tilde{\phi})\right),
\end{equation}
with
\begin{equation}
	\tilde{U}( \tilde{\phi}) = 2 \tilde{\phi} +2\Lambda e^{- \tilde{\phi}}.
\end{equation}
The model will dependen nontrivially on $\alpha$ and $\Lambda$ independently once quantum corrections are taken into account. It is interesting to note from this that $1-\alpha$ plays the role of $\hbar$, so $\alpha \to 1$ is a natural classical limit to study.

Now, we will describe the classical thermodynamics of the model following \cite{Grumiller:2007ju,Witten:2020ert}. First, we describe the qualitative features of the potential, which determine the classical stability. There are three different parameter regimes, shown in figure \ref{fig:shapedilpot}. For $\Lambda<0$, $U$ is decreasing, so has a single zero. For $\Lambda>0$, the potential $U$ has a minimum at
\begin{equation}
	\phi_{\min} = \frac{\log \Lambda }{2 \pi  (1-\alpha )},~~~~U(\phi_{\min}) = \frac{\log  \Lambda +1}{\pi(1 -  \alpha) }
\end{equation}
For $\Lambda$ larger than a critical value, this minimum is positive, 
\begin{equation}
	\Lambda >\Lambda_c=e^{-1} \implies U(\phi_{\min})>0,
\end{equation}
so for $\Lambda >\Lambda_c$ the potential does not have any zeros.
For $\Lambda<\Lambda_c$, the largest zero of $U$ is at
\begin{equation}
	\phi_0 = \frac{W (-\Lambda)}{2 \pi  (1-\alpha)},
\end{equation}
where $W$ is the Lambert function or product log. For $0<\Lambda<\Lambda_c$ there is a second zero, and the potential increases exponentially as $\phi\to -\infty$. The classical Euclidean solutions of the theory are uniquely determined  by the minimum value $\phi_h$ of the dilaton, as
\begin{equation}
	ds^2 = \gamma^{-2}f(r)dt^2 + \frac{dr^2}{f(r)}, \quad\phi=r, \quad f(\phi) = \int_{\phi_h}^\phi dr\, U(r),
\end{equation}
where we have chosen the gauge $\phi=r$, and $\phi>\phi_h$. This gives a Euclidean metric as long as $f(\phi)$ is always positive, so we require
\begin{equation}
	\int_{\phi_h}^\phi U(r)dr >0 \qquad \forall \phi>\phi_h.
\end{equation}

For $\Lambda<0$,  our solutions correspond to $\phi_h>\phi_0$. For $0<\Lambda<\Lambda_c$, we have one branch of solutions for $\phi_h>\phi_0$, but also a second branch for sufficiently negative $\phi_h$. For $\Lambda>\Lambda_c$ we have solutions for all $\phi_h$.

Now recall our boundary conditions at the asymptotic boundary with large positive dilaton and
\begin{equation}
	\frac{L_\partial}{\phi_\partial} = \frac{\beta}{\gamma},
\end{equation}
where $\beta$ is the period of the $t$ coordinate. The proper length of the boundary (at large $r$) is $L_\partial = \beta \gamma^{-1}\sqrt{f(\phi_\partial)}\sim \beta\gamma^{-1} \phi_\partial$, so we have chosen the correct normalization of the $t$ coordinate. The temperature is determined by smoothness at the origin, where $f(r)\sim U(r_h)(r-r_h)$; if we write $r-r_h\sim \frac{U(r_h)}{4}\rho^2$ our metric goes like $d\rho^2 +\rho^2 d(\frac{U(r_h)}{2\gamma}t)^2$, so $\beta = \frac{4\pi\gamma}{U(r_h)}$. The entropy is given by the value of $\phi_h$, and the energy is read off from the constant term when we expand $f$ at infinity. Putting this together, our thermodynamic quantities are given by
\begin{align}
	T&=\frac{U(\phi_h)}{4\pi \gamma}, \\
	S&=S_0+2\pi \phi_h, \\
	E&= \tfrac{1}{2\gamma}\phi_h^2 - \frac{\Lambda}{(2\pi(1-\alpha))^2\gamma}e^{-2\pi(1-\alpha)\phi_h},
\end{align}
and it is manifest that we have the expected relation $dE=TdS$.

As a rudimentary comparison between the dilaton model and the defects, we now compare the thermodynamics to linear order in $\Lambda$ with the effect of a single defect. We note that the result will not only apply for small $\Lambda$, but also at sufficiently large energy that the exponential term in the potential is a small correction.

To do this, write $\phi_h = \sqrt{2\gamma E} + O(\Lambda)$, where the term of order $\Lambda$ is chosen such that the energy remains fixed at $E$ to order $\Lambda$. From this, we read off the first order variation in the entropy from including the exponential term in the potential:
\begin{equation}
	\delta S = \frac{\Lambda}{2\pi(1-\alpha)^2}   \frac{e^{-2\pi(1-\alpha)\sqrt{2\gamma E}}}{\sqrt{2\gamma E}}
\end{equation}

Now, from the defects we find the variation in the entropy by taking the logarithm of the density of states \eqref{eq:rhonaive} including a single defect:
\begin{equation}
	\delta S = \frac{2\pi\lambda}{\sqrt{2\gamma E}} \frac{\cosh\left(2\pi\alpha\sqrt{2\gamma E}\right)}{\sinh\left(2\pi\sqrt{2\gamma E}\right)} \sim 2\pi\lambda\frac{e^{-2\pi(1-\alpha)\sqrt{2\gamma E}}}{\sqrt{2\gamma E}},
\end{equation}
where we have taken $\gamma E\gg 1$ so that we can expect the classical thermodynamics to be applicable. The two results match as a function of energy, and give us relation 
\begin{equation}
	\Lambda = (2\pi(1-\alpha))^2 \lambda
\end{equation}
between the parameters in the defect calculations and the dilaton potential.

\mciteSetMidEndSepPunct{}{\ifmciteBstWouldAddEndPunct.\else\fi}{\relax}
\bibliographystyle{utphys}
{\small \bibliography{references.bib}{}}

\providecommand{\href}[2]{#2}\begingroup\raggedright\begin{mcitethebibliography}{10}

\bibitem{Maldacena:1997re}
J.~M. Maldacena, ``{The Large N limit of superconformal field theories and
  supergravity},'' \href{http://dx.doi.org/10.1023/A:1026654312961}{{\em Int.
  J. Theor. Phys.} {\bfseries 38} (1999) 1113--1133},
  \href{http://arxiv.org/abs/hep-th/9711200}{{\ttfamily arXiv:hep-th/9711200}}.

\bibitem{Heemskerk:2009pn}
I.~Heemskerk, J.~Penedones, J.~Polchinski, and J.~Sully, ``{Holography from
  Conformal Field Theory},''
  \href{http://dx.doi.org/10.1088/1126-6708/2009/10/079}{{\em JHEP} {\bfseries
  10} (2009) 079}, \href{http://arxiv.org/abs/0907.0151}{{\ttfamily
  arXiv:0907.0151 [hep-th]}}.

\bibitem{Heemskerk:2010ty}
I.~Heemskerk and J.~Sully, ``{More Holography from Conformal Field Theory},''
  \href{http://dx.doi.org/10.1007/JHEP09(2010)099}{{\em JHEP} {\bfseries 09}
  (2010) 099}, \href{http://arxiv.org/abs/1006.0976}{{\ttfamily arXiv:1006.0976
  [hep-th]}}.

\bibitem{ElShowk:2011ag}
S.~El-Showk and K.~Papadodimas, ``{Emergent Spacetime and Holographic CFTs},''
  \href{http://dx.doi.org/10.1007/JHEP10(2012)106}{{\em JHEP} {\bfseries 10}
  (2012) 106}, \href{http://arxiv.org/abs/1101.4163}{{\ttfamily arXiv:1101.4163
  [hep-th]}}.

\bibitem{Fitzpatrick:2012cg}
A.~Fitzpatrick and J.~Kaplan, ``{AdS Field Theory from Conformal Field
  Theory},'' \href{http://dx.doi.org/10.1007/JHEP02(2013)054}{{\em JHEP}
  {\bfseries 02} (2013) 054}, \href{http://arxiv.org/abs/1208.0337}{{\ttfamily
  arXiv:1208.0337 [hep-th]}}.

\bibitem{Camanho:2014apa}
X.~O. Camanho, J.~D. Edelstein, J.~Maldacena, and A.~Zhiboedov, ``{Causality
  Constraints on Corrections to the Graviton Three-Point Coupling},''
  \href{http://dx.doi.org/10.1007/JHEP02(2016)020}{{\em JHEP} {\bfseries 02}
  (2016) 020}, \href{http://arxiv.org/abs/1407.5597}{{\ttfamily arXiv:1407.5597
  [hep-th]}}.

\bibitem{Benjamin:2019stq}
N.~Benjamin, H.~Ooguri, S.-H. Shao, and Y.~Wang, ``{Light-cone modular
  bootstrap and pure gravity},''
  \href{http://dx.doi.org/10.1103/PhysRevD.100.066029}{{\em Phys. Rev.}
  {\bfseries D100} no.~6, (2019) 066029},
\href{http://arxiv.org/abs/1906.04184}{{\ttfamily arXiv:1906.04184 [hep-th]}}.

\bibitem{Saad:2019lba}
P.~Saad, S.~H. Shenker, and D.~Stanford, ``{JT gravity as a matrix integral},''
\href{http://arxiv.org/abs/1903.11115}{{\ttfamily arXiv:1903.11115 [hep-th]}}.

\bibitem{Maloney:2007ud}
A.~Maloney and E.~Witten, ``{Quantum Gravity Partition Functions in Three
  Dimensions},'' \href{http://dx.doi.org/10.1007/JHEP02(2010)029}{{\em JHEP}
  {\bfseries 02} (2010) 029},
\href{http://arxiv.org/abs/0712.0155}{{\ttfamily arXiv:0712.0155 [hep-th]}}.

\bibitem{Keller:2014xba}
C.~A. Keller and A.~Maloney, ``{Poincare Series, 3D Gravity and CFT
  Spectroscopy},'' \href{http://dx.doi.org/10.1007/JHEP02(2015)080}{{\em JHEP}
  {\bfseries 02} (2015) 080}, \href{http://arxiv.org/abs/1407.6008}{{\ttfamily
  arXiv:1407.6008 [hep-th]}}.

\bibitem{Alday:2019vdr}
L.~F. Alday and J.-B. Bae, ``{Rademacher Expansions and the Spectrum of 2d
  CFT},'' \href{http://arxiv.org/abs/2001.00022}{{\ttfamily arXiv:2001.00022
  [hep-th]}}.

\bibitem{Maldacena:1998bw}
J.~M. Maldacena and A.~Strominger, ``{AdS(3) black holes and a stringy
  exclusion principle},''
  \href{http://dx.doi.org/10.1088/1126-6708/1998/12/005}{{\em JHEP} {\bfseries
  12} (1998) 005}, \href{http://arxiv.org/abs/hep-th/9804085}{{\ttfamily
  arXiv:hep-th/9804085}}.

\bibitem{Banados:1992wn}
M.~Banados, C.~Teitelboim, and J.~Zanelli, ``{The Black hole in
  three-dimensional space-time},''
  \href{http://dx.doi.org/10.1103/PhysRevLett.69.1849}{{\em Phys. Rev. Lett.}
  {\bfseries 69} (1992) 1849--1851},
  \href{http://arxiv.org/abs/hep-th/9204099}{{\ttfamily arXiv:hep-th/9204099}}.

\bibitem{Benjamin:2020mfz}
N.~Benjamin, S.~Collier, and A.~Maloney, ``{Pure Gravity and Conical
  Defects},'' \href{http://arxiv.org/abs/2004.14428}{{\ttfamily
  arXiv:2004.14428 [hep-th]}}.

\bibitem{Maxfield:2019hdt}
H.~Maxfield, ``{Quantum corrections to the BTZ black hole extremality bound
  from the conformal bootstrap},''
\href{http://arxiv.org/abs/1906.04416}{{\ttfamily arXiv:1906.04416 [hep-th]}}.

\bibitem{MTwip}
H.~Maxfield and G.~J. Turiaci, ``work in progress,''.

\bibitem{Jackiw:1984je}
R.~Jackiw, ``{Lower Dimensional Gravity},''
\href{http://dx.doi.org/10.1016/0550-3213(85)90448-1}{{\em Nucl. Phys.}
  {\bfseries B252} (1985) 343--356}.

\bibitem{Teitelboim:1983ux}
C.~Teitelboim, ``{Gravitation and Hamiltonian Structure in Two Space-Time
  Dimensions},''
\href{http://dx.doi.org/10.1016/0370-2693(83)90012-6}{{\em Phys. Lett.}
  {\bfseries 126B} (1983) 41--45}.

\bibitem{Almheiri:2014cka}
A.~Almheiri and J.~Polchinski, ``{Models of AdS$_{2}$ backreaction and
  holography},'' \href{http://dx.doi.org/10.1007/JHEP11(2015)014}{{\em JHEP}
  {\bfseries 11} (2015) 014},
\href{http://arxiv.org/abs/1402.6334}{{\ttfamily arXiv:1402.6334 [hep-th]}}.

\bibitem{Jensen:2016pah}
K.~Jensen, ``{Chaos in AdS$_2$ Holography},''
  \href{http://dx.doi.org/10.1103/PhysRevLett.117.111601}{{\em Phys. Rev.
  Lett.} {\bfseries 117} no.~11, (2016) 111601},
\href{http://arxiv.org/abs/1605.06098}{{\ttfamily arXiv:1605.06098 [hep-th]}}.

\bibitem{Maldacena:2016upp}
J.~Maldacena, D.~Stanford, and Z.~Yang, ``{Conformal symmetry and its breaking
  in two dimensional Nearly Anti-de-Sitter space},''
  \href{http://dx.doi.org/10.1093/ptep/ptw124}{{\em PTEP} {\bfseries 2016}
  no.~12, (2016) 12C104},
\href{http://arxiv.org/abs/1606.01857}{{\ttfamily arXiv:1606.01857 [hep-th]}}.

\bibitem{Engelsoy:2016xyb}
J.~Engelsoy, T.~G. Mertens, and H.~Verlinde, ``{An investigation of AdS$_{2}$
  backreaction and holography},''
  \href{http://dx.doi.org/10.1007/JHEP07(2016)139}{{\em JHEP} {\bfseries 07}
  (2016) 139},
\href{http://arxiv.org/abs/1606.03438}{{\ttfamily arXiv:1606.03438 [hep-th]}}.

\bibitem{Grumiller:2016dbn}
D.~Grumiller, J.~Salzer, and D.~Vassilevich, ``{Aspects of AdS$_2$ holography
  with non-constant dilaton},''
  \href{http://dx.doi.org/10.1007/s11182-017-0978-x}{{\em Russ. Phys. J.}
  {\bfseries 59} no.~11, (2017) 1798--1803},
  \href{http://arxiv.org/abs/1607.06974}{{\ttfamily arXiv:1607.06974
  [hep-th]}}.

\bibitem{Cvetic:2016eiv}
M.~Cvetic and I.~Papadimitriou, ``{AdS$_{2}$ holographic dictionary},''
  \href{http://dx.doi.org/10.1007/JHEP12(2016)008}{{\em JHEP} {\bfseries 12}
  (2016) 008}, \href{http://arxiv.org/abs/1608.07018}{{\ttfamily
  arXiv:1608.07018 [hep-th]}}. [Erratum: JHEP 01, 120 (2017)].

\bibitem{Mertens:2017mtv}
T.~G. Mertens, G.~J. Turiaci, and H.~L. Verlinde, ``{Solving the Schwarzian via
  the Conformal Bootstrap},''
  \href{http://dx.doi.org/10.1007/JHEP08(2017)136}{{\em JHEP} {\bfseries 08}
  (2017) 136},
\href{http://arxiv.org/abs/1705.08408}{{\ttfamily arXiv:1705.08408 [hep-th]}}.

\bibitem{Lam:2018pvp}
H.~T. Lam, T.~G. Mertens, G.~J. Turiaci, and H.~Verlinde, ``{Shockwave S-matrix
  from Schwarzian Quantum Mechanics},''
  \href{http://dx.doi.org/10.1007/JHEP11(2018)182}{{\em JHEP} {\bfseries 11}
  (2018) 182},
\href{http://arxiv.org/abs/1804.09834}{{\ttfamily arXiv:1804.09834 [hep-th]}}.

\bibitem{Harlow:2018tqv}
D.~Harlow and D.~Jafferis, ``{The Factorization Problem in Jackiw-Teitelboim
  Gravity},'' \href{http://dx.doi.org/10.1007/JHEP02(2020)177}{{\em JHEP}
  {\bfseries 02} (2020) 177}, \href{http://arxiv.org/abs/1804.01081}{{\ttfamily
  arXiv:1804.01081 [hep-th]}}.

\bibitem{Blommaert:2018oro}
A.~Blommaert, T.~G. Mertens, and H.~Verschelde, ``{The Schwarzian Theory - A
  Wilson Line Perspective},''
  \href{http://dx.doi.org/10.1007/JHEP12(2018)022}{{\em JHEP} {\bfseries 12}
  (2018) 022},
\href{http://arxiv.org/abs/1806.07765}{{\ttfamily arXiv:1806.07765 [hep-th]}}.

\bibitem{Kitaev:2018wpr}
A.~Kitaev and S.~J. Suh, ``{Statistical mechanics of a two-dimensional black
  hole},'' \href{http://dx.doi.org/10.1007/JHEP05(2019)198}{{\em JHEP}
  {\bfseries 05} (2019) 198},
\href{http://arxiv.org/abs/1808.07032}{{\ttfamily arXiv:1808.07032 [hep-th]}}.

\bibitem{Yang:2018gdb}
Z.~Yang, ``{The Quantum Gravity Dynamics of Near Extremal Black Holes},''
  \href{http://dx.doi.org/10.1007/JHEP05(2019)205}{{\em JHEP} {\bfseries 05}
  (2019) 205},
\href{http://arxiv.org/abs/1809.08647}{{\ttfamily arXiv:1809.08647 [hep-th]}}.

\bibitem{Iliesiu:2019xuh}
L.~V. Iliesiu, S.~S. Pufu, H.~Verlinde, and Y.~Wang, ``{An exact quantization
  of Jackiw-Teitelboim gravity},''
\href{http://arxiv.org/abs/1905.02726}{{\ttfamily arXiv:1905.02726 [hep-th]}}.

\bibitem{Ghosh:2019rcj}
A.~Ghosh, H.~Maxfield, and G.~J. Turiaci, ``{A universal Schwarzian sector in
  two-dimensional conformal field theories},''
\href{http://arxiv.org/abs/1912.07654}{{\ttfamily arXiv:1912.07654 [hep-th]}}.

\bibitem{Dabholkar:2014ema}
A.~Dabholkar, J.~Gomes, and S.~Murthy, ``{Nonperturbative black hole entropy
  and Kloosterman sums},''
  \href{http://dx.doi.org/10.1007/JHEP03(2015)074}{{\em JHEP} {\bfseries 03}
  (2015) 074},
\href{http://arxiv.org/abs/1404.0033}{{\ttfamily arXiv:1404.0033 [hep-th]}}.

\bibitem{Sorkin:1983ns}
R.~Sorkin, ``{Kaluza-Klein Monopole},''
  \href{http://dx.doi.org/10.1103/PhysRevLett.51.87}{{\em Phys. Rev. Lett.}
  {\bfseries 51} (1983) 87--90}.

\bibitem{Townsend:1995kk}
P.~Townsend, ``{The eleven-dimensional supermembrane revisited},''
  \href{http://dx.doi.org/10.1016/0370-2693(95)00397-4}{{\em Phys. Lett. B}
  {\bfseries 350} (1995) 184--187},
  \href{http://arxiv.org/abs/hep-th/9501068}{{\ttfamily arXiv:hep-th/9501068}}.

\bibitem{Mertens:2019tcm}
T.~G. Mertens and G.~J. Turiaci, ``{Defects in Jackiw-Teitelboim Quantum
  Gravity},''
\href{http://arxiv.org/abs/1904.05228}{{\ttfamily arXiv:1904.05228 [hep-th]}}.

\bibitem{Stanford:2019vob}
D.~Stanford and E.~Witten, ``{JT Gravity and the Ensembles of Random Matrix
  Theory},'' \href{http://arxiv.org/abs/1907.03363}{{\ttfamily arXiv:1907.03363
  [hep-th]}}.

\bibitem{Mirzakhani:2006fta}
M.~Mirzakhani, ``{Simple geodesics and Weil-Petersson volumes of moduli spaces
  of bordered Riemann surfaces},''
\href{http://dx.doi.org/10.1007/s00222-006-0013-2}{{\em Invent. Math.}
  {\bfseries 167} no.~1, (2006) 179--222}.

\bibitem{Mirzakhani:2006eta}
M.~Mirzakhani, ``{Weil-Petersson volumes and intersection theory on the moduli
  space of curves},''
\href{http://dx.doi.org/10.1090/S0894-0347-06-00526-1}{{\em J. Am. Math. Soc.}
  {\bfseries 20} no.~01, (2007) 1--24}.

\bibitem{tan2004generalizations}
S.~P. Tan, Y.~L. Wong, and Y.~Zhang, ``Generalizations of mcshane's identity to
  hyperbolic cone-surfaces,''
  \href{http://arxiv.org/abs/math/0404226}{{\ttfamily arXiv:math/0404226
  [math.GT]}}.

\bibitem{do2006weilpetersson}
N.~Do and P.~Norbury, ``Weil-petersson volumes and cone surfaces,''
  \href{http://arxiv.org/abs/math/0603406}{{\ttfamily arXiv:math/0603406
  [math.AG]}}.

\bibitem{do2011moduli}
N.~Do, ``Moduli spaces of hyperbolic surfaces and their weil-petersson
  volumes,'' \href{http://arxiv.org/abs/1103.4674}{{\ttfamily arXiv:1103.4674
  [math.GT]}}.

\bibitem{Cotler:2019nbi}
J.~Cotler, K.~Jensen, and A.~Maloney, ``{Low-dimensional de Sitter quantum
  gravity},''
\href{http://arxiv.org/abs/1905.03780}{{\ttfamily arXiv:1905.03780 [hep-th]}}.

\bibitem{mtms}
T.~G. Mertens and G.~J. Turiaci, ``{Liouville quantum gravity -- holography, JT
  and matrices},'' \href{http://arxiv.org/abs/2006.07072}{{\ttfamily
  arXiv:2006.07072 [hep-th]}}.

\bibitem{Eynard:2007kz}
B.~Eynard and N.~Orantin, ``{Invariants of algebraic curves and topological
  expansion},'' \href{http://dx.doi.org/10.4310/CNTP.2007.v1.n2.a4}{{\em
  Commun. Num. Theor. Phys.} {\bfseries 1} (2007) 347--452},
  \href{http://arxiv.org/abs/math-ph/0702045}{{\ttfamily
  arXiv:math-ph/0702045}}.

\bibitem{Witten:2020wvy}
E.~Witten, ``{Matrix Models and Deformations of JT Gravity},''
  \href{http://arxiv.org/abs/2006.13414}{{\ttfamily arXiv:2006.13414
  [hep-th]}}.

\bibitem{Achucarro:1993fd}
A.~Achucarro and M.~E. Ortiz, ``{Relating black holes in two-dimensions and
  three-dimensions},'' \href{http://dx.doi.org/10.1103/PhysRevD.48.3600}{{\em
  Phys. Rev.} {\bfseries D48} (1993) 3600--3605},
\href{http://arxiv.org/abs/hep-th/9304068}{{\ttfamily arXiv:hep-th/9304068
  [hep-th]}}.

\bibitem{Coussaert:1994tu}
O.~Coussaert and M.~Henneaux, ``{Selfdual solutions of (2+1) Einstein gravity
  with a negative cosmological constant},'' in {\em {The Black Hole 25 Years
  After}}, pp.~25--39.
\newblock 1, 1994.
\newblock \href{http://arxiv.org/abs/hep-th/9407181}{{\ttfamily
  arXiv:hep-th/9407181}}.

\bibitem{Brown:1986nw}
J.~D. Brown and M.~Henneaux, ``{Central Charges in the Canonical Realization of
  Asymptotic Symmetries: An Example from Three-Dimensional Gravity},''
\href{http://dx.doi.org/10.1007/BF01211590}{{\em Commun. Math. Phys.}
  {\bfseries 104} (1986) 207--226}.

\bibitem{Eynard:2004mh}
B.~Eynard, ``{Topological expansion for the 1-Hermitian matrix model
  correlation functions},''
  \href{http://dx.doi.org/10.1088/1126-6708/2004/11/031}{{\em JHEP} {\bfseries
  11} (2004) 031}, \href{http://arxiv.org/abs/hep-th/0407261}{{\ttfamily
  arXiv:hep-th/0407261}}.

\bibitem{Banks:1990mk}
T.~Banks and M.~O'Loughlin, ``{Two-dimensional quantum gravity in Minkowski
  space},'' \href{http://dx.doi.org/10.1016/0550-3213(91)90547-B}{{\em Nucl.
  Phys. B} {\bfseries 362} (1991) 649--664}.

\bibitem{ITZYKSON_1992}
C.~Itzykson and J.-B. Zuber, ``Combinatorics of the modular group ii: the
  kontsevich integrals,''
  \href{http://dx.doi.org/10.1142/s0217751x92002581}{{\em International Journal
  of Modern Physics A} {\bfseries 07} no.~23, (Sep, 1992) 5661–5705}.

\bibitem{Brezin:1990rb}
E.~Brezin and V.~Kazakov, ``{Exactly Solvable Field Theories of Closed
  Strings},'' \href{http://dx.doi.org/10.1016/0370-2693(90)90818-Q}{{\em Phys.
  Lett. B} {\bfseries 236} (1990) 144--150}.

\bibitem{Douglas:1989ve}
M.~R. Douglas and S.~H. Shenker, ``{Strings in Less Than One-Dimension},''
  \href{http://dx.doi.org/10.1016/0550-3213(90)90522-F}{{\em Nucl. Phys. B}
  {\bfseries 335} (1990) 635}.

\bibitem{Gross:1989vs}
D.~J. Gross and A.~A. Migdal, ``{Nonperturbative Two-Dimensional Quantum
  Gravity},'' \href{http://dx.doi.org/10.1103/PhysRevLett.64.127}{{\em Phys.
  Rev. Lett.} {\bfseries 64} (1990) 127}.

\bibitem{Banks:1989df}
T.~Banks, M.~R. Douglas, N.~Seiberg, and S.~H. Shenker, ``{Microscopic and
  Macroscopic Loops in Nonperturbative Two-dimensional Gravity},''
  \href{http://dx.doi.org/10.1016/0370-2693(90)91736-U}{{\em Phys. Lett. B}
  {\bfseries 238} (1990) 279}.

\bibitem{Ambjorn:1990ji}
J.~Ambjorn, J.~Jurkiewicz, and Y.~Makeenko, ``{Multiloop correlators for
  two-dimensional quantum gravity},''
  \href{http://dx.doi.org/10.1016/0370-2693(90)90790-D}{{\em Phys. Lett. B}
  {\bfseries 251} (1990) 517--524}.

\bibitem{Moore:1991ir}
G.~W. Moore, N.~Seiberg, and M.~Staudacher, ``{From loops to states in 2-D
  quantum gravity},''
  \href{http://dx.doi.org/10.1016/0550-3213(91)90548-C}{{\em Nucl. Phys. B}
  {\bfseries 362} (1991) 665--709}.

\bibitem{Okuyama:2019xbv}
K.~Okuyama and K.~Sakai, ``{JT gravity, KdV equations and macroscopic loop
  operators},'' \href{http://dx.doi.org/10.1007/JHEP01(2020)156}{{\em JHEP}
  {\bfseries 01} (2020) 156},
\href{http://arxiv.org/abs/1911.01659}{{\ttfamily arXiv:1911.01659 [hep-th]}}.

\bibitem{Okuyama:2020ncd}
K.~Okuyama and K.~Sakai, ``{Multi-boundary correlators in JT gravity},''
\href{http://arxiv.org/abs/2004.07555}{{\ttfamily arXiv:2004.07555 [hep-th]}}.

\bibitem{Johnson:2019eik}
C.~V. Johnson, ``{Non-Perturbative JT Gravity},''
  \href{http://arxiv.org/abs/1912.03637}{{\ttfamily arXiv:1912.03637
  [hep-th]}}.

\bibitem{Eynard:2015aea}
B.~Eynard, T.~Kimura, and S.~Ribault, ``{Random matrices},''
  \href{http://arxiv.org/abs/1510.04430}{{\ttfamily arXiv:1510.04430
  [math-ph]}}.

\bibitem{Eynard:2014zxa}
B.~Eynard, ``{A short overview of the "Topological recursion"},''
  \href{http://arxiv.org/abs/1412.3286}{{\ttfamily arXiv:1412.3286 [math-ph]}}.

\bibitem{Eynard:2007fi}
B.~Eynard and N.~Orantin, ``{Weil-Petersson volume of moduli spaces,
  Mirzakhani's recursion and matrix models},''
  \href{http://arxiv.org/abs/0705.3600}{{\ttfamily arXiv:0705.3600 [math-ph]}}.

\bibitem{higherdim}
L.~V. Iliesiu and G.~J. Turiaci, ``{The statistical mechanics of near-extremal
  black holes},''
\href{http://arxiv.org/abs/2003.02860}{{\ttfamily arXiv:2003.02860 [hep-th]}}.

\bibitem{CotlerJensenDoubleTrumpet}
J.~Cotler and K.~Jensen, ``{AdS$_3$ gravity and random CFT},''
  \href{http://arxiv.org/abs/2006.08648}{{\ttfamily arXiv:2006.08648
  [hep-th]}}.

\bibitem{Kitaev:2017awl}
A.~Kitaev and S.~J. Suh, ``{The soft mode in the Sachdev-Ye-Kitaev model and
  its gravity dual},'' \href{http://dx.doi.org/10.1007/JHEP05(2018)183}{{\em
  JHEP} {\bfseries 05} (2018) 183},
\href{http://arxiv.org/abs/1711.08467}{{\ttfamily arXiv:1711.08467 [hep-th]}}.

\bibitem{Almheiri:2019qdq}
A.~Almheiri, T.~Hartman, J.~Maldacena, E.~Shaghoulian, and A.~Tajdini,
  ``{Replica Wormholes and the Entropy of Hawking Radiation},''
  \href{http://dx.doi.org/10.1007/JHEP05(2020)013}{{\em JHEP} {\bfseries 05}
  (2020) 013},
\href{http://arxiv.org/abs/1911.12333}{{\ttfamily arXiv:1911.12333 [hep-th]}}.

\bibitem{Penington:2019kki}
G.~Penington, S.~H. Shenker, D.~Stanford, and Z.~Yang, ``{Replica wormholes and
  the black hole interior},''
\href{http://arxiv.org/abs/1911.11977}{{\ttfamily arXiv:1911.11977 [hep-th]}}.

\bibitem{Coleman:1988cy}
S.~R. Coleman, ``{Black Holes as Red Herrings: Topological Fluctuations and the
  Loss of Quantum Coherence},''
  \href{http://dx.doi.org/10.1016/0550-3213(88)90110-1}{{\em Nucl. Phys. B}
  {\bfseries 307} (1988) 867--882}.

\bibitem{Giddings:1988cx}
S.~B. Giddings and A.~Strominger, ``{Loss of Incoherence and Determination of
  Coupling Constants in Quantum Gravity},''
  \href{http://dx.doi.org/10.1016/0550-3213(88)90109-5}{{\em Nucl. Phys. B}
  {\bfseries 307} (1988) 854--866}.

\bibitem{Giddings:1988wv}
S.~B. Giddings and A.~Strominger, ``{Baby Universes, Third Quantization and the
  Cosmological Constant},''
  \href{http://dx.doi.org/10.1016/0550-3213(89)90353-2}{{\em Nucl. Phys. B}
  {\bfseries 321} (1989) 481--508}.

\bibitem{Marolf:2020xie}
D.~Marolf and H.~Maxfield, ``{Transcending the ensemble: baby universes,
  spacetime wormholes, and the order and disorder of black hole information},''
  \href{http://arxiv.org/abs/2002.08950}{{\ttfamily arXiv:2002.08950
  [hep-th]}}.

\bibitem{Maloney:2020nni}
A.~Maloney and E.~Witten, ``{Averaging Over Narain Moduli Space},''
  \href{http://arxiv.org/abs/2006.04855}{{\ttfamily arXiv:2006.04855
  [hep-th]}}.

\bibitem{Afkhami-Jeddi:2020ezh}
N.~Afkhami-Jeddi, H.~Cohn, T.~Hartman, and A.~Tajdini, ``{Free partition
  functions and an averaged holographic duality},''
  \href{http://arxiv.org/abs/2006.04839}{{\ttfamily arXiv:2006.04839
  [hep-th]}}.

\bibitem{Cardy:2017qhl}
J.~Cardy, A.~Maloney, and H.~Maxfield, ``{A new handle on three-point
  coefficients: OPE asymptotics from genus two modular invariance},''
  \href{http://dx.doi.org/10.1007/JHEP10(2017)136}{{\em JHEP} {\bfseries 10}
  (2017) 136}, \href{http://arxiv.org/abs/1705.05855}{{\ttfamily
  arXiv:1705.05855 [hep-th]}}.

\bibitem{Collier:2019weq}
S.~Collier, A.~Maloney, H.~Maxfield, and I.~Tsiares, ``{Universal Dynamics of
  Heavy Operators in CFT$_2$},''
\href{http://arxiv.org/abs/1912.00222}{{\ttfamily arXiv:1912.00222 [hep-th]}}.

\bibitem{Belin:2020hea}
A.~Belin and J.~de~Boer, ``{Random Statistics of OPE Coefficients and Euclidean
  Wormholes},'' \href{http://arxiv.org/abs/2006.05499}{{\ttfamily
  arXiv:2006.05499 [hep-th]}}.

\bibitem{Saad:2019pqd}
P.~Saad, ``{Late Time Correlation Functions, Baby Universes, and ETH in JT
  Gravity},''
\href{http://arxiv.org/abs/1910.10311}{{\ttfamily arXiv:1910.10311 [hep-th]}}.

\bibitem{Pollack:2020gfa}
J.~Pollack, M.~Rozali, J.~Sully, and D.~Wakeham, ``{Eigenstate Thermalization
  and Disorder Averaging in Gravity},''
  \href{http://arxiv.org/abs/2002.02971}{{\ttfamily arXiv:2002.02971
  [hep-th]}}.

\bibitem{Collier:2018exn}
S.~Collier, Y.~Gobeil, H.~Maxfield, and E.~Perlmutter, ``{Quantum Regge
  Trajectories and the Virasoro Analytic Bootstrap},''
\href{http://arxiv.org/abs/1811.05710}{{\ttfamily arXiv:1811.05710 [hep-th]}}.

\bibitem{Murugan:2017eto}
J.~Murugan, D.~Stanford, and E.~Witten, ``{More on Supersymmetric and 2d
  Analogs of the SYK Model},''
  \href{http://dx.doi.org/10.1007/JHEP08(2017)146}{{\em JHEP} {\bfseries 08}
  (2017) 146},
\href{http://arxiv.org/abs/1706.05362}{{\ttfamily arXiv:1706.05362 [hep-th]}}.

\bibitem{Berkooz:2016cvq}
M.~Berkooz, P.~Narayan, M.~Rozali, and J.~Simon, ``{Higher Dimensional
  Generalizations of the SYK Model},''
  \href{http://dx.doi.org/10.1007/JHEP01(2017)138}{{\em JHEP} {\bfseries 01}
  (2017) 138},
\href{http://arxiv.org/abs/1610.02422}{{\ttfamily arXiv:1610.02422 [hep-th]}}.

\bibitem{Turiaci:2017zwd}
G.~Turiaci and H.~Verlinde, ``{Towards a 2d QFT Analog of the SYK Model},''
  \href{http://dx.doi.org/10.1007/JHEP10(2017)167}{{\em JHEP} {\bfseries 10}
  (2017) 167},
\href{http://arxiv.org/abs/1701.00528}{{\ttfamily arXiv:1701.00528 [hep-th]}}.

\bibitem{Cotler:2018zff}
J.~Cotler and K.~Jensen, ``{A theory of reparameterizations for AdS$_3$
  gravity},'' \href{http://dx.doi.org/10.1007/JHEP02(2019)079}{{\em JHEP}
  {\bfseries 02} (2019) 079},
\href{http://arxiv.org/abs/1808.03263}{{\ttfamily arXiv:1808.03263 [hep-th]}}.

\bibitem{Witten:2007kt}
E.~Witten, ``{Three-Dimensional Gravity Revisited},''
  \href{http://arxiv.org/abs/0706.3359}{{\ttfamily arXiv:0706.3359 [hep-th]}}.

\bibitem{Beasley:2005vf}
C.~Beasley and E.~Witten, ``{Non-Abelian localization for Chern-Simons
  theory},'' {\em J. Diff. Geom.} {\bfseries 70} no.~2, (2005) 183--323,
  \href{http://arxiv.org/abs/hep-th/0503126}{{\ttfamily arXiv:hep-th/0503126}}.

\bibitem{Blau:2013oha}
M.~Blau and G.~Thompson, ``{Chern-Simons Theory on Seifert 3-Manifolds},''
  \href{http://dx.doi.org/10.1007/JHEP09(2013)033}{{\em JHEP} {\bfseries 09}
  (2013) 033}, \href{http://arxiv.org/abs/1306.3381}{{\ttfamily arXiv:1306.3381
  [hep-th]}}.

\bibitem{Kusuki:2018wpa}
Y.~Kusuki, ``{Light Cone Bootstrap in General 2D CFTs and Entanglement from
  Light Cone Singularity},''
  \href{http://dx.doi.org/10.1007/JHEP01(2019)025}{{\em JHEP} {\bfseries 01}
  (2019) 025}, \href{http://arxiv.org/abs/1810.01335}{{\ttfamily
  arXiv:1810.01335 [hep-th]}}.

\bibitem{Grumiller:2007ju}
D.~Grumiller and R.~McNees, ``{Thermodynamics of black holes in two (and
  higher) dimensions},''
  \href{http://dx.doi.org/10.1088/1126-6708/2007/04/074}{{\em JHEP} {\bfseries
  04} (2007) 074}, \href{http://arxiv.org/abs/hep-th/0703230}{{\ttfamily
  arXiv:hep-th/0703230}}.

\bibitem{Kyono:2017jtc}
H.~Kyono, S.~Okumura, and K.~Yoshida, ``{Deformations of the
  Almheiri-Polchinski model},''
  \href{http://dx.doi.org/10.1007/JHEP03(2017)173}{{\em JHEP} {\bfseries 03}
  (2017) 173}, \href{http://arxiv.org/abs/1701.06340}{{\ttfamily
  arXiv:1701.06340 [hep-th]}}.

\bibitem{Witten:2020ert}
E.~Witten, ``{Deformations of JT Gravity and Phase Transitions},''
  \href{http://arxiv.org/abs/2006.03494}{{\ttfamily arXiv:2006.03494
  [hep-th]}}.

\bibitem{Anninos:2017hhn}
D.~Anninos and D.~M. Hofman, ``{Infrared Realization of dS$_2$ in AdS$_2$},''
  \href{http://dx.doi.org/10.1088/1361-6382/aab143}{{\em Class. Quant. Grav.}
  {\bfseries 35} no.~8, (2018) 085003},
  \href{http://arxiv.org/abs/1703.04622}{{\ttfamily arXiv:1703.04622
  [hep-th]}}.

\bibitem{Anninos:2018svg}
D.~Anninos, D.~A. Galante, and D.~M. Hofman, ``{De Sitter Horizons \&
  Holographic Liquids},'' \href{http://dx.doi.org/10.1007/JHEP07(2019)038}{{\em
  JHEP} {\bfseries 07} (2019) 038},
  \href{http://arxiv.org/abs/1811.08153}{{\ttfamily arXiv:1811.08153
  [hep-th]}}.

\bibitem{Maldacena:2019cbz}
J.~Maldacena, G.~J. Turiaci, and Z.~Yang, ``{Two dimensional Nearly de Sitter
  gravity},''
\href{http://arxiv.org/abs/1904.01911}{{\ttfamily arXiv:1904.01911 [hep-th]}}.

\bibitem{Iliesiu:2020zld}
L.~V. Iliesiu, J.~Kruthoff, G.~J. Turiaci, and H.~Verlinde, ``{JT gravity at
  finite cutoff},'' \href{http://arxiv.org/abs/2004.07242}{{\ttfamily
  arXiv:2004.07242 [hep-th]}}.

\end{mcitethebibliography}\endgroup

\end{document}